\documentclass[%
 reprint,
 amsmath,amssymb,
 aps, prx
]{revtex4-2}

\usepackage{graphicx}
\usepackage{float, booktabs}
\usepackage{dcolumn}
\usepackage{bm}

\begin{document}

\preprint{APS/123-QED}

\title{Self organisation of invasive breast cancer driven by the interplay of active and passive nematic dynamics}
\author{Pablo Gottheil$^{1,\dagger}$}
\author{Saraswat Bhattacharyya$^{2,\dagger}$}
\author{Philip Friedrich$^{1}$}
\author{Kilian Roth$^{1,3}$}
\author{Mario Merkel$^{1}$}
\author{Kolya Lettl$^{1}$}
\author{Kudor Harb$^{4}$}
\author{Salvador Rivera-Moreno$^{1}$}
\author{Assal Daneshgar$^{3}$}
\author{Jonas Wieland$^{3}$}
\author{Bahriye Aktas$^{5}$}
\author{Jens-Uwe Blohmer$^{6}$}
\author{Hans Kubitschke$^{1}$}
\author{Anne-Sophie Wegscheider$^{7}$}
\author{Christine Sers$^{4}$}
\author{Igor M. Sauer$^{3}$}
\author{Armin Jarosch$^{4}$}
\author{Axel Niendorf$^{7}$}
\author{Julia M. Yeomans$^{2,\ddagger}$}
\author{Josef A. Käs$^{1,\ddagger}$}
\thanks{$^{\dagger}$ These authors contributed equally. $^{\ddagger}$Correspondence: jkaes@uni-leipzig.de, julia.yeomans@physics.ox.ac.uk. }
\makeatletter
\renewcommand{\frontmatter@makefnmark}{}
\makeatother

\affiliation{$^{1}$Peter Debye Institute for Soft Matter Physics, University of Leipzig\\ Linnéstraße 5, 04103, Leipzig, Saxony, Germany}

\affiliation{$^{2}$Rudolf Peierls Centre for Theoretical Physics, University of Oxford\\ Parks Road, Oxford, OX1 3PU, Oxfordshire, United Kingdom}

\affiliation{$^{3}$ Department of Surgery, Experimental Surgery, CCM|CVK, Charit\'{e} – Universitätsmedizin Berlin, Augustenburger Platz 1, 13353, Berlin, Germany}

\affiliation{$^{4}$ Charité – Universitätsmedizin Berlin, Institute of Pathology, Charitéplatz 1, 10117 Berlin, Germany ($^{+}$corporate member of Freie Universität Berlin and Humboldt-Universität zu Berlin)}

\affiliation{$^{5}$Department of Gynaecology, University Clinic Leipzig\\ Liebigstraße 20a, 04103, Leipzig, Saxony, Germany}

\affiliation{$^{6}$ Gynäkologie mit Brustzentrum, Charité-Universitätsmedizin Berlin, Berlin, Germany}

\affiliation{$^{7}$Pathology Hamburg-West, Medical Care Center\\ Lornsenstraße 4, 22767, Hamburg, Hamburg, Germany}

\begin{abstract}
In invasive breast carcinomas, active cancer cell clusters of varying sizes and
shapes are embedded in a nematic extracellular matrix (ECM) of aligned fibres.
Although the tissue patterns are regularly used in diagnosis, the physical mechanisms by which these mesoscale structures emerge on long time scales remain unexplored, partly because in-vitro or in-vivo experiments over the timescale of cancer progression ($\sim$ months-years) are not feasible. To overcome this fundamental problem in studying tumour dynamics, we model the cancer clusters as active nematic aggregates moving through the ECM, which acts as a confining, passive nematic phase. The approach is verified by comparison to histological images showing hundreds of thousands of cancer clusters and their micro-environments from over 2,000 breast cancer patients. We predict that collective cancer cell motion leads to cancer cluster splitting and coalescence, which is manifested in specific scaling behaviours for cancer cluster shape and size distributions. Active stresses from cancer cluster movement give rise to topological defects in the ECM and tend to align the passive fibres parallel to the cluster boundary. We further show that cluster shapes and defect distributions as measures of increased activity are associated with poorer survival probabilities for breast cancer patients. Our work provides a novel way to study human breast cancer by linking long-timescale, data-driven simulations with large-scale patient histology.
\end{abstract}

\maketitle

\section{\label{sec:level1} Introduction}

\begin{figure*} [htp]
    \centering   
    \includegraphics[width=1.0\textwidth]{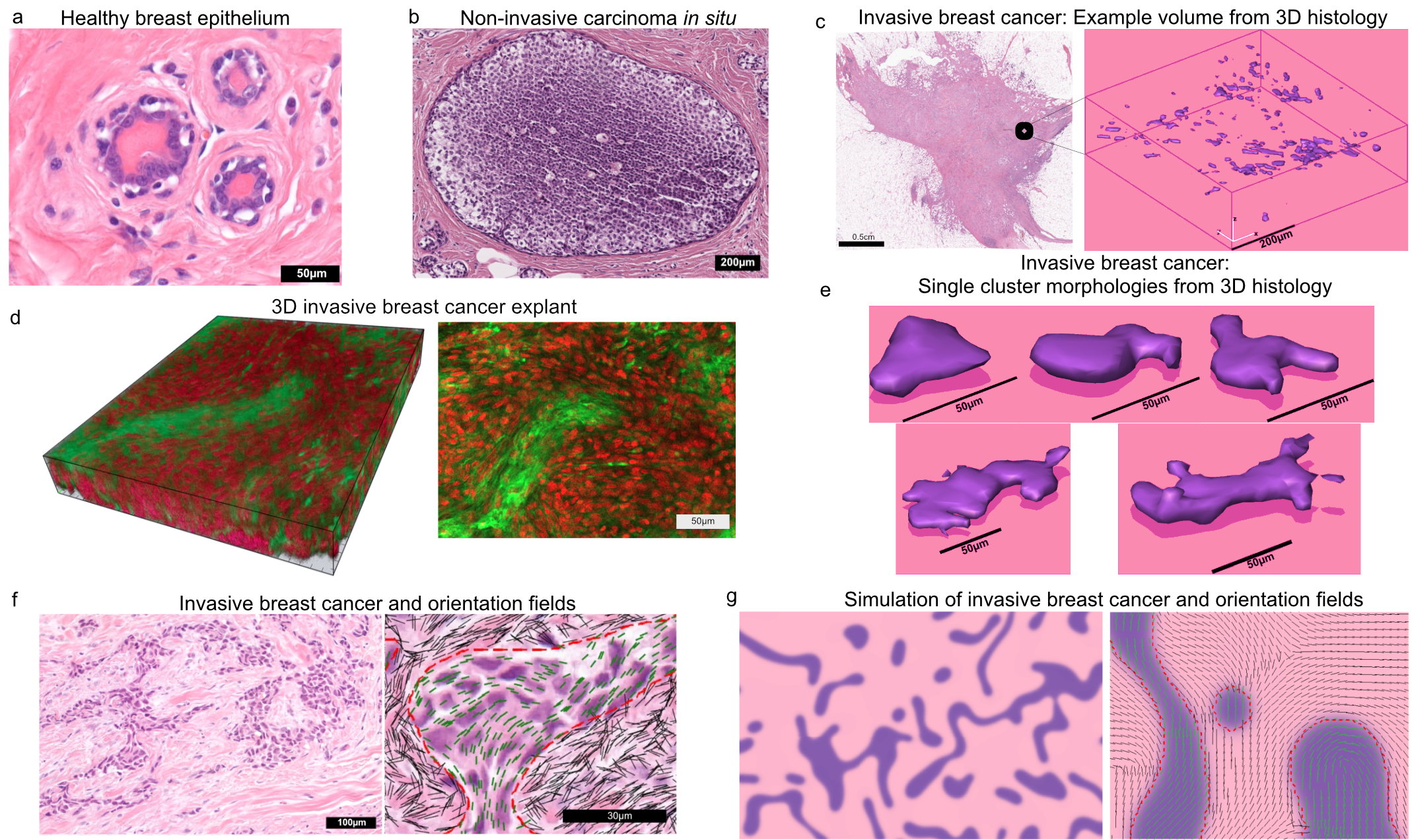}
    \caption{\textbf{Structure of solid breast tumours.} Haematoxylin- and Eosin- (H\&E-) staining shows cell nuclei in purple and extracellular matrix (ECM) in pink. {\bf a} Healthy breast epithelium confined by ECM fibres forming a basement membrane.  {\bf b} The ductal carcinoma \textit{in situ} (DCIS) is an early non-invasive tumour defined by uncontrolled growth of neoplastic epithelial cells only in the (milk) duct, still confined by an intact basement membrane. {\bf c} Macroscopic view of a 2D histological slide of invasive breast cancer (left side, complete tumour ca. 2.5cm) and exemplary 3D cancer cluster reconstruction of a small window (right side). Purple surface renderings indicate the 3D (disconnected) clusters of invasive breast cancer ($600 \mu \text{m} \times 600 \mu \text{m} \times 180 \mu \text{m}$). {\bf d} 3D confocal images of primary breast cancer ($316 \mu \text{m} \times 253 \mu \text{m} \times 40 \mu \text{m}$) stained for DNA (red) and actin (green) exhibits 3D cell nematic order (Appendix \ref{app:cell_dimensionality}). {\bf e} Renderings of single clusters found from reconstructing 3D histology from serial sectioning. The clusters exhibit varying sizes and shapes. {\bf f left:} Consistent with the 3D observations, 2D histological view of invasive breast cancer exhibiting more complex structures than ductal \textit{in situ} carcinoma: Histological image (H\&E stained) of the typical structure of an invasive primary breast tumour: Many smaller, highly irregular cancer cell clusters with well-defined boundaries are embedded in the ECM. \textbf{f} \textbf{right} : Zoom into an invasive breast cancer lesion demonstrating the highly aligned state of elongated motile cells embedded into passive aligned fibres of the ECM. {\bf g} The assembly of this heterogenous tumour structure can be understood by simulations that rely on liquid crystal physics describing an active nematic phase embedded in a passive nematic matrix.  \textbf{g} \textbf{left} : simulation showing cancer clusters (purple) and the confining ECM (pink). \textbf{g} \textbf{right} : Zooming into the respective orientation fields in the simulated clusters, the cells form aligned streams of moving cells (green lines). The surrounding ECM consists of regions of collagen fibres with local alignment (black lines). Note the +1/2 topological defect (Fig. \ref{fig:fingerprintsECM}b) in the cluster in the lower right corner in sub-figure g.}\label{qualitativeClusterECM}
\end{figure*}

Healthy breast tissue is composed of two primary components, the glandular tissue which contains the milk ducts where the production and excretion of milk take place, and the surrounding connective tissue which is comprised of adipose tissue and collagen-rich extra-cellular matrix (ECM) (Fig.~\ref{qualitativeClusterECM}a). Breast cancer typically originates within the ducts. In its earliest stage cancerous cells proliferate to form a growing cell mass, but are confined to the duct by an intact and highly crosslinked ECM layer, called the basement membrane. The confined tumour mass is referred to as ductal carcinoma in situ (DCIS) \cite{burstein2004ductal} (Fig.~\ref{qualitativeClusterECM}b), and corresponds to an early, and less aggressive, stage of the disease.\\

The transition to invasive breast cancer is defined by the breach of the basement membrane, allowing cancer cells to leave the duct and invade the surrounding ECM. This step is accompanied by the cells undergoing an epithelial-to-mesenchymal transition through which epithelial cancer cells acquire an elongated, motile, mesenchymal-like phenotype \cite{kroger2019hybrid}. In breast cancer this transition is typically only partial \cite{kroger2019hybrid}, so that cells become migratory while remaining mechanically coupled enough to move as cohesive groups rather than scattering as single cells. The resulting invasive lesion has a markedly different architecture to the DCIS (Fig.~\ref{qualitativeClusterECM}c): instead of a single confined mass, many smaller, heterogeneous clusters of aligned cancer cells are dispersed through a dense, fibrotic ECM of locally aligned collagen fibres \cite{insua2016extracellular} (Fig.~\ref{qualitativeClusterECM}d,e,f). This spatial pattern of the clusters and extracellular matrix in invasive breast cancer - cluster size, shape, and the organisation of the surrounding fibres - is routinely read by pathologists as a diagnostic of invasiveness and informs their tumour assessment \cite{mallon2000basic}. Yet how these mesoscale patterns dynamically emerge, and what physical processes govern them, is not understood.\\

We have recently shown that cells within a cluster can undergo a collective cell motility transition \cite{gottheil2023state} in addition to the processes of cell proliferation and ECM production. However, how self organisation dynamically generates the heterogeneous and complex tumour structure, which serves as a clinical signature for the onset of malignancy \cite{mallon2000basic}, has remained unclear. This is because the long time scales of  \textit{in vivo} dynamics relevant to disease progression remain inaccessible for direct observation by techniques such as vital imaging in patient derived tumour explants \cite{grosser2021cell}, mice models \cite{chepizhko2025confined}, organoids \cite{kang2021novel} and other tumour models.\\

 Specifically, ductal carcinomas \textit{in situ} take median time spans of 6-11 years to progress to (clinically observable) invasive breast cancer \cite{wallis2012effect}, although relevant individual processes may ``only'' take weeks to months. Here we present an approach to solve this dilemma by combining extensive image analysis of tumour patient data, acting as static snapshots of real tumours, and dynamical simulations using realistic symmetries and material parameters obtained from direct measurements of vital  primary tumours after surgery. With this approach we aim to reveal the processes that self-organize invasive lesions.
We compare extensive histological images of invasive breast cancer patients to continuum modelling, showing that the mechanical interplay that leads to early invasive lesions can be understood using concepts from liquid crystal physics. In particular, the local alignment of ECM fibres is characteristic of nematic ordering, where the term nematic denotes a preferred alignment axis \cite{degennes_book}. Moreover, recent work has shown that cellular motility transitions can lead to streams of cells in the clusters \cite{fuhs2022rigid, ilina2020cell}. The collectively moving cells can be described as an active nematic, where \textit{active} indicates that the cells are motile \cite{doostmohammadi2018active, SimhaRamaswamy2002, saw2018biological}. As the cells move, they generate active stresses across cluster boundaries, resulting in a dynamical steady state of clusters deforming, breaking up and merging. Hence the distinctive pattern of invasive tumours highlights the role of collective cell motility.\\

Our results establish a quantitative bridge between active matter physics and clinical oncology. We describe tumour-wide organisation of invasive breast cancer across the years-long progression - showing how nematic cell collectives self-organize and influence the surrounding ECM. This single physically-grounded model reproduces multiple independent features of invasive breast cancer across a cohort of more than 2,000 patients and, at the same time, suggests clinically relevant prognostic markers.
\section{Histological images reveal nematic order in cancer cell aggregates and surrounding ECM\label{sec2}}
To quantitatively assess the range of tissue structures found in tumours we analysed digital histological slides of 2170 breast cancer patients ($N\approx 1.8 \times 10^5$ clusters). Such a large patient collective guarantees reliable data on cancer cluster size distribution and shape, as well as nematic properties of the cells and ECM, rendering it an ideal starting point for the comparison to our continuum modelling. We also included 87 patients with ductal \textit{in situ} carcinoma from an open histology database \cite{brancati2022bracs} ($N\approx 5\times 10^2$ clusters), and 32 samples from healthy tissue regions that serve as references to demonstrate that invasive tumours differ. Although we show histological data from the non-invasive ductal \textit{in situ} carcinoma for instructive comparisons, our modelling efforts only target invasive breast cancer diagnosed after the basement membrane has been breached.\\


We analyse haematoxylin- and eosin- (H\&E-) stained tumour cutouts (see Methods, Sec.~\ref{sec:Methods_HnE}), extracting the nematic order of the ECM from the eosin channel (fibres) and of hundreds of thousands of cell clusters from the haematoxylin channel (nuclei) \cite{gottheil2023state}. In Appendix \ref{app:dimensionality}, we show that the structure and order parameters of 2D cross sections used in histology and simulations directly connect to the physical behaviour of the respective nematic structure (cells and ECM) in 3D tumours. A detailed histological reconstruction of a 3D breast tumour with distinct 3D clusters is given in Appendix \ref{app:3DClusterShapeSize}. The fibres of the densely packed ECM in invasive cancer are clearly aligned forming polymeric, passive nematic domains with an average order parameter of $0.52 \pm 0.07$.  Clusters of invasive breast cancer contain elongated motile cells \cite{grosser2021cell, gottheil2023state} that align to an active nematic phase (compare Fig.~\ref{qualitativeClusterECM}f) with an average nematic order parameter of $0.46 \pm 0.22$ in the histological sections. The nematic order parameter varies between 1 in a fully aligned state and 0 if there is no alignment. The order parameter is a measure of how well the local fibres are aligned. It is defined in detail in Sec.~\ref{methods:imageAnalyses} for 2D and in Appendix \ref{app:dimensionality} for 3D.
As visible in Figs.~\ref{qualitativeClusterECM}f  and ~\ref{fig:fingerprintsECM}a, the cancer cell clusters in invasive cancer display a sharp boundary indicating that cell-cell attraction prevails over cell-ECM adhesion. This generates surface tension between cell clusters and ECM. Nevertheless, the clusters have a non-spherical, irregular shape (Fig.~\ref{qualitativeClusterECM}), which suggests that mechanical stress which results from cell motility deforms the cluster shape and exerts forces on the ECM.
\section{Invasive breast tumour clusters can be described as active nematic materials}

These findings motivate the hypothesis that the collective behaviour of invasive cancer can be described as active nematic cancer clusters which exert deforming stresses on a surrounding viscoelastic nematic phase of ECM fibres. Active nematic physics has been successful in capturing local, short-term collective cell motility \cite{saw2018biological, doostmohammadi2018active, chepizhko2025confined, ilina2020cell}, and biphasic active-passive nematics are established in soft matter \cite{blow2014biphasic}. While previous works have studied cancer spheroid morphologies in ECM \cite{kang2021novel}, our work makes two major advancements. Firstly, we model the nematic order in the ECM, and how it couples to forces generated by the cancer clusters. Therefore, we can capture the distribution of topological defects in the ECM, and the alignment of the ECM at the cluster boundaries. The second is in scale: our results focus on the statistics of the long-term dynamical steady state, reached years after the basement membrane has been breached, and memory of the initial tumour mass is lost. In contrast to single spheroid studies \cite{kang2021novel}, our work focusses on the self-organization of the whole tumour, spanning millimetres across. To gain detailed insight into the dynamical mechanisms controlling the nonequilibrium distribution and evolution of clusters in tumours, we simulate a viscoelastic nematic phase, representing the ECM, coupled to an active nematic phase, representing the cancer cell clusters, described in detail below. \\

In simulations, we distinguish between the cancer clusters and ECM using a phase-field $\phi$, where $\phi=1$ labels the cancer clusters (Fig. \ref{qualitativeClusterECM}g purple), and $\phi=0$ labels the ECM (Fig. \ref{qualitativeClusterECM}g pink). The dynamics of $\phi$ are given by the Cahn-Hilliard equation
\begin{equation}
    \frac{\partial\phi}{\partial t} + (\mathbf{u}\cdot\nabla)\phi = \Gamma_\phi \nabla^2 \mu \: ,
\end{equation}
where $(\mathbf{u}\cdot\nabla)\phi $ describes advection by the flow field $\mathbf{u}$ , and $\Gamma_\phi \nabla^2 \mu $  describes the relaxation to a free energy minimum - which accounts for surface tension. Both cancer clusters and the ECM have their own nematic order tensor $\mathbf{Q}^{i} = 2 S^{i}\,\Big( \mathbf{n}^i\, \mathbf{n}^i - \frac{\mathbb{I}}{2} \Big)$, where the index $i=1$ represents the cancer cells and $i=0$ represents the ECM. The dynamics of the $\mathbf{Q}$ fields are given by the Beris-Edwards equation \cite{beris1994thermodynamics}
\begin{equation}
    D_t \mathbf{Q}^{i} -\mathbf{\mathcal{W}^{i}} = \Gamma \mathbf{H}^{i}\:.
\end{equation}
Here, the terms on the left-hand side represent, respectively, advection and co-rotation of the orientation field by the flow field. The right-hand side represents the tendency of the orientation field to relax to its free energy minimum, where nematic elasticity prefers a perfectly aligned state. The flow field $\mathbf{u}$ is incompressible ($\nabla\cdot\mathbf{u} = 0$). Force balance yields
\begin{equation}
    \rho \left( \partial_{t} + \mathbf{u} \cdot \mathbf{\nabla} \right) \mathbf{u} = \mathbf{\nabla} \cdot \mathbf{\Pi} \: ,
\end{equation}
where $\rho$ is the density of the fluid, and $\mathbf{\Pi}$ denotes the sum of all stresses acting on the fluid, composed of viscoelastic, active nematic, nematic backflow, and capillary terms. \\

The viscoelastic stress is modelled by the incompressible Kelvin-Voigt model \cite{thomson1865iv, voigt1892ueber, rajagopal2009note} with a viscous contribution $2\eta^i \mathbf{E}$ and an elastic contribution $\mathbf{\Pi}^{i, elastic} =  E^i \, \mathbf{T}$. Here, $\eta^i$ and $E^i$ are the viscous and elastic moduli of species $i$, and $\mathbf{E}$ and $\mathbf{T}$ are the strain-rate tensor and strain tensor in the fluid respectively. The cancer cells produce active nematic stresses along their long axis, given by \cite{doostmohammadi2018active}
\begin{equation}
    \mathbf{\Pi}^{1, act} = - \zeta \mathbf{Q^{1}}\:,
\end{equation}
where $\zeta$ is the activity characterizing the magnitude of mechanical stresses exerted by the cells. These stresses set the typical speeds of motile cells in cancer clusters. Finally, the capillary stress $ \mathbf{\Pi}^{cap}$ models the stresses from surface tension acting on the fluid at the cancer-ECM boundary \cite{lazaro2015phase}. Further details are presented in Sec.~\ref{methods:computational_model}.\\

Mapping from simulation to real physical parameters in tumours follows from the choice of a length, time and stress scale. Typical cluster velocities based on our vital cell tracking data  in patient-derived tumour explants \cite{gottheil2023state} are $\sim 3\mu \mbox{m} \, \mbox{h}^{-1}$ and typical cell radii $\sim 5\mu \mbox{m}$. The typical bulk modulus of a breast tumour sample dominated by ECM is $\sim 500$Pa as our AFM measurements of primary human breast cancer (Appendix \ref{app:AFM} and Sec.~\ref{methods:AFM}) and prior publications \cite{seewaldt2014ecm, acerbi2015human} confirm. Traction force microscopy shows that the cancer cells typically exercise mechanical stresses of a few 100 Pa on the ECM \cite{legant2010measurement, kraning2012cellular}.  These experimental values determined our choice of realistic parameter values used in the simulations (see Sec.~\ref{methods:computational_model} and Appendix \ref{app:robustness}) providing a direct connection to physiological tumour conditions. Considering tumour progression and heterogeneity we have probed varying ECM stiffness caused by desmoplasia and deforming stress caused by cancer cell motility in the clusters in our simulations. These two variables correspond to the ECM elasticity and activity respectively. \\

\begin{figure*} [htp]
    \centering    
    \includegraphics[width=\textwidth]{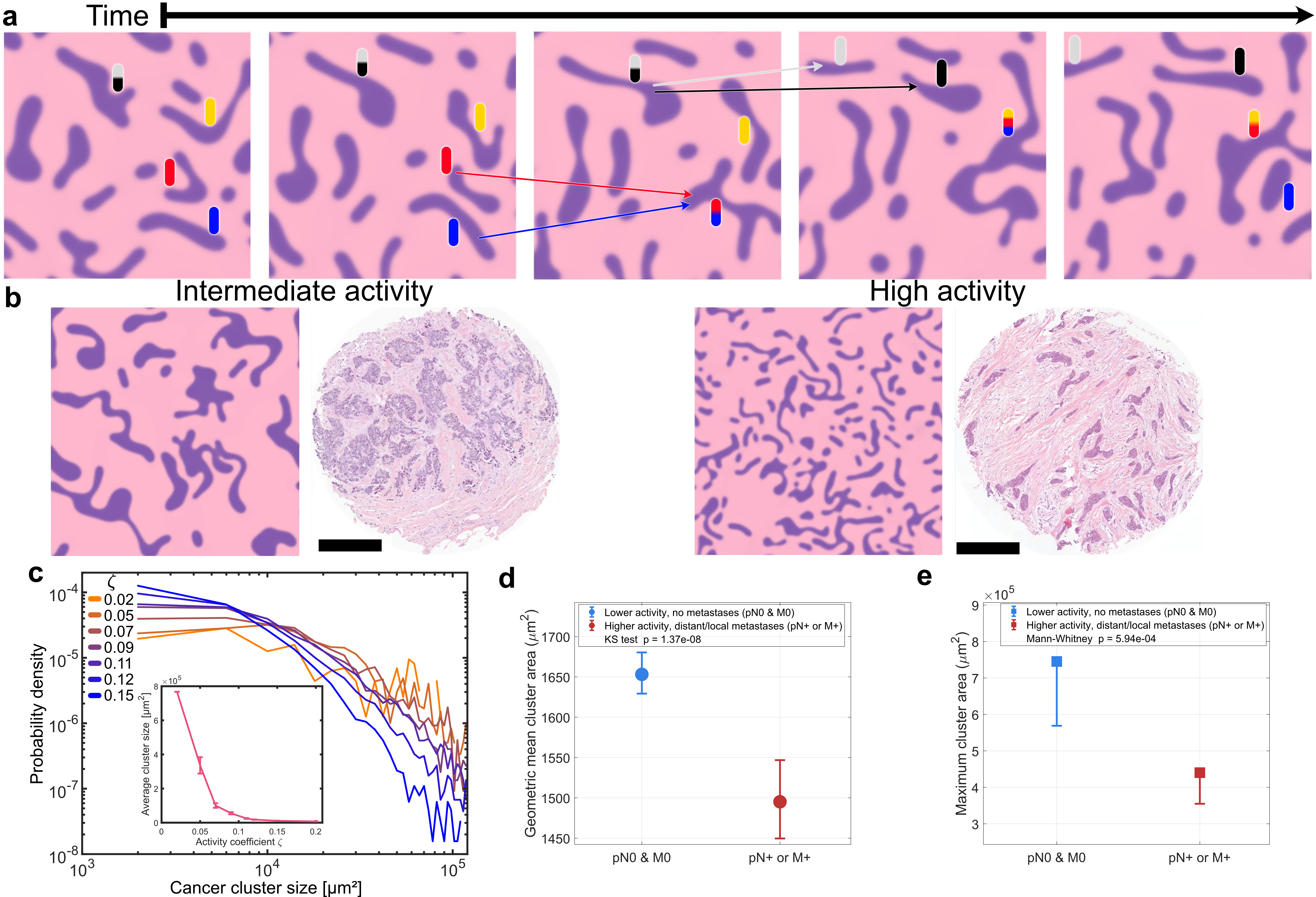}
 \caption{\textbf{Mechanical activity in the cancer cell clusters caused by cell motility determines the shape and size of the clusters.} 
 {\bf a} Time sequence from simulations showing the cluster dynamics, with droplets moving, splitting, and merging. We indicate splitting and merging by marking clusters with the respective colour(s).
 {\bf b} With increasing activity the clusters become smaller. Simulations are compared to examples from the histological images chosen to show similar structures. {\bf c} Probability density functions of the cancer cluster size distributions in simulations for  different activity levels. All distributions are significantly different to each other with more small clusters at higher activities. All two-sided Kolmogorov–Smirnov tests exhibit $p<10^{-9}$. Inset: Average cluster sizes as a function of the activity coefficient $\zeta$. The cluster sizes show a negative Spearman correlation with activity ($R = -0.37$; $p<10^{-6}$), meaning that the average cluster size decreases with increasing activity (compare Methods, Sec.~ \ref{methods:StatisticalTests}). {\bf d , e}: As clinical proxy for cluster activity, the patient collective was divided into a lower-activity group of patients who developed neither distant metastases nor lymph-node involvement (pN0 \& M0), and a higher-activity group who developed distant metastases and/or lymph-node invasion (pN+ or M+). Clusters in the higher-activity (pN+ or M+) group are {\bf d} on average smaller ($1495$ vs. $1653\,\mu$m$^2$; two-sided Kolmogorov-Smirnov $p \approx 1.4\times10^{-8}$) and {\bf e} reach $41\%$ smaller maximum sizes (Mann-Whitney $p<10^{-3}$) than in the lower-activity (pN0 \& M0) group. This is consistent with the simulation results, showing that cluster sizes decrease with increasing activity.}
    \label{fig:activity}
\end{figure*}

 The 2D simulations show a large number of moving, tortuous cancer cell clusters within the ECM (Fig.~\ref{qualitativeClusterECM}g) that are propelled and stretched by active flows caused by collective cell motility. The mechanical forces due to activity are able to extend the clusters and overcome surface tension, so that the clusters break up, move around, collide and re-form \cite{giomi2014spontaneous,singh2019hydrodynamically, blow2014biphasic}. This dynamic is displayed in the time sequence in Fig.~\ref{fig:activity}a and in Supplementary Movie S1 \cite{SM}.\\ 
 
 Representative snapshots of the cluster pattern and distribution for two different activities agree with histological data in Fig.~\ref{fig:activity}b. Fig.~\ref{fig:activity}c gathers simulation data on the variation of cluster size with activity. In our simulations, the cluster size is reduced by higher activities, which promote the clusters to extend and break up. We emphasise here that the cluster size is set by the balance of active forces, surface tension, and elastic confinement of the ECM. It is therefore more accurate to speak of the relative weight of cluster activity against the ECM elasticity, rather than of activity in isolation.\\

 We next tested the prediction of smaller clusters for higher activity tumours against histological patient data and their clinical follow-up (discussed in more detail in section \ref{sec:Prognosis} below and Appendix \ref{app:patientData}). We use clinical outcome, that is documented for each patient/histological image (details in Appendix \ref{app:patientData}), as an independent proxy for cluster activity, on the assumption that tumour invasion (lymph nodes or organs) correlates at least coarsely with motile cluster activity. We divide the invasive patient collective into a lower-activity group of patients who developed neither distant metastases nor lymph-node involvement (pN0 \& M0), and a higher-activity group who developed distant metastases and/or lymph-node invasion (pN+ or M+).\\

The cluster sizes in the higher-activity (pN+ or M+) group are on average smaller ($1495$ vs. $1653\,\mu$m$^2$; two-sided Kolmogorov-Smirnov $p \approx 1.4\times10^{-8}$; Fig.~\ref{fig:activity}d) and reach $41\%$ smaller maximum sizes (Mann-Whitney $p<10^{-3}$; Fig.~\ref{fig:activity}e) than in the lower-activity (pN0 \& M0) group. This is consistent with the largest clusters being unable to persist under a higher-activity distribution, supporting motility-driven fission rather than growth as the origin of the invasive size distribution.\\

Finally, in order to check in the simulations that we were considering a dynamical steady state independent of initial conditions, we ran the model starting from both a single large tumour mass (Supplementary Movie S1 \cite{SM}) or a large number of tiny tumour clusters (Supplementary Movie S2 \cite{SM}), with the cells occupying an area fraction $\phi=0.25$. In both cases, the clusters reached the same steady-state  distribution and patterns. While this self-organization is new in tumours, similar steady states have been observed in other active systems \cite{tiribocchi2015modelh, Adkins2022}. We have also performed 3D simulations and studied 2D cross-sections from them as well as 3D explants from resected tumours as reference for histological sectioning (Appendices \ref{app:dimensionality} and \ref{app:3D_Sims}) to make sure that our 2D histological and simulation data faithfully reproduce a robust representation of the tumour behaviour in 3D.
\section{Cancer cluster area and shape distributions}
\label{sec:SizeShapeDistro_Main}

These results on whole cluster dynamics were unexpected as current tumour biology only reports that motility facilitates cancer cell escape from the clusters into the ECM by single cells or small cell aggregates breaking away \cite{ilina2020cell, kang2021novel} rather than processes where cluster fission and fusion impact the size distribution. We next obtained the cluster area and shape distributions over several orders of magnitude from the histological images and confirmed the distributions obtained in our simulations. 
In our simulations, we assigned the clusters a range of different activities according to the motility distributions of our previous experiments \cite{gottheil2023state, fuhs2022rigid}  that indicate tumour heterogeneity.  \\

\begin{figure*} [htp]
    \centering    
    \includegraphics[width=0.9\textwidth]{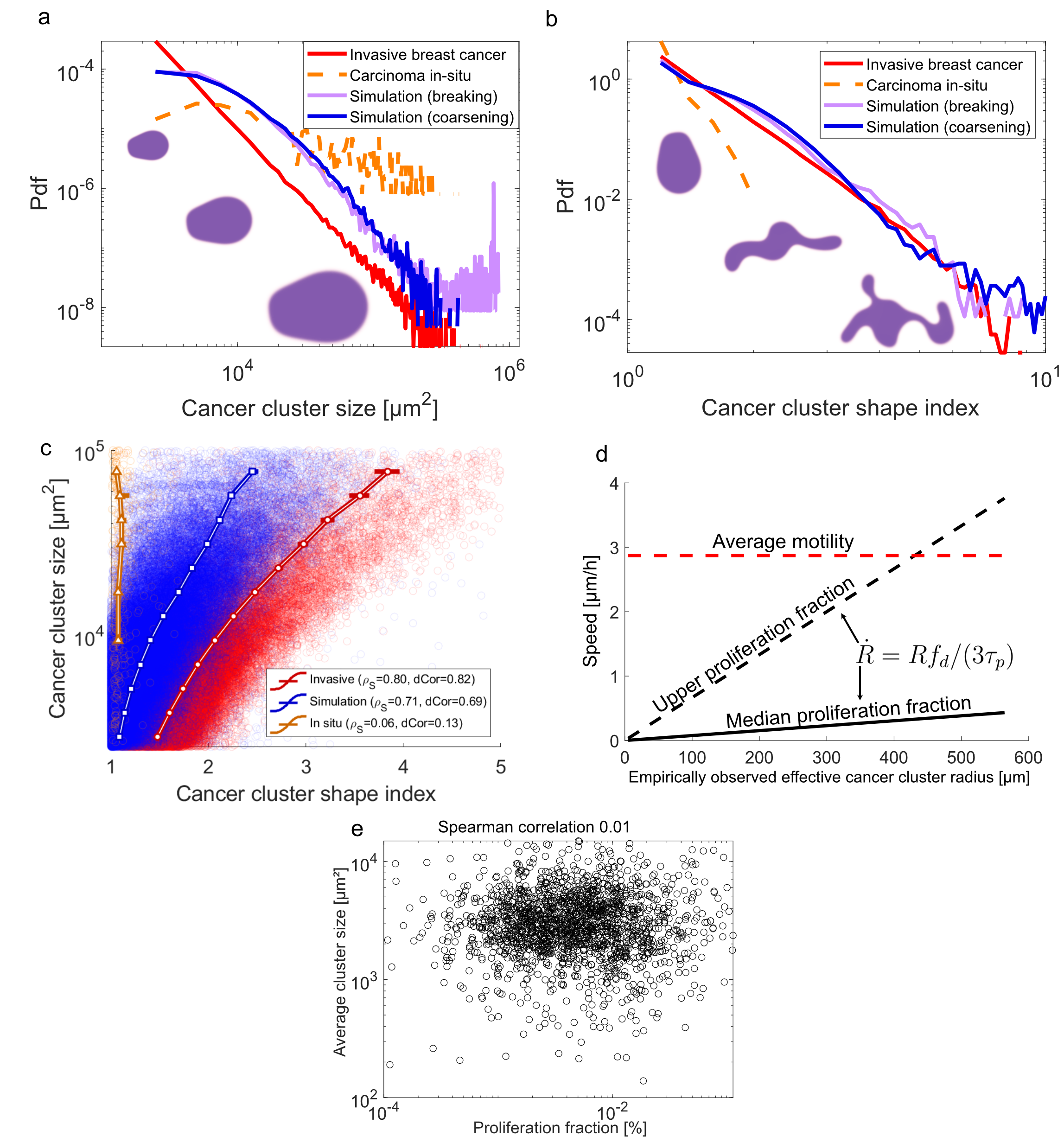}
 \caption{\textbf{Scaling of shape and size in breast tumour cell clusters.}  Distribution of {\bf a} cluster sizes. {\bf b} cluster shapes ($SI=P/\sqrt{4 \pi A}$, where $A$ is the cluster area and $P$ the cluster perimeter). 
 In each graph, the red line displays the result from analyzing the histological slides from 2012 patients. 
 Data from our simulations, with an initial condition of one large tumour cluster or many tiny clusters, are shown in deep blue and violet, respectively. The yellow dashed line shows results for \textit{in situ} breast tumours that do not show a scaling behaviour. Inset cartoons give a pictorial guide to changing sizes and shapes. {\bf c} The size and shape of invasive lesions are strongly correlated in both histological data from breast cancer patients (red symbols; Spearman $R=0.80; p<10^{-6}$, distance correlation $R = 0.82$) and simulations (blue symbols; Spearman $R=0.71; p<10^{-6}$, distance correlation $R = 0.69$ for clusters $>2000\mu \mbox{m}^2$). We found no significant correlation for \textit{in situ} breast tumours (yellow symbols; Spearman $R=0.06; p = 0.17$, distance correlation $R = 0.13$). {\bf d} Movement of the clusters can be caused by cell motility or proliferation: Only in very large clusters can proliferation dominate. In 99\% of all clusters cell motility is the dominant factor. The red dotted line represents the average cancer cell motility measured in vital primary human breast cancer explants (for details compare Appendix \ref{app:dominant_activity} and Ref.~\cite{gottheil2023state}). The black dotted lines represent proliferation-induced flow speeds of cancer clusters that were computed using the estimate $v_{\text{proliferation}} \sim \frac{R f_d}{3 \tau_p}$, which uses the measured fractions of proliferating cells $f_d$ in our patient collective, $\tau_p$ is the duration of mitosis and $R$ the effective cancer cluster radius. The line for the upper proliferation fraction (dotted black) shows the estimated proliferation-induced flow speeds for proliferation fractions of the $99$th percentile of the values found in our patient collective. Details can be found in the section \ref{methods:Proliferation}. {\bf e} We find vanishing correlations (Spearman $R = 0.01; p = 0.001$) between cluster sizes and proliferation fractions. This is consistent with the hypothesis that cell motility is responsible for cluster size distribution.}
    \label{fig:sizeshape}
\end{figure*}

The probability density distribution of cancer cluster areas in histological images of invasive breast cancer as a measure of cluster size ($N \sim 10^5$ clusters) is plotted on a log-log scale in Fig.~\ref{fig:sizeshape}a  (red line), showing a clear power law distribution of cluster sizes $P(A) \sim A^\alpha$ with $\alpha= -2.38$ with $95\%$ confidence interval $[-2.43, -2.33]$ ($R^2 > 0.99$) over a range of areas from $ \sim 3 \times 10^2 \mu$m$^2 - 3 \times 10^5 \mu$m$^2$ with an average value $\langle A \rangle= 3.7 \times 10^3 \mu$m$^2$. The simulations obtain the same power law exponent from $\sim 10^3 \mu$m$^2 - 10^5\mu$m$^2$ ($R^2 = 0.98$, Fig.~\ref{fig:sizeshape}a, blue lines). Note that the scaling exponents do not change significantly under changes of the activity distribution (Appendix \ref{app:robustness} and Fig.~\ref{fig:sizeshape}), which is what is studied in the dynamical steady state. The first few bins from the simulation data do not exhibit the power law scaling because of the difficulty of resolving the smallest clusters in simulations. This is due to the finite width of the phase field at an interface.\\

To confirm that these cross-sectional distributions reflect genuinely three-dimensional objects rather than sections through a single connected mass, we reconstructed 3D volumes from serial histological sections of an invasive tumour (Fig.~\ref{qualitativeClusterECM}c,e; Methods Sec.~\ref{methods:3Dreconstruction} and~\ref{methods:3DClusterAnalyses}). Segmenting clusters as connected components in 3D yields $N \approx 3 \cdot 10^4$ individual, disconnected clusters, confirming that invasive lesions consist of separate objects rather than fingers of one percolated structure. The 3D cluster-volume distribution follows a power law $P(V) \sim V^{\alpha_{3D}}$ with $\alpha_{3D} = -2.49$, while the 2D cross-sectional areas measured on the same reconstructed volume give $\alpha_{2D} = -2.46$ (more details in Appendix ~\ref{app:3DClusterShapeSize}); both are close to the value of the patient collective $\alpha = -2.38$ and consistent with the stereological result that a 3D power law is preserved under random sectioning with the same exponent (Appendices~\ref{SI:clusterMapping} and~\ref{app:3DClusterShapeSize}).\\

 The clusters observed in both patient images and simulation have tortuous and highly irregular shapes, which we attribute to the cluster motion within the ECM. To quantify these shapes, we measured the perimeter $P$ of each cluster within the ECM and defined the cluster shape index $SI = P/\sqrt{4 \pi A}$. It is 1 for a circular cluster and increases from unity as the perimeter increases more strongly than the area for more tortuous cluster shapes (See Appendix \ref{app:AR_ir} for a discussion of this shape metric and a comparison to cluster aspect ratios). The probability distribution of $SI$ for the histological images is shown in Fig.~\ref{fig:sizeshape}b (red line), together with the distribution of $SI$ for simulated clusters (blue lines). The mean value of $SI$ of the invasive breast cancer clusters is 1.38. Due to the small range of $SI$, we are unable to distinguish between a power law distribution ($\sim (SI)^\beta$; $\beta= -5.65 \pm 0.29$ (fit $\pm $95\% CI), $R^2_{\text{histology}} = 0.98$, $R^2_{\text{simulation}} = 0.96$) and an exponential distribution ($\sim e^{b (SI)}$; $b= -3.47 \pm 0.11$ (fit $\pm $95\% CI), $R^2_{\text{histology}} = 1.00$, $R^2_{\text{simulation}} = 0.96$). \\
 
The size and shape distributions are closely related to each other, as can be seen by the direct correlation of the two quantities (Fig.~\ref{fig:sizeshape}c). The corresponding Spearman coefficient for invasive breast cancer clusters is $R = 0.8$ and for simulations $R = 0.71$, reflecting strong positive correlations between size and shape - clusters with larger area are more tortuous in shape. Details and additional correlation metrics can be found in Sec.~\ref{methods:StatisticalTests} and Appendix \ref{app:sizeshapecorrelation}.\\

The close agreement between data and simulations gives further confidence in the physical interpretation of the distribution of the cancer clusters as a dynamical steady state arising from a balance between active cluster break up and fusion. \\

In the histological data, the average cross-sectional areas of invasive lesions are $\sim 20$-fold smaller when compared to average ductal carcinoma \textit{in situ} lesions. Consequently, the 3D volumes are on the order of $\sim 100$-fold smaller. This demands that a process that decreases cluster size such as cluster breaking must have taken place to achieve these drastically smaller clusters - especially considering that invasive tumours occur later in disease progression and have had more time to proliferate.\\

As a reference, we analysed the shape and size distributions of clusters of ductal \textit{in situ} carcinomas. For these \textit{in situ} clusters, whose sizes are only determined by proliferation, we do not find a power law behaviour as shown by the yellow, dashed curves in Figs.~\ref{fig:sizeshape}a-c. A striking difference to invasive cancer is the round regular cluster shape. The cluster shape index never exceeds 2 and over $80\%$ of clusters exhibit indices of less than 1.2 (dashed line, Fig.~\ref{fig:sizeshape}b). In contrast to invasive clusters, there is no significant correlation between cluster shapes and sizes (Fig.~\ref{fig:sizeshape}c). Moreover, the distribution of cluster sizes is weighted away from small sizes towards larger sizes when compared to the later invasive lesions  (Fig.~\ref{fig:sizeshape}a). This confirms the difference between \textit{in situ} tumours which are just a growing cell mass confined by the basement membrane which is unable to break and the highly motile, later stage, invasive clusters. \\

In growing tumours, cell division can result in active (nematic) flows \cite{saw2018biological} of a similar nature to cell motility. We now estimate the relative contributions of proliferation and motility to the activity of the clusters in the invasive lesions. Approximating the clusters by spheres of radius $R$, the average spreading speed caused by cell proliferation is $\dot{R} = R  f_d/(3 \tau_p)$, where $f_d$ is the fraction of mitotic, i.e. actively dividing cells and $\tau_p$ is the typical duration of mitosis. We assumed  $\tau_p \sim 2$h \cite{ibrahim2022assessment} and measured the fraction of mitotic cells in the patient collective using the proliferation marker Phospho-histone H3 (see Methods, Sec.~\ref{methods:Proliferation}), giving a median value of $f_d \sim 0.46 \%$ and an extreme value of $\sim 4 \%$ for the 99th percentile of all patients. To compare these values to cell motility we have performed cell tracking in vital primary breast cancer explants derived after surgery (Appendix  \ref{app:dominant_activity}), where cell nuclei were stained and tracked. We find that $31 \%$ of cells change neighbours due to cell motility within time frames that are comparable to the time of mitosis $\sim 2h$ (see Fig.~\ref{fig:SI_vital}). Given that motility and proliferation generate active stresses of the same order of magnitude, measured by traction force microscopy \cite{jahan2017traction, liu2021quantifying, nam2021cellular, rossetti2024optogenetic, peschetola2013time, trepat2009physical, kraning2012cellular}, we conclude that the active energy input from cell motility is substantially larger than the one from cell proliferation (compare $\sim 31\%$ to $\sim 0.46\%$), see Appendix \ref{app:dominant_activity}. This is also confirmed by simple visual inspection of a movie of a vital primary breast tumour explant shown in Supplemental Movie S5 \cite{SM}, where we do not see any mitotic events but many migrating cells. \\

This mechanical argument can be further confirmed by considering the dynamics: We estimate the proliferation speed (for median and extreme proliferation fractions $f_d$) for clusters of different radius, and compare them to the average speed of unjammed cells inside breast tumour explants ($\sim 3 \mu$m/h, obtained from our previous vital cell tracking experiments \cite{gottheil2023state} and reconfirmed in Appendix \ref{app:dominant_activity}) in Fig.~\ref{fig:sizeshape}d. The cluster speed caused by proliferation for most of the cluster sizes is smaller than the movement caused by cell motility.  The lack of any significant correlation between cluster size and proliferation fraction (Fig.~\ref{fig:sizeshape}e) further suggests that motility, instead of growth, is responsible for the observed distribution of cluster sizes.  Uncontrolled proliferation already occurs in DCIS. The large clusters found in DCIS and their size and shape distribution confirm that proliferation cannot drive cluster breakup. Although these are estimates, they provide evidence that motility dominates proliferation in driving the motion of almost all cancer clusters.\\

\begin{figure*} [htp]
    \centering    
    \includegraphics[width=0.95\textwidth]{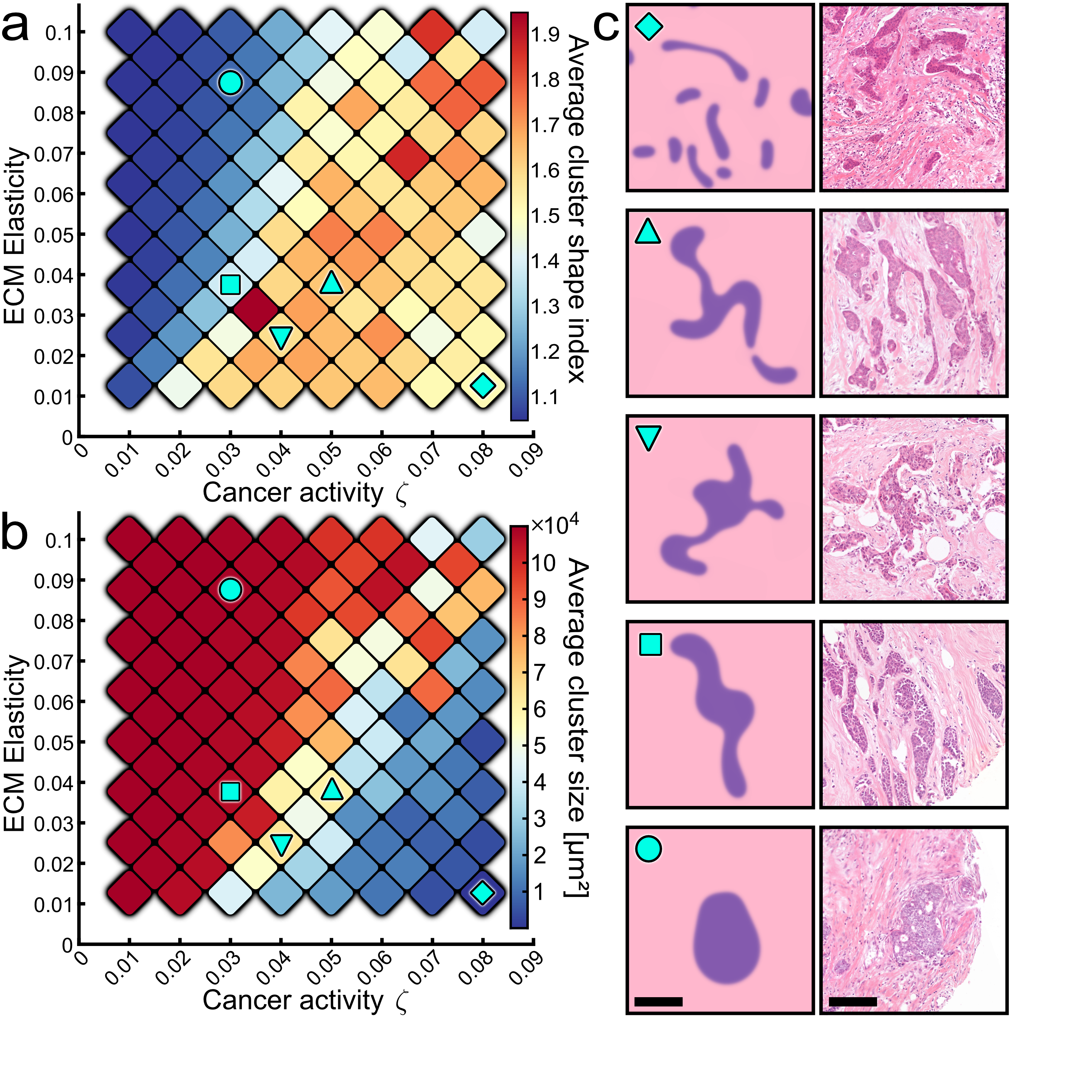}
 \caption{\textbf{Stability of tumour cell clusters confined by nematic extracellular matrix.} The state diagram shows \textbf{a} the cluster shape index (heat map, defined as $SI = P/ {\sqrt{4\pi A}}$, where $P$ is the perimeter and $A$ the area) and \textbf{b} cluster areas (heat map) as a function of the mechanical resistance of the ECM and the activity of the cluster. Going from the upper left to the lower right, a single active cell cluster breaks apart into multiple clusters. Droplet splitting dynamics are chaotic - and the exact parameter values of splitting are dependent on initial conditions and simulation noise. However, the broad transition from individual smooth droplets via irregularly shaped droplets to multiple  smaller droplets is consistent with decreasing ECM elasticity and increasing activity across different simulations.
 For the points marked by symbols in the diagram, we compare typical cluster shapes from simulations and histological images in sub-figure \textbf{c}. Scale bars for simulation and histology correspond to 300 $\mu$m in sub-figure \textbf{c}.} 
\label{fig:simulations}
\end{figure*}
\section{Breaking up of a single cluster}
To investigate further the cluster breakup, we performed simulations starting from a single cluster and varying the range of activities and the mechanical resistance of the ECM within the experimentally observed values. The resulting state diagrams for shape indices and sizes are shown in Fig.~\ref{fig:simulations}a--b respectively. In the upper-left corner of the state diagram, for smaller activities and larger ECM resistance, the cluster remains stable and intact, with a low shape index corresponding to few fingering protrusions. When the activity and elasticity are comparable, active forces distort the cluster to a fluctuating and elongated shape, with high shape index corresponding to many protrusions. In the bottom-right corner of the state diagram, the high activity overcomes surface tension and elastic confinement and pulls the tumour apart into smaller clusters (see Fig.~\ref{fig:simulations}b,  Fig.~\ref{fig:activity}a), each with a lower shape index (corresponding to few finger-like protrusions).\\

Collapsing the state diagram onto the single control parameter $\zeta/E$ - the ratio of cluster activity to ECM elasticity - shows that the average cluster shape index is maximal near $\zeta/E \approx 1$, where the active stresses locally balance the restoring forces of surface tension and ECM elasticity: clusters are then maximally deformed but not yet broken. At higher $\zeta/E$ the clusters fragment into smaller, individually rounder daughters and the average shape index falls (Appendix~\ref{app:shape_ratio}).\\

For exemplary points in the state diagram, we display a typical simulated cell cluster together with a similar histological reference image to confirm that the different cluster shapes are observed in real tumour tissue. The motion of these representative clusters is shown in Supplementary Movies S3a--S3e \cite{SM}. It is important to note here that Fig.~\ref{fig:simulations} concerns the effect of activity and elasticity on a single small cluster confined by an elastic ECM. In the rest of the paper, we focus on characterizing the properties of the dynamical steady state, which consists of a very large number of clusters derived from a broader distribution of activity.

\section{Mechanical cancer activity leads to topological defects in the surrounding nematic ECM}
\begin{figure*}[!ht]
    \centering    
\includegraphics[width=0.85\textwidth]{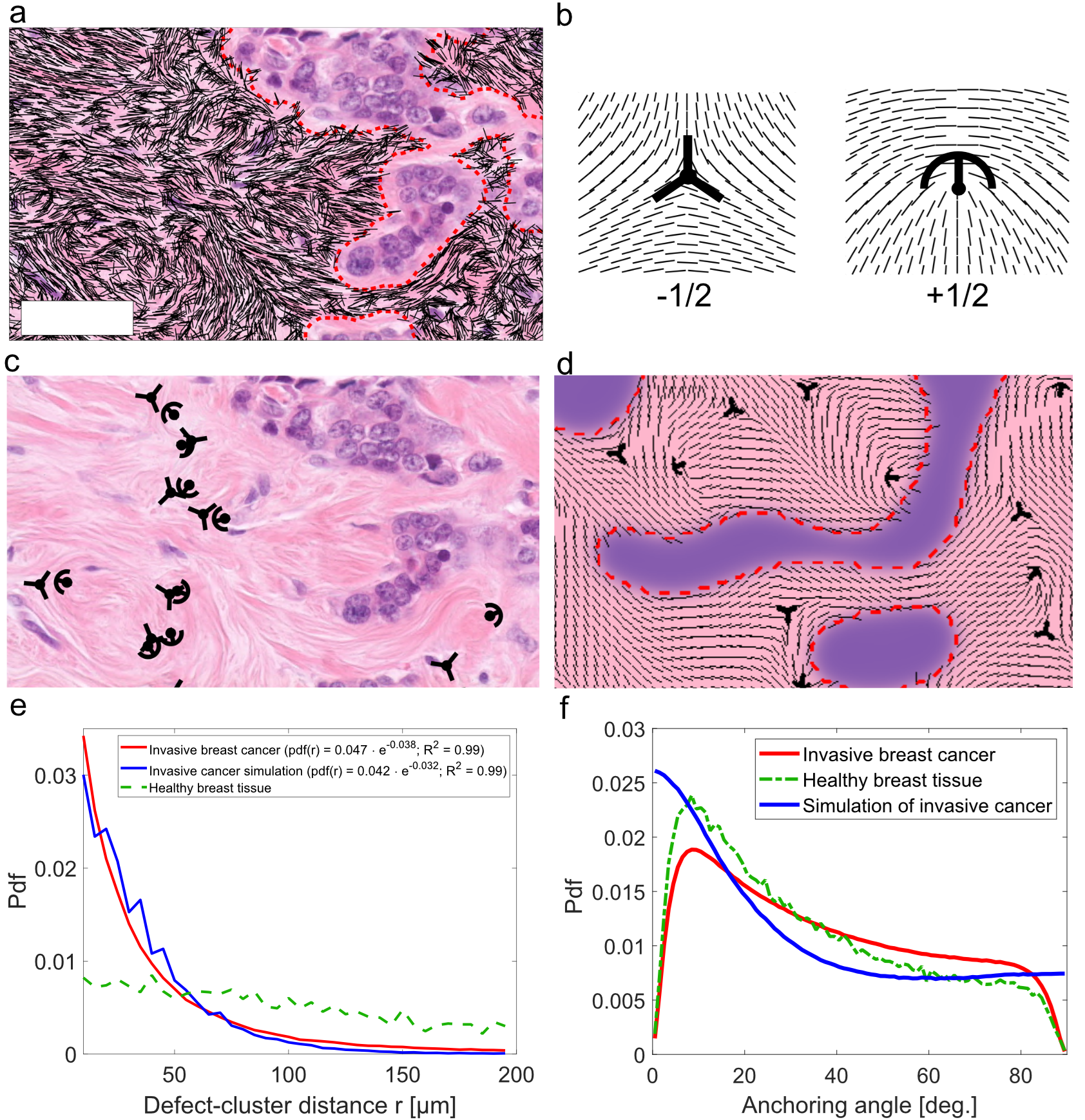}
 \caption{\textbf{Effects of cancer-induced active stresses on the ECM.} \textbf{a} Nematic ordering in the ECM director field, shown by the black lines. The scale bar is $100 \mu$m and also applies for c, d.
 {\bf b} Alignment around a -1/2 (trefoil) and +1/2 (comet) topological defect. {\bf c,d} Topological defects in the ECM surrounding breast cancer lesions
in \textbf{c} a histological slide and \textbf{d} a simulation snapshot. {\bf e} Distribution of the shortest distance between defects in the ECM and a cluster boundary. The number of defects decreases exponentially with the distance from the cluster boundaries in invasive breast cancer tissues (red) and active nematic simulations (blue) which is fundamentally different from healthy tissues (green dotted) where the distribution is relatively uniform. Exponential fitting: $\text{pdf}(r)=0.047e^{-0.038r}$ (histology, invasive) / $\text{pdf}(r) = 0.042e^{-0.032r}$ (simulations, invasive), $R^2=0.99$.  {\bf f} Distributions of the anchoring angle (0 degree ECM parallel to cancer cluster tangent, 90 degree orthogonal). The nematic ECM in simulation and histology of invasive breast cancer predominantly aligns parallel to the active cluster boundary. The parallel anchoring seen in healthy ducts is due to the physiological structure of the intact basement membrane.}
    \label{fig:fingerprintsECM}
\end{figure*}

We next investigate how moving clusters interact with the surrounding ECM in the activity-averaged dynamical steady state (Fig.~\ref{fig:fingerprintsECM}a). We compare the nematic ordering of the fibres within the ECM between healthy breast tissue and invasive lesions (Appendix \ref{app:compareHealthyCancer}). Although both show a wide variability, invasive lesions do not show higher nematic order parameters despite cancer-associated fibroblasts generating additional fibrotic ECM by desmoplasia \cite{kaushik2016transformation}. Local disruptions in the alignment of the nematic ECM generate topological defects, points where the ordering of the ECM locally takes the form of a comet (+1/2 defect) or trefoil (-1/2 defect) (Fig.~\ref{fig:fingerprintsECM}b). Figs.~\ref{fig:fingerprintsECM}c,d provide visual examples of defects in the director field for the histological images and simulations, respectively. The 2D defects in the images correspond to disclinations in the 3D ECM (Appendices \ref{app:defect_dimen} and \ref{app:disclinations}). The active stresses generated by the cancer clusters are transmitted into the passive matrix, bending the ECM director and nucleating the topological defects that concentrate near the cluster boundary. The number of defects decreases exponentially with the distance from the cluster boundaries (Fig.~\ref{fig:fingerprintsECM}e) in invasive breast cancer tissues (red, $\text{pdf}(r)=0.047e^{-0.038r}, \; R^2 = 0.99$) and active nematic simulations (blue, $\text{pdf}(r) = 0.042e^{-0.032r}, \; R^2 = 0.99$). This is fundamentally different from healthy tissues (green dotted) where the distribution is relatively uniform throughout the ECM. \\

Previous works have shown that collagen fibre orientation at a tumour boundary plays an important role in suppressing or enhancing the escape of single cancer cells from a cluster into the surrounding ECM. Parallel-aligned ECM fibres at the tumour boundary (denoted tumour-associated collagen signature 2, TACS-2) constrain its volume and mechanically hinder cells from leaving it. When the fibres are perpendicularly aligned to the tumour boundary, denoted TACS-3, escape is instead promoted, with cells migrating outward along the fibres \cite{Provenzano2006, Ray2021}. TACS-3 has been reported as a signature of aggressiveness in human invasive breast cancer, and is a local rather than a perimeter-wide feature: it occurs at discrete sites around a tumour region, where invasion is localised. In Conklin et al. \cite{conklin2011aligned}, TACS-3 is scored as present or absent within selected regions of interest. In contrast, we study the distribution of ECM alignment angles along the perimeter, and attribute this to collective active flows.\\

Strikingly, across our patient collective the ECM is predominantly aligned \textit{parallel} to the cancer-cluster boundaries--the configuration that hinders rather than promotes single-cell escape--contrary to a radial alignment expected from cell traction. We compared the orientation of the nematic director in the ECM at the cluster boundary between histological images and simulations, and in both the director preferentially aligns parallel to the boundaries. This is quantified in Fig.~\ref{fig:fingerprintsECM}f, which plots the angle between the boundary tangent and the nematic direction, showing a clear maximum at $\sim$$0^\circ$ and qualitative agreement of the simulations with the data. The healthy ducts also peak at parallel alignment, but more sharply: the anchoring order parameter $\langle\cos 2\theta\rangle$ (with $1$, $0$ and $-1$ denoting perfectly parallel, random and perfectly radial anchoring) drops from $0.32$ in healthy tissue to $0.21$ in invasive cancer, i.e. the invasive distribution is parallel on average but markedly broader. This broadening is the signature expected if the two parallel alignments have different origins. The healthy basement membrane is a thin, highly cross-linked sheet running parallel to the epithelial boundary, which enforces sharp parallel alignment. Once the cancer cells have breached it, this organising sheet is gone, and cell traction would then be expected to align the fibres radially at the cluster boundary \cite{conklin2011aligned}--yet the histology consistently shows predominantly parallel alignment. Our simulations suggest a resolution for this: the parallel alignment is dynamically maintained by the flow field of the collectively moving clusters, and is parallel on average but noisier, since the same active stresses that align the fibres also continually distort them. \\

Crucially, our measurement is coarse-grained over the full population of cluster boundaries, capturing the majority alignment of the ECM set by the collective flow of whole clusters. This does not contradict the more local mechanism by which individual cells may still escape through radially aligned fibres at the single-cell scale: the two operate at different scales, with collective cluster flows organising the bulk of the ECM while single-cell traction can locally realign fibres to enable escape.\\

Together with the boundary-localised topological defects and the elevated nematic order near the clusters (Fig.~\ref{fig:fingerprintsECM}e and Appendix~\ref{app:nematic_screening}), this points to the ECM being organised by active cluster flows rather than by a passive, basement-membrane-imposed structure. This is the simplest explanation consistent with all of these observations.

\section{Mechanical cancer activity influences tumour progression\label{sec:Prognosis}}
Active forces generated by cancer clusters confined by ECM can explain the self-organisation that leads to the characteristic structures found in invasive breast cancer. The question remains whether this process and thus nematic physics are important enough to clinically impact tumour progression, which can be best tested by determining if the process is connected to patient outcomes.\\

As measures of the active stresses generated by cancer clusters and that drive the formation of invasive lesions, we used the cluster shape index (see Fig.~\ref{fig:sizeshape}b), and the normalized maximum defect-cluster distance, Sec.~\ref{methods:survival}. We applied the Kaplan-Meier estimator \cite{dudley2016introduction} to determine whether the survival function for breast cancer patients depended on each of these parameters. Our breast cancer patient collective was divided into a training group (600 patients) to find the optimal threshold in our activity estimates to distinguish between low and high risk and a test group (remaining  1412 patients) to validate our prognostic stratification criteria. Details of the parameters used for risk assessment and thresholds can be found in Sec.~\ref{methods:survival} and 4-fold cross validation analyses is shown in Appendix \ref{app:crossval}. Complete information on the patient collectives used for the retrospective study can be found in Appendix \ref{app:patientData}. \\

\begin{figure*} [htp]
    \centering    
    \includegraphics[width=0.85\textwidth]{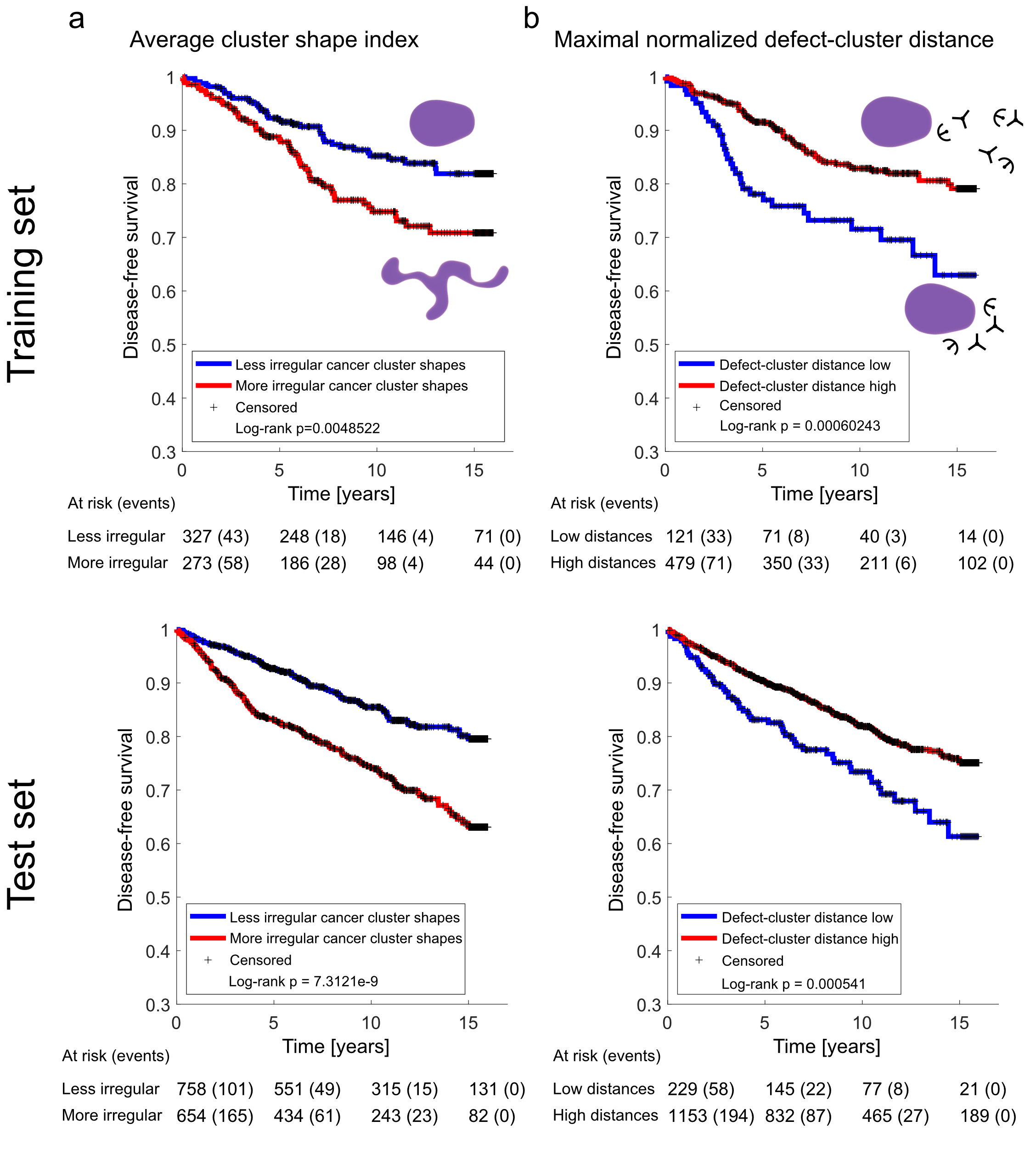}
 \caption{
 Survival curves of invasive breast cancer patients are shown for a random training (N = 600) and  (N=1412) test set (detailed patient statistics in Appendix \ref{app:patientData}). In the training set, we determine the threshold that maximizes risk group separation according to the log-rank statistic. The threshold is subsequently applied to the test set patients. Disease-free survival probabilities are plotted according to Kaplan-Meier estimators. \textbf{a} The average cluster shape index was used for risk group separation in training (N = 600) and test (N=1412) collectives. The log-rank p-value in the training set is $p \approx 0.005$ and in the test set $p < 0.001$, respectively.
 \textbf{b} The maximum defect-cancer distance normalized to the ECM length scale was used for risk group separation in training (N = 600) and test ($N=1382$; $N<1412$ because a few patients did not exhibit topological defects in the ECM) collectives. The log-rank p-value in the training set is $p < 0.001$ and in the test set $p < 0.001$, respectively.
\textit{At risk} indicates the total number of patients who have not been censored, developed distant metastasis or died by a specific follow-up time.
In parentheses after the \textit{At risk} is the cumulative number of individuals that developed distant metastases or died later than indicated. Prognosis was validated by 4-fold cross validation, see Appendix \ref{app:crossval}}
    \label{fig:survivalcurves}
\end{figure*}

The Kaplan-Meier plots (Fig.~\ref{fig:survivalcurves}a) show that a higher average cluster shape index correlates with a worsened prognosis. A higher shape index corresponds to higher activity (stronger active forces) inside the cancer cluster or a lower ECM elasticity near the clusters (Fig.~\ref{fig:simulations}). The next Kaplan-Meier plots (Fig.~\ref{fig:survivalcurves}b), emphasizing the role of cluster activity in tumour progression, show that a low defect-cluster distance also corresponds to a worsened prognosis. We interpret this as an interplay of active forces and  ECM stiffness, where topological defects created by active forces of the cancer clusters remain close to the cluster boundary when the stiffness of the ECM leads to a higher resistance.  \\

 Our results demonstrate that nematic physics is not limited to describe local, short term processes in cellular tissues, which are also increasingly recognized processes in other cancer types and model systems \cite{ilina2020cell, barberis2024self, chepizhko2025confined}, and can quantify  clinically relevant, long term and large scale tumour dynamics. Nematic physics translates tumour dynamics into tumour geometry and morphology providing a pathomechanical link that may lead to new prognostic tumour markers in digital pathology. Our findings permit us to quantitatively understand the interplay between motility, proliferation and fibrosis and may ultimately inspire new therapeutic approaches that target this interplay.\\
 
Current breast cancer diagnosis uses histopathology in grading that assesses the abnormal look of cancer cells, the receptor status using immunohistochemistry and the detection of lymph node metastases in tissue samples. Quantitative measures of cancer cluster shape and surrounding ECM do not exist despite the importance of rising cell motility and desmoplasia for tumour progression. Pathology  reports  occasionally note invasion front texture but not individual cluster irregularity within the primary tumour. Lymphovascular invasion serves as the only motility marker used in routine diagnosis, which only detects these cells after they have left the primary tumour. Cluster shape may provide a measure of the degree of cancer cell motility in the primary tumour. With the measure \textit{defect-cluster distance} we may have a first quantitative tumour marker that concerns the mechanical state of the ECM. Both prognostic metrics have the potential to fill an important gap in digital histopathology in terms of assessing metastatic risk and overall survival (Fig.~\ref{fig:survivalcurves}).
Moreover, our nematic quantification of the fibrotic ECM may be used as a first marker to consider desmoplasia in diagnostics. \\

We caution, however, that the feedback between clusters and the ECM is likely to be complex: cluster instabilities and breakage increase the total surface between cell clusters and ECM. Small clusters foster exchange with the microenvironment, affecting metabolic supply, and chemical signalling \cite{pickup2014extracellular}. Moreover,  mechanotransductive signalling on stiffer ECM can increase active forces exerted by cells at the boundary, further enhancing tumour invasion \cite{Ghibaudo2008, Levental2009}. There is also a suggestion that cluster boundaries increase the selective advantage of new and more aggressive cancer cell phenotypes generated by genetic instability at the boundaries compared to new cell phenotypes emerging in the bulk of the cell clusters \cite{lavrentovich2015survival}. Conversely, the caging effects by the ECM structure around the inclusions hinders cancer cell escape from the clusters into the ECM, since most ECM fibres are aligned parallel to the boundary, and constrain the cancer cells \cite{mierke2018two}.
\section{Summary}
Invasive breast tumours are characterised by many cancerous cell clusters embedded within the ECM \cite{zegers2025soft, mallon2000basic, ilina2020cell}. The evolution of breast cancer takes place over a time scale of months to years, so an experimental investigation of the tumour progression is out of reach. Therefore, to understand the cluster dynamics at the scale of the whole tumour,  we compared histological slides from standard medical investigations of over 2,000 patients to a physically grounded model, which describes the clusters as active nematic inclusions embedded in the ECM, modelled as a viscoelastic nematic matrix. The active-passive nematic model simultaneously reproduces several distinct observables—including cluster-size and shape statistics, correlations between size and shape, ECM organisation, and the spatial distribution of ECM topological defects—using parameters determined independently from experimental measurements of tissue mechanics and cell motility.\\

The comparison of histology and simulations allows us to demonstrate that the pattern of tumour cell clusters within the ECM that characterises invasive breast cancer is the result of a dynamical process whereby clusters break up, move, collide and re-form, driven by the motility of the cancer cells. The same active stresses are transmitted into the surrounding matrix, where they bend the nematic director and nucleate the topological defects that concentrate near the cluster boundary. The flow generated by the moving clusters aligns the fibres parallel to that boundary and raises the ECM nematic order over a short screening length.\\

The physical quantities also prove to be prognostically informative. We find that both the cluster shape index and the ECM defect–cluster distance stratify disease-free survival in a large patient collective. Both quantities can be measured from the digitised H\&E sections, which are already produced in routine practice. Moreover we demonstrate a correlation between metastasis, cluster size and activity. This suggests that concepts from active nematic physics can provide new prognostic markers for disease progression.\\

In future research, it will be interesting to investigate the progression between the initial breach of the basement membrane \cite{burstein2004ductal, mallon2000basic, chang2017force} to the formation of the multi-cluster structure which characterises the invasive stage of breast cancer. Moreover, the onset of cancer motility is often driven by the collective motion of elongated cells triggered by upregulation of RhoA/ROCK and Rac1 \cite{nobes1995rho, sahai2002rho} or RAB5A \cite{malinverno2017endocytic}. Thus, active nematic theories provide a generic description of motile cells, constrained by fibrotic ECM and may therefore be broadly relevant across solid tumours. This is suggested by very recent findings on cluster size distributions in pancreatic tumours \cite{kiemen2024power} and nematic structures in glioma tumours \cite{argento2025three}. Since nematic physics covers all scales from local events to long term transport phenomena it may ultimately provide a general description of mechanical tumour progression. 

\section{Materials and Methods \label{methods}}

\subsection{Histological tissue sections}
\label{sec:Methods_HnE}
Imaging and staining of patient breast cancer tissue was performed following standard protocols. H$\&$E-staining was carried out using the Leica ST4040 automat. The formalin-fixated breast cancer tissue sections were then optically inspected and digitized using the Panoramic 1000 Flash IV Scanner of 3DHISTECH with a Zeiss Plan Apochromat 40x objective. Each patient is represented by a tissue micro-array core of diameter $\sim 1000 \mu$m which was sampled from within the centre of the primary tumour. Patients were informed and gave ethical, informed consent, approved by the ethics committee of the Ärztekammer Hamburg ('Medical Council Hamburg') with processing number PV2946.

\subsection{Nematic image analysis \label{methods:imageAnalyses}}

Following a custom colour-deconvolution algorithm based on the work of Macenko et al. \cite{macenko2009method}, we use the eosin-intensities to characterize ECM ordering and haematoxylin-intensities for nuclei, respectively. Cancer cell clusters are identified via an algorithm described in our previous work \cite{gottheil2023state}. Using this intensity image $I$, we compute the structure tensor $J$ by convolving (denoted by $*$) with a Gaussian kernel $G(\sigma)$ \cite{bigun1987optimal},
\begin{equation}
    J = G(\sigma)* (\nabla I)(\nabla I)^{T}\:.
\end{equation}
The kernel size is $2 \cdot 3 \sigma$, ignoring low-weight contributions greater or lower than $+3 \sigma$ and $-3\sigma$, respectively. We use $\sigma \approx 1 \mu m$ for the ECM fibres, and $\sigma \approx 5 \mu m$ for the cancer cells.

For each pixel $\textbf{p}$, we need to calculate the local ECM or cell orientation $\textbf{n}=(\cos\theta,\sin\theta)$. We do this by solving the eigenvalue problem $\textbf{J}(\textbf{p})\textbf{v} = \lambda \textbf{v}$. The eigenvector $\textbf{v}_{min}$ corresponding to the smaller eigenvalue $\lambda_{min}$ describes the local ECM or cell orientation $\textbf{n}$ at that pixel \cite{guillamat2022integer}. We ignore regions where the orientation field has low coherency ($\frac{\lambda_{max} - \lambda_{min}}{\lambda_{max} + \lambda_{min}} < 0.1$) or low energy ($|\lambda_{max}| + |\lambda_{min}| < 0.01$).

We calculate the coarse-grained nematic tensor \textbf{Q} from the orientation field $\textbf{n}$ as
\begin{align}
    Q_{\alpha \beta} = 2 \langle n_\alpha n_\beta - \frac{1}{2} \delta_{\alpha \beta} \rangle \:.
\end{align}
Here, $\delta_{\alpha \beta}$ is the Kronecker-Delta, and $\langle ... \rangle$ denotes taking a Gaussian average over space with standard deviation $\sigma$. Here, we use a standard deviation of $5 \mu$m for ECM fibres and $25 \mu$m for cancer cells. The nematic order parameter can be found by diagonalizing $Q_{\alpha \beta}$, concretely $S = \sqrt{Q_{xx}^2 + Q_{xy}^2}$.

We find topological defects by evaluating the winding number 
\begin{equation}
     \text{Winding number} = \frac{1}{2\pi} \oint_\gamma d\theta
     \approx  \frac{1}{2\pi} \left( \frac{\partial \theta}{\partial x} * \dot{\gamma}_x +   \frac{\partial \theta}{\partial y} * \dot{\gamma}_y \right)\:,
\end{equation}
where we integrate over paths $\gamma$ which are rings with a radius of 4 pixels, and $\dot{\mathbf{\gamma}}(s)=(\dot{\gamma_x}(s),\dot{\gamma_y}(s))$ is the derivative of the path $\gamma(s)$ with respect to the parametrization $s$.

\subsection{Fixed and Live Tumour Sample Imaging and Tracking Workflow}
\label{sec:Methods:ImagingStainingWorkflow}

Primary breast cancer tumour samples were obtained immediately after surgical excision and processed under sterile conditions. Tumour tissues were sectioned into smaller fragments using sterile scalpels and either fixed for immunofluorescence analysis or prepared for live-cell imaging, as detailed below.

\subsubsection*{Fixed Sample Preparation and Staining}

For fixed imaging, tissue fragments were immersed in 4\% (wt/vol) paraformaldehyde in phosphate-buffered saline (PBS) for 30 minutes at room temperature. Following fixation, samples were washed three times with PBS (10 minutes each) to eliminate residual fixative. Permeabilization was achieved by overnight incubation in 1\% Triton X-100 in PBS at 4\textdegree C, followed by additional PBS washes.

Fixed tissues were stained overnight at 4\textdegree C in PBS containing 1\% bovine serum albumin (BSA) with 1~$\mu$M Alexa Fluor 488-Phalloidin (Cat.No: A12379, Thermo Fisher Scientific) to visualize F-actin, and 1~$\mu$M SPY650-DNA (Cat.No: SC501, Spirochrome) for nuclear labeling.

After staining, samples were mounted in ibidi IMM mounting medium (Cat.No: 50001, ibidi; refractive index $\approx$ 1.445) and transferred into $\mu$-Slide 8 Well chambers (Cat.No: 80821, ibidi). Confocal imaging was performed using a Zeiss LSM 980 microscope equipped with AiryScan2. A Zeiss LD LCI Plan-Apochromat 25$\times$/0.8 Imm Korr DIC M27 objective was used with a glycerol-water immersion medium (refractive index $\approx$ 1.45) to match the refractive index of the mounting medium. Z-stack images were acquired at 1~$\mu$m intervals for high-resolution 3D reconstruction.

\subsubsection*{Live-Cell Imaging of Tumour Tissue and Tracking Analyses}

For live imaging of cell dynamics (see Supplemental Movie S5 \cite{SM}, Appendix \ref{app:dominant_activity} and main text), freshly sectioned tumour fragments were transferred into $\mu$-Slide 6 Well Glass Bottom chambers (Cat.No: 80637, ibidi) containing DMEM with 4.5~g/L glucose (Cat.No: DMEM-HA, Capricorn Scientific), supplemented with 0.2~$\mu$M SPY650-DNA (Cat.No: SC501, Spirochrome) to label nuclei in viable cells. Samples were incubated overnight at 37\textdegree C in 5\% CO\textsubscript{2} to ensure sufficient dye uptake and cell viability.

Live imaging was carried out over an 18-hour period with a time-lapse interval of 15 minutes. Samples were imaged directly on the Zeiss LSM 980 confocal microscope, which is equipped with an integrated Zeiss incubation chamber to maintain physiological temperature (37\textdegree C), 5\% CO\textsubscript{2}, and humidity throughout the imaging session. A Zeiss EC Plan-Neofluar 20$\times$/0.5 Ph 2 M27 objective was used, and z-stacks were acquired at 0.8~$\mu$m intervals to capture 3D cell migration and behavior over time.

We subsequently track the cells using custom Python scripts.
We processed the sequence of 3D image stacks of the time-lapse volumetric microscopy data. For each time point the corresponding 3D image is loaded, cell nuclei (elliptical structures) are detected using the Laplacian of Gaussian (LoG) method (using in-built functions from \textit{skimage} python package), with specified parameters for minimum ($\approx 1.5 \mu$m) and maximum ($\approx 8 \mu$m) nucleus (blob) effective radii, number of probed length scales ($N = 8$), and allowed spot overlap ($30\%$). For blobs that are detected, their spatial coordinates (z, x, y) are recorded, along with the time frame index. The detected blobs across all frames are combined into a single data frame and tracked with in-built methods from \textit{trackpy}.
We performed a 3D Voronoi tesselation around the tracked coordinates in order to calculate the effective cell radii $R_{eff} = (3 \cdot V / (4 \cdot \pi))^{1/3}$, where $V$ is the volume of the Voronoi cell, and to find adjacent cells as the Voronoi neighbours. We find average effective cell radii of (median $\pm$ median absolute deviation) $\langle R_{eff} \rangle = 8.4\mu$m $ \pm 5.9 \mu $m and average number of next nearest neighbours of $N=14.1$. Next, we calculate the cage-relative mean squared displacements of cell \textit{i} as $MSD_{CR}^{i}(t) = | MSD^{i}(t) - \langle MSD^{i}_{nn}(t) \rangle|$, where $MSD^{i}$ is the $MSD$ of cell $i$, $MSD^{i}_{nn}(t)$ are the $MSDs$ at time $t$ of the neighbours of cell $i$ (we ignore tracks that are shorter than  $\approx 160$ minutes for this analyses; compare Fig. \ref{fig:SI_vital}b). We calculate the fraction of cells (tracks) that have at least one neighbour change event by evaluating whether the absolute cage-relative displacement exceeds a typical effective cell radius for each track $\sqrt{MSD^{i}_{CR}(t)} > \langle R_{eff} \rangle$. We calculate that $31 \%$ of cell tracks at least once changed neighbours at the cage-change-times $t$, where the condition $\sqrt{MSD^{i}_{CR}(t)} > \langle R_{eff} \rangle$ is satisfied (see Fig. \ref{fig:SI_vital}c).

\subsection{Proliferation fractions and proliferation-induced flows \label{methods:Proliferation}}
A patient's tumour tissue is stored in paraffin blocks from which sections (3-5 $\mu$m) are cut and stained. Additional sections of the primary tumours were made approximately 10 $\mu$m below the section plane of the analyzed HE-stained sections and stained with the mitotic marker Phospho-histone H3.

The quantitative analysis starts with a custom colour-deconvolution algorithm based on the work of Macenko et al. \cite{macenko2009method}. This operation separates the colour channels of the positive stain (brownish) and negative stain (blueish). Within the separate colour channels we employ the StarDist model \cite{schmidt2018cell} (pre-trained model for fluorescence images) to segment positive (mitotic activity) and negative (no mitotic activity) nuclei. The digital segments are filtered for areas $> (1.5 \mu m)^2 \cdot \pi$ and $< (6 \mu m)^2 \cdot \pi$ for reliable segments. The proliferation fraction $f_p$ is then calculated as $f_p = N_{\text{positive}}/(N_{\text{positive}} + N_{\text{negative}})$ per patient.

In the main text and Fig. \ref{fig:sizeshape} we estimate proliferation-induced flow speeds under the assumption of spherical clusters as:
\begin{align*}
    v_{\text{proliferation}} &\sim \frac{dR}{dt} = \frac{1}{4\cdot R^2} \frac{dV}{dt}\\
    &= \frac{1}{4\cdot R^2} \frac{f_d n_C V_C}{\tau_p} = \frac{R f_d}{3 \tau_p}
\end{align*}
where $V\sim \frac{4}{3}\pi R^3$ is the effective cancer cluster volume, $f_d$ is the measured proliferation fraction, $\tau_p$ is the time scale of mitosis, $V_C$ is the cell volume, $n_C$ is the number of cells in a clusters and $R$ is the effective cluster radius that we empirically measure in the histology sections as $R = \sqrt{A/\pi}$, where $A$ is the cross-sectional area of the cancer cluster.

\subsection{Mechanical probing of primary human breast cancer tissue\label{methods:AFM}}

Primary human cancer samples taken during breast tumour surgeries were temporarily stored in a phosphate-buffered saline (PBS) solution at 4$^\circ$C and measured within a few hours after removal. To prepare the tissue samples for AFM measurements, they were cut into thin slices of about 400$\mu$m thickness using sharp scalpels and a McIlwain tissue chopper. The slices were then glued on a thin glass slide using Histoacryl tissue adhesive (B.Braun Surgical, Rubí, Spain) and mounted in a 40-mm Petri dish (TPP, Trasadingen, Switzerland) filled with a few millilitres of modified ringer solution (B. Braun, Melsungen, Germany) to protect them from drying out. All measurements were carried out in a liquid environment at room temperature. 
Samples were measured using a Zeiss upright AxioZoom.V16 fluorescence microscope (Carl Zeiss Microscopy GmbH, Jena, Germany) equipped with the NanoWizard 4 XP setup and a 300 $\mu$m HybridStage (Bruker Nano GmbH, Berlin, Germany). A modified Pointprobe-CONT cantilever (0.2 N/m) (Nanoworld AG, Neuchâtel, Switzerland) with an attached polystyrene bead (6 $\mu$m diameter) was used as a spherical indenter for an increased contact area. Several thousand force indentation curves were measured across each sample surface with a set force of 2nN, covering an area of about $600 \mu \text{m} \times 600 \mu \text{m}$. For each force curve, the cantilever was held at constant height as soon as the set force was reached and sinusoidal oscillations were performed in a range from 10Hz-100Hz. After the mechanical probing was completed, Hoechst 33258 nucleic acid dye (Thermo Fisher Scientific, Waltham, USA) was added to the dish (mixing ratio 1:1000) to stain for DNA visualizing the nuclei positions to use for further tissue classification. The camera and cantilever positions were calibrated and after $\sim$30 min of incubation, the samples were imaged in tiles.
Young's moduli were obtained from a standard Hertz model fit \cite{hertz1881beruhrung, sneddon1965relation} with a Poisson's ratio of 0.5 using the JPK data processing software V7.1.18 (Bruker Nano GmbH, Berlin, Germany).  The viscoelastic properties, i.e. shear storage and shear loss, were calculated from the vibration data using a Sneddon model fit \cite{sneddon1965relation}.  Self-written MATLAB scripts (MathWorks, Natick, USA) were then used to statistically analyze the data and to stitch the individual image tiles. 

\subsection{Fitting Cluster Sizes and Shapes}

Cancer cell clusters in the histological images are identified via an algorithm described in our previous work \cite{gottheil2023state}. Subsequently, MATLAB's \textit{regionprops} function is used to measure areas and perimeters. We filter empirical cluster areas below $(10 \mu m)^2 \cdot \pi \approx 314 \mu m^2$ since below this area it is not clear if the cluster contains more than one cell. Also, areas above $1.153 \cdot 10^5 \mu m^2$ are ignored due to data fluctuations. 
In the simulations, we identified individual clusters by using the inbuilt contour plot function in MATLAB with a single intermediate level. We then calculated the area and perimeter of the clusters from the contour matrix data, using the inbuilt polygon area and perimeter functions in MATLAB. The width of cluster boundaries (where $0.1 \leq \phi \leq 0.9$) is approximately 5 LB $=25\mu$m, and we filter clusters smaller than this when extracting the power law exponents.

\subsection{Survival analyses, prognostic observables\label{methods:survival}}

We divide the patient collective presented in Table \ref{tab:All} into a random training set (Table \ref{tab:trainingSet}) and test set (Table \ref{tab:testSet}). As follow-up events, we take the \textit{death from any cause} and \textit{distant metastasis}. We censor all patients above 15 years of follow-up time to avoid events not related to cancer. Disease-free survival probabilities are estimated via the Kaplan-Meier estimator $\hat{S}(t) = \Pi_{i: t_i \leq t} \left( 1 - \frac{d_i}{N_i} \right)$, where $t_i$ are time steps where at least one event happened, $d_i$ is the number of events at $t_i$ and $N_i$ is the number of individuals known to have no event until time $t_i$. This estimates the probability of not exhibiting certain events over time. For risk group separation we search for the maximum separation in the training set based on the log-rank statistic \cite{Cardillo2024}, $\text{logrank} = \sum \limits_{i=1}^{m} \frac{(O_i - E_i)^2}{E_i}$, where $m$ is the number of groups, $O_i$ the total number of observed events and $E_i$ the total number of expected events. Additionally, each risk group must contain at least 15\% of patients in the training set to avoid unreliably small patient groups. The threshold found is then applied to the test set.

As observables we use the \textit{average cluster shape index $\langle SI \rangle$} per patient and the \textit{normalized maximum defect-cancer distances $\max(\Delta_{\text{DC}})/\xi_{\text{ECM}}$} per patient. The maximum distances $\max(\Delta_{\text{DC}})$ are normalized with the effective length scale of the ECM of the respective patient $\xi_{\text{ECM}}$. $\xi_{\text{ECM}}$ is determined by considering the radially averaged spatial pixel autocorrelations $C(r)$ of the binary ECM mask using the \textit{Wiener-Khinchin Theorem} \cite{wienerTheorem}. The length at $C = 1/e$ is then identified as $\xi_{\text{ECM}}$. This normalization is necessary since ECM length scales vary across patients and thus affect the distributions $\Delta_{\text{DC}}$. Thus, $\max(\Delta_{\text{DC}})/\xi_{\text{ECM}}$ presents a dimensionless, standardized measure without confounding effects of varying ECM sizes and associated length scales.

The thresholds found for risk group separation shown in the main text are $\langle SI \rangle = 1.62$ and $\max(\Delta_{\text{DC}})/\xi_{\text{ECM}} = 0.81$.

\subsection{Computational Model \label{methods:computational_model}} 

We distinguish the cancer clusters and ECM using a phase-field $\phi$, where $\phi=1$ labels the cancer clusters, and $\phi=0$ labels the ECM. This phase field relaxes to the minimum of the Landau-Ginzburg free energy
\begin{equation}
    \mathcal{F}_{\phi} = \frac{1}{2}A_\phi \,\phi^2 (1-\phi)^2 + \frac{1}{2} \kappa_\phi\, (\nabla\phi)^2  \:,
\end{equation}
where $A_\phi$ and $\kappa_\phi$ are material parameters describing the surface tension between the species, and the sharpness of the interface. In this formalism, the surface tension between the two phases is $\gamma = \frac{1}{6} \sqrt{\kappa_\phi A_\phi}$ \cite{kusumaatmaja2010lattice}. In the absence of growth, the dynamics of $\phi$ are given by the Cahn-Hilliard equation
\begin{equation}
    \frac{\partial\phi}{\partial t} + (\mathbf{u}\cdot\nabla)\phi = \Gamma_\phi \nabla^2 \mu \: ,
    \label{eq:CahnHilliard}
\end{equation}
where $\mathbf{u}$ is the flow field discussed later, $\mu = \frac{\delta \mathcal F_\phi}{\delta \phi}$ is the chemical potential, and $\Gamma_\phi$ is a mobility parameter. 

We model the cancer clusters and ECM fibres inside each region using a nematic order tensor and a flow field, labelled by an index $i$. We choose $i=1$ for the cancerous cells and $i=0$ for the ECM. We define the concentration of phase $i$ as $\phi^i$, where $\phi^1 = \phi$ and $\phi^0=1-\phi$. The nematic order of species $i$ is described by the symmetric and traceless rank-2 tensor $\mathbf{Q}^{i} = 2 S^{i}\,\Big( \mathbf{n}^i\, \mathbf{n}^i - \frac{\mathbb{I}}{2} \Big)$, where $\mathbf{n}^i$ is the vector orientation field denoting the average orientation, and $S^i$ is the magnitude of the nematic order parameter capturing the average degree of alignment. Each $\mathbf{Q^i}$ tensor relaxes to the minimum of the Landau de-Gennes free energy 
\begin{equation}
    \mathcal{F}_{LdG}^{i} = \frac{1}{2}\mathcal{C}^{i} \: \Big( ( S_{0}^{i}\phi^i)^{2} - \frac{\mbox{Tr}(\mathbf{Q}^{i\,2})}{2} \Big)^2 + \frac{K^{i}_{LC}}{2} \left( \mathbf{\nabla} \mathbf{Q} ^i\right)^{2} \:,
\end{equation}
where, $C^{i}$, $S^i_0$ and $K_{LC}^{i}$ are material parameters representing the bulk free energy, average nematic order, and nematic elasticity respectively for each species. We define the molecular field 
\begin{equation}
    \mathbf{H^{i}} = -\left( \frac{\delta \mathcal{F}_{LdG}^{i}}{\delta \mathbf{Q^{i}}} - \frac{\mathbf{I}}{2} \text{Tr}\left( \frac{\delta \mathcal{F}_{LdG}^{i}}{\delta \mathbf{Q^{i}}} \right) \right) \: .
    \label{eq:MolecularField}
\end{equation}

The flow field is modelled by a velocity field $\mathbf{u}$, which satisfies the following equations:
\begin{equation}
    \nabla\cdot\mathbf{u} = 0 \: ,
    \label{eq:FluidContinuity}
\end{equation}
\begin{equation}
    \rho \left( \partial_{t} + \mathbf{u} \cdot \mathbf{\nabla} \right) \mathbf{u} = \mathbf{\nabla} \cdot \mathbf{\Pi} \: .
    \label{eq:NavierStokes}
\end{equation}
Here, $\rho$ is the density of the fluid, and $\mathbf{\Pi}$ denotes the sum of all stresses acting on the fluid, composed of viscoelastic, active, nematic backflow, and capillary terms. Inside each phase ($\phi=0$ or $1$), this velocity field represents the mean-coarse grained flow field of the phase. Modelling the flow as a single continuous velocity field across the interface physically corresponds to imposing no-slip (continuity of velocity due to frictional forces between the cells and the ECM) and force-balance conditions at the interfacial region. 

The viscoelastic stress is modelled by the incompressible Kelvin-Voigt model \cite{thomson1865iv, voigt1892ueber, rajagopal2009note} 
\begin{equation}
     \mathbf{\Pi}^{i, visc-el} = - p \mathbf{1}+ 2 \rho \nu^{i} \, \mathbf{E} \: + E^i \, \mathbf{T},
\end{equation} 
where $p$ is an isotropic pressure term chosen to ensure incompressibility,  $\nu^{i}$ is the kinematic viscosity, $E^i$ is the elastic modulus, $\mathbf{E} = [(\nabla\mathbf{u})+(\nabla\mathbf{u})^T]/2$ is the strain rate tensor and $\mathbf{T} = \int \mathbf{E}(t) \, dt $ is the strain tensor. The total viscoelastic stress is
\begin{equation}
    \mathbf{\Pi}^{visc-el} = \sum_{i = 0}^1 \phi^i\, \mathbf{\Pi}^{i,  visc-el}
\end{equation}

The nematic director field generates passive stresses upon stretching and distortion of the director field, called the nematic backflow. This is given by
\begin{align}
    \mathbf{\Pi}^{i, nem} &= 2\lambda^{i} \big(\mathbf{Q^{i}} + \mathbb{I}/d \big) (\mathbf{Q^{i} : H^{i}})-  \lambda^{i} \mathbf{H^{i}}\cdot \big( \mathbf{Q^{i}} + \mathbb{I}/d \big)    \nonumber\\
& \qquad -\lambda^{i} \big( \mathbf{Q^{i}}  + \mathbb{I}/d \big)\cdot\mathbf{H^{i}} -\big( \nabla \mathbf{Q^{i}} \big)\, . \frac{\partial \mathcal F^{i}}{\partial \nabla \mathbf{Q^{i}}} \nonumber \\ & \qquad + \mathbf{Q^{i} \cdot H^{i}} - \mathbf{H^{i} \cdot Q^{i}} \:,
\end{align}
where $\lambda^i$ is the flow-alignment parameter that describes the strength of alignment of the director axis with the strain rate in the fluid, and $\mathcal{F}$ describes the total free energy. 

Each species can produce active nematic stresses along the long axis of the director field, modelled by an active stress \cite{doostmohammadi2018active}
\begin{equation}
    \mathbf{\Pi}^{i, act} = - \zeta^i \mathbf{Q^{i}}
\end{equation}
where $\zeta$, the activity, characterizes the typical mechanical stresses exerted by each species on its environment. In our model, only the cancer clusters are active and hence $\zeta^0 = 0$.

The capillary stress models the stresses on the fluid across the boundary \cite{lazaro2015phase}. This is given by
\begin{equation}
    \mathbf{\Pi}^{cap} = \big( \mathcal{F}_\phi -  \mu \phi \big)- \nabla\phi \Bigg( \frac{\partial \mathcal{F}_\phi}{\partial \,(\nabla\phi) }\Bigg) \:.
\end{equation}
The total stress tensor acting on the fluid is then given by
\begin{equation}
     \mathbf{\Pi} = \mathbf{\Pi}^{visc-el} + \mathbf{\Pi}^{cap} + \sum_{i=0}^1 (\mathbf{\Pi}^{i, \,nem} + \mathbf{\Pi}^{i, \,act}) \: .
\end{equation}

The dynamics of the director field is modelled by a Beris-Edwards equation \cite{beris1994thermodynamics}
\begin{equation}
    D_t \mathbf{Q}^{i} -\mathbf{\mathcal{W}^{i}} = \Gamma \mathbf{H}^{i}
    \label{eq:BerisEdwards}
\end{equation}
Here, the first term of the left-hand side is the convected time derivative, which describes the rate of change of $\mathbf{Q}^{i}$, which is being advected by the flow $\mathbf{u}$. The second term $\mathcal{W}^{i}$ is the co-rotation term modelling the response of the director field to strain rate and vorticity in the flow. This is given by
\begin{align}
    \mathbf{\mathcal{W}^{i}} &= \left( \lambda^{i} \mathbf{E} + \mathbf{\Omega} \right) \Big( \mathbf{Q}^{i} + \frac{\mathbb I}{2} \Big) + \Big( \mathbf{Q}^{i} + \frac{\mathbb I}{2} \Big) \left( \lambda^{i} \mathbf{E}-\mathbf{\Omega} \right) \nonumber \\ & \qquad \qquad- 2\lambda^{i} \Big( \mathbf{Q}^{i} + \frac{\mathbb I}{2} \Big) (\mathbf{Q}^{i}:\mathbf{E}) 
\end{align}
where $\mathbf{\Omega} = [(\nabla\mathbf{u})-(\nabla\mathbf{u})^T]/2$  is the vorticity tensor.  The final term on the right-hand side of Eq. \eqref{eq:BerisEdwards} is the molecular field (Eq. \eqref{eq:MolecularField}), where $\Gamma$ describes the rate of relaxation to the free-energy minimum. Finally, the model assumes free boundary conditions on each nematic fields $\mathbf{Q}^{i}$ at the cancer-ECM interface, where they are free to respond to the underlying flow field. This setup for the cancer clusters is analogous to the model used to study active anchoring \cite{blow2014biphasic}. In our model, the ECM is a passive nematic driven by active flows.

\subsubsection*{Simulation Parameters} 

We choose simulation parameters by first estimating the typical scales of the system.

\textbf{Elastic and viscous moduli:} We estimate the order of magnitude of the viscous $G''$ [Pa] moduli within cancer clusters and ECM via frequency-dependent AFM measurements on a primary human breast cancer sample from the University Hospital in Leipzig. At low frequencies ($f = 0.01$Hz), the viscous moduli for both cancer tumours and ECM is around $G''\sim 10 \, Pa$, resulting in an effective viscosity of $\eta_{1}= \eta_2= 160 \text{Pa} \cdot \text{s}$ for the entire system. Cancer clusters fluidize the ECM in their vicinity \cite{alcaraz2011collective}, making it hard to accurately measure the elastic shear modulus in the system. Typical scales of the ECM shear elastic modulus can be estimated between 100 Pa - 1 kPa \cite{acerbi2015human}. For a breast cancer sample measured with AFM, we estimate a shear modulus of the ECM $\sim 500$ Pa (see Appendix \ref{app:AFM}).

\textbf{Lengths and velocities:} Typical cell motility velocities are measured to be roughly $v_0 \sim 3 \mu m/h$ in vital primary human breast cancer, compare reference \cite{gottheil2023state} and Appendix \ref{app:dominant_activity}. Cancer cell tumours span a wide range of length scales, but typical cluster radii are on the scale of $R \sim 10-100 \mu m$ (see size distribution in main text Fig.~\ref{fig:sizeshape}a).

\textbf{Calibrating surface tension:} We estimate this from cultured breast cancer cell spheroid fusion experiments reported in Ref. \cite{grosser2021cell}, which show that touching cancer clusters merge in $t \sim 36$hr.

\textbf{Active forces:} Activity models the typical forces exerted by the cancer cells on their environment, which can include traction forces due to motility, and forces due to proliferation. The typical magnitude of these forces is around $\zeta \sim 400 \,\text{Pa}$ \cite{kraning2012cellular}. We caution that there will be considerable variation in these quantities between different biological samples.  \\

With our choice of scales, we can set normalized simulation units, hereafter referred to as `LB units'. We choose the typical length, time and stress units to be $\Delta x = 5\,\mu m$, $\Delta t = 30\,s$, and $\Sigma = 5\,\text{kPa}$ respectively. We use the elastic moduli $E^0 = 0.1$ $=500\,\text{Pa}$ (see Appendix \ref{app:AFM}), $E^1= 0.001$ $=5\,\text{Pa}$, and activities in the range $\zeta^1 \in [0.007, 0.2]$ $= [35\,\text{Pa}, 1000\,\text{Pa}]$ (See \ref{app:robustness}), and $\zeta^0 = 0$. We further use $\rho = 40$,  $\nu^1=\nu^0 = 1$, $A_\phi=K_\phi = 1$, and $\Gamma_\phi=0.33$ for the remaining fluid parameters. 

The nematic parameters for the cancer clusters are chosen to be  $\mathcal{C}^1 = 0.1$, $\mathcal{K}_{LC}^1 = 0.1$, $S^1_0 = 1$, $\lambda^1 = 0.1$, and $\Gamma=0.33$. These are typical parameters for a nematic liquid crystal in an ordered phase, where the correlation length of the nematic orientation field is typically a few lattice spacings ($\sim10-20\,\mu m$). 

For the ECM, we use $\mathcal{C}^0 = 0.02$, $\mathcal{K}_{LC}^0 = 0.01$, $S_0^0 = 0$, and $\lambda^0 = 0.1$ for the nematic field instead. With these parameters, the 
Landau de-Gennes free energy has a weak effect on ECM dynamics (Eq.\eqref{eq:BerisEdwards}), and topological defects and nematic alignment of the ECM are driven by active flows in the ECM.

With these parameter values, the typical active velocity is $\sim 6\times10^{-3}$ (corresponding to $3.6\, \mu m \, hr^{-1}$ ). Two droplets numerically initialized adjacent to each other merge, and the notch between two droplets of radii $20$ ($=100 \mu m$) disappears in approximately $6\times10^3$ timesteps (corresponding to $\sim 50$ hr, see Supplementary Movie S4 \cite{SM}, which is in agreement with breast cancer spheroid fusion experiments \cite{grosser2021cell}). More details about the parameter choice, and the robustness of the simulation results to changes in parameter values, are included in the Methods and Appendix \ref{app:robustness}.

We ran simulations in a periodic box of size $400\times400$ for $400,000$ timesteps, with data recorded every $2000$ timesteps. We either initialized the system with a circular droplet with radius $100$, covering $\sim 20\%$ of the simulation box (Supplementary Movie S1 \cite{SM}), or with a uniform value of $\phi=0.25$ everywhere, resulting in many small droplets of $\phi=1$ covering $25\%$ of the simulation domain (Supplementary Movie S2 \cite{SM}). Both cases resulted in the same steady state. To obtain the results in Fig. 4, we ran simulations starting from a circular cluster of radius $R = 40$ in a periodic box of size $200\times200$ for $84,000$ timesteps, with data collected every $2000$ timesteps. In all cases, the nematic director field was initialized randomly and simulated for 100 pre-initialization steps without activity. Supplementary Movie S7 \cite{SM} shows a simulated cancer cluster breaking apart, along with ECM nematic order, velocity field and vorticity field. 

Although both Tiribocchi et al.~\cite{tiribocchi2015modelh} and Adkins et al.~\cite{Adkins2022} observe arrested phase separation and irregular cluster shapes, these papers model different physical systems. They differ in how and where the nematic forces act. Adkins et al.~\cite{Adkins2022} have active nematic forces in the background phase instead of in the cellular droplets, and Tiribocchi et al.~\cite{tiribocchi2015modelh} add dipolar active stresses only at the droplet interface. Our choice of model reflects that there is activity throughout the cancer clusters, and the surrounding ECM is viscoelastic.

\subsubsection*{Averaging over different activities}

Different cancer clusters, within and across samples, show a spread of motility values \cite{grosser2021cell, gottheil2023state}. To compare simulation results to clinical data, we average over a distribution of activity values. The robustness of our results to different activity sampling schemes is discussed in Appendix \ref{app:robustness}.

\subsection{Anchoring distributions}
In the histology slides, we estimate boundary tangents from digitally interpolated (Matlabs in-built Savitzky-Golay-Filter \textit{sgolayfilt}) pixel boundary segments. Angles between the respective tangents and ECM orientations in a square region ($5 \mu m \times 5 \mu m$) in front of the boundary are calculated and the local median is used as the anchoring angle.

In the simulations, we calculate anchoring angles by measuring the angle between the ECM orientation and the maximum gradient of $\phi$ in regions close to the cancer-ECM boundary ($0.05 \leq \phi \leq 0.2$).

\subsection{Decellularized 3D ECM imaging}
\label{sec:methods:dECM_Imaging}
In Appendix \ref{app:ECM_dimensionality}, we present findings on the relationship between 2D and 3D nematic ordering within the context of 3D ECM. The imaging of samples was conducted as follows:\\ 
Samples of human liver-derived decellularized extracellular matrix (dECM) were thawed at +8°C and subsequently transferred to a 40 mm petri dish (TPP, Trasadingen, Switzerland) containing cooled phosphate buffered saline solution (PBS)(Life Technologies Limited, Paisley, United Kingdom). A major sample piece of dECM was cut using a scalpel such that a minor piece of $\sim 3 \times 3$ mm size was obtained. A PBS solution containing 0.5\% of Tetramethylrhodamine (TAMRA) was prepared. The dECM slice was submerged into the solution and incubated for $\sim$24 hours at 37°C. After the labeling was complete, the sample was washed with PBS in a total of 3 washing steps with a 5 minute halt in between. The sample was subsequently placed in a $\mu$-slide 18 well plate (Ibidi, Gräfelfing, Germany) and submerged in 500 $\mu$l glycerol. 
For the microscopy, an Axio observer Z1/7 (Zeiss, Jena, Germany) equipped with a Yokogawa CSU-X1A 5000 spinning disk confocal scanning unit was used in fluorescence mode. The objective in use was an LCI Plan-Neo-fluar 25x/08 Imm (Zeiss, Jena, Germany). The staining excitation at 500 nm wavelength was performed using a CSU-X1 laser at 27\% intensity. The image capturing was performed using a Hamamatsu Orca flash 4.0. The $\mu$-slide containing the sample was placed on the holder stage and the objective was slowly approached to the sample until the glycerol droplet connected both the objective and sample vessel. We used the ZEN 2.6 (Zeiss, Jena, Germany) built-in image tessellation application and tessellated the sample into 8 equal-sized areas. The upper and lower z-boundary of the sample within the tessellated area was estimated by changing the focus. We divided the range from the lower to the upper boundary ($\sim$405 $\mu$m) yielding a total of 196 images with a step size of 2$\mu$m in z-direction. The total recorded volume of 811 $\mu$m x 1070 $\mu$m x 405 $\mu$m was processed by using ZEN 2.6 built-in image convolution and background subtraction algorithm.
\subsection{3D reconstruction from serial histology \label{methods:3Dreconstruction}}

Whole slide images of an invasive breast cancer patient from collaborators from the \textit{Institute of Pathology} at Charité – Universitätsmedizin Berlin (approved ethics vote number 2026-213-CBF) were serially sectioned at $4\,\mu$m $z$-spacing, H\&E-stained, and digitized at
$0.27\,\mu$m/px. The section stack was reconstructed in 3D using the TIAToolbox
whole-slide-image framework~\cite{pocock2022tiatoolbox}, following the
coarse-to-fine scheme common to serial-histology 3D reconstruction
workflows~\cite{kiemen2022coda}.

Registration proceeded in two stages, both aligning every moving section to a
common central reference slice $z_{\text{ref}}$. First, a \textit{global} stage
operating on strongly downsampled images ($30\,\mu$m/px) estimated a
rigid/affine transform by deep-feature-based registration (DFBR) and manually segmented masks of the WSI, followed by a
coarse B-spline (grid spacing $\sim\!12$~mm) to correct large-scale
translation, rotation and gross tissue distortion. Second, at $8\,\mu$m/px refined each section against its nearest
already-registered neighbour with a finer affine and B-spline (grid spacing
$\sim\!1$~mm), recovering the smaller, spatially varying deformations
introduced by sectioning. The global and local transforms were composed into a
single effective displacement field per section that maps it directly onto the
$z_{\text{ref}}$ frame; these fields were stored at the $8\,\mu$m/px local
registration resolution. Sections with badly torn or folded tissue were
excluded to limit error accumulation.\\
We exported the stack as an aligned 3D volume at $4\,\mu$m/px working resolution used for all cluster analysis and added a second more local registration procedure. Coarsely-aligned images, were refined against the corresponding $z_{\text{ref}}$ tile by a residual affine transform, estimated from LoFTR feature correspondences~\cite{sun2021loftr}, and low-pass filtered along $z$, followed by a B-spline (TIAToolbox, maximizing the Mattes mutual information).\\
Because some sections were excluded or missing, the stack was split at these gaps into well-registered subvolumes, which were analysed individually.

\subsection{3D cluster segmentation and morphological analysis \label{methods:3DClusterAnalyses}}
Within each subvolume we performed colour deconvolution~\cite{macenko2009method}
and analysed the haematoxylin channel. The volumes were at
$4\,\mu$m voxels. Cancer-cell clusters were segmented by thresholding the
haematoxylin intensity, removing objects smaller than a single cell
($(4/3)\pi(10\,\mu \text{m})^3$), and
filling holes. Clusters were then identified as connected
components in 3D (26-connectivity). Clusters touching any face of the subvolume
were discarded, since it is unknown whether they extend beyond the observable
volume; clusters below $\tfrac{4}{3}\pi(10\,\mu\text{m})^3$ were also excluded
as sub-cellular.\\
For each 3D cluster we measured its volume $V$ as the voxel count, and its 3D
shape index $SI_{3D}=A_{s}^{3/2}/(6\sqrt{\pi}\,V)$, which equals $1$ for a sphere
and increases with surface irregularity. The surface area $A_{s}$ was computed
from a smoothed isosurface mesh rather than the raw voxel boundary, to suppress
the staircase artefact of the discretised surface. To relate the 3D
morphology to the 2D histological measurements of the main text, we also
extracted, on every section plane of each subvolume, the cross-sectional
cluster areas $A$ and the 2D shape index $SI_{2D}=P/\sqrt{4\pi A}$, using the
same definition as in Sec.~\ref{methods:imageAnalyses}.\\
Analyses of this data are mentioned in the main text and shown in Appendix \ref{app:3DClusterShapeSize}.

\subsection{Statistical tests and correlation \label{methods:StatisticalTests}}

\textbf{Kolmogorov–Smirnov test:} This is a non-parametric hypothesis test that is commonly used to test if two distributions are statistically different. It evaluates the difference between the cumulative distribution functions of the distributions of the two sample distributions. The test statistic is defined as $D^{*} = \max_x \{ | F_1(x) - F_2(x) |\}$ where $F_1(x)$ ($F_2(x)$) is the proportion of $x_1$ ($x_2$) values less than or equal to $x$. We use an in-built MATLAB function to perform this test.  \\
\textbf{Spearman correlation:} The \textit{Spearman rank correlation coefficient} is a non-parametric measure of the strength and direction of association between two ranked variables, defined as:
$  \rho = 1 - \frac{6 \sum d_i^2}{n(n^2 - 1)} $, where \( d_i \) is the difference between the ranks of the corresponding values of the two variables, \( n \) is the number of data points. It evaluates how well the relationship between the observables can be found using a monotonic function. The coefficient ranges from \( -1 \) (perfect negative correlation) to \( +1 \) (perfect positive correlation), with \( 0 \) indicating no monotonic relationship.
We use an in-built MATLAB function to calculate this coefficient. In the context of activity and cluster sizes each value of each observable is given a rank and the negative correlation value means that as the rank of the activity parameter increases the rank of the cluster sizes decreases, that is, the clusters become smaller. \\
\textbf{Kendall correlation:} The \textit{Kendall rank correlation coefficient} $\tau$ is a non-parametric measure of association based on the number of concordant and discordant pairs, $\tau = (n_c - n_d)/\binom{n}{2}$, where $n_c$ ($n_d$) is the number of concordant (discordant) pairs among the $n$ observations. Like the Spearman coefficient it ranges from $-1$ to $+1$ and captures monotonic association. We use an in-built MATLAB function to calculate it. \\
\textbf{Distance correlation:} The \textit{distance correlation} \cite{szekely2007measuring} quantifies statistical dependence between two variables and, unlike the Pearson, Spearman, or Kendall coefficients, is zero if and only if the variables are independent, thereby also detecting non-monotonic dependence. It is defined as $\text{dCor}(X,Y) = \text{dCov}(X,Y)/\sqrt{\text{dVar}(X)\,\text{dVar}(Y)}$, where the distance covariance $\text{dCov}$ and distance variances $\text{dVar}$ are computed from the doubly-centred matrices of pairwise distances between observations \cite{szekely2007measuring}. It takes values in $[0,1]$. We evaluate it on the logarithm of the cluster size, and, due to its quadratic complexity in the number of samples, on a random subsample of the data.\\
\textbf{Mann-Whitney:} The two-sided Mann-Whitney $U$ test (Wilcoxon rank-sum test) was used to compare per-patient cluster size measures between groups. All observations from both groups are pooled and ranked, and the test statistic is $U = R_1 - n_1(n_1+1)/2$, where $R_1$ is the sum of ranks in the first group and $n_1$ its sample size.

{\bf Data availability}: All data to reproduce the conclusions of our study are attached in the online contents or are available upon request, with the following exception: Raw digitized images of H$\&$E-stained invasive breast cancer and their respective raw follow-up data cannot be publicly shared due to privacy concerns and existing material and data transfer agreements. Qualified researchers can apply for access by sending an initial request to the Pathology Hamburg-West GmbH. Subsequently, the researcher submits a brief research proposal and a standard form describing the project, data/materials requested, applicable ethics, and purpose. Requests will be reviewed and discussed based on scientific merit. The time-frame of response to an initial request is about 1-2 months. After approval, the parties will agree on the conditions of a data access/sharing agreement and restrictions of use, which may increase the total time-frame to around 6 months.\\
Patients were informed and gave ethical, informed consent, approved by the ethics committee of the Ärztekammer Hamburg (Medical Council Hamburg) with Processing No. PV2946.

{\bf Code availability}: Code to reproduce the findings can be found in the following repositories: github.com/pgotth/Cancer-Nematics
and github.com/saraswatb/CancerNematics.\\

{\bf Acknowledgements}: We thank Thomas Suchanek, 
Jacques Prost, Pascal Silberzan, Carles Blanch-Mercader, Steffen Grosser and Maxim O. Lavrentovich for helpful discussions. S.B. acknowledges funding from the Crewe Graduate Award (Lincoln College, Oxford) and the Rudolf Peierls Centre for Theoretical Physics. J.M.Y. acknowledges support from ERC Advanced Grant Act-Bio (funded as UKRI Frontier Research Grant EP/Y033981/1). J.A.K, I.M.S, K.L. and K.R. acknowledge funding from the German Research Foundation FOR5628. A.D. received funding from the Berlin Institute of Health at Charité – Universitätsmedizin Berlin, BIH Biomedical Innovation Academy, BIH Charité Clinician Scientist Program, Berlin, Germany. J.A.K acknowledges support from the Greve Prize from the German National Academy of Sciences Leopoldina.

{\bf Author contributions}: J.M.Y., J.A.K, S.B. and P.G. wrote the manuscript. J.M.Y., J.A.K, S.B. and P.G. conceptualized the study. S.B. performed computer simulations, data analyses and data visualization. P.G. and K.L. performed quantitative analyses of histology slides. S.B. and P.G. compared clinical images and simulations and performed respective data analyses and visualizations. P.F. performed AFM measurements and P.G. analysed the respective data. H.K. assisted data analysis, data interpretation and data visualization. I.S., A.D., J.W. prepared and provided 3D dECM. K.R. and S.R.M. imaged 3D dECM. P.G. analysed dECM. B.A. performed surgery and provided 3D breast cancer tissue. M.M. prepared and imaged 3D (vital) primary breast cancer. P.G. analysed (vital) 3D primary breast cancer. K.H., J.-U.B., C.S. A-S.W., A.J. and A.N. discussed/generated histology and patient data and contributed to conceptualization of the study. J.M.Y. and J.A.K. supervised the study. All authors provided feedback and gave approval for the final version of the manuscript.

\clearpage
\appendix

\section{Dimensionality analyses (2D-3D relations) \label{app:dimensionality}}

\subsection{ECM nematic order dimensionality\label{app:ECM_dimensionality}}
\begin{figure}[htp]
\includegraphics[width=0.5 \textwidth]{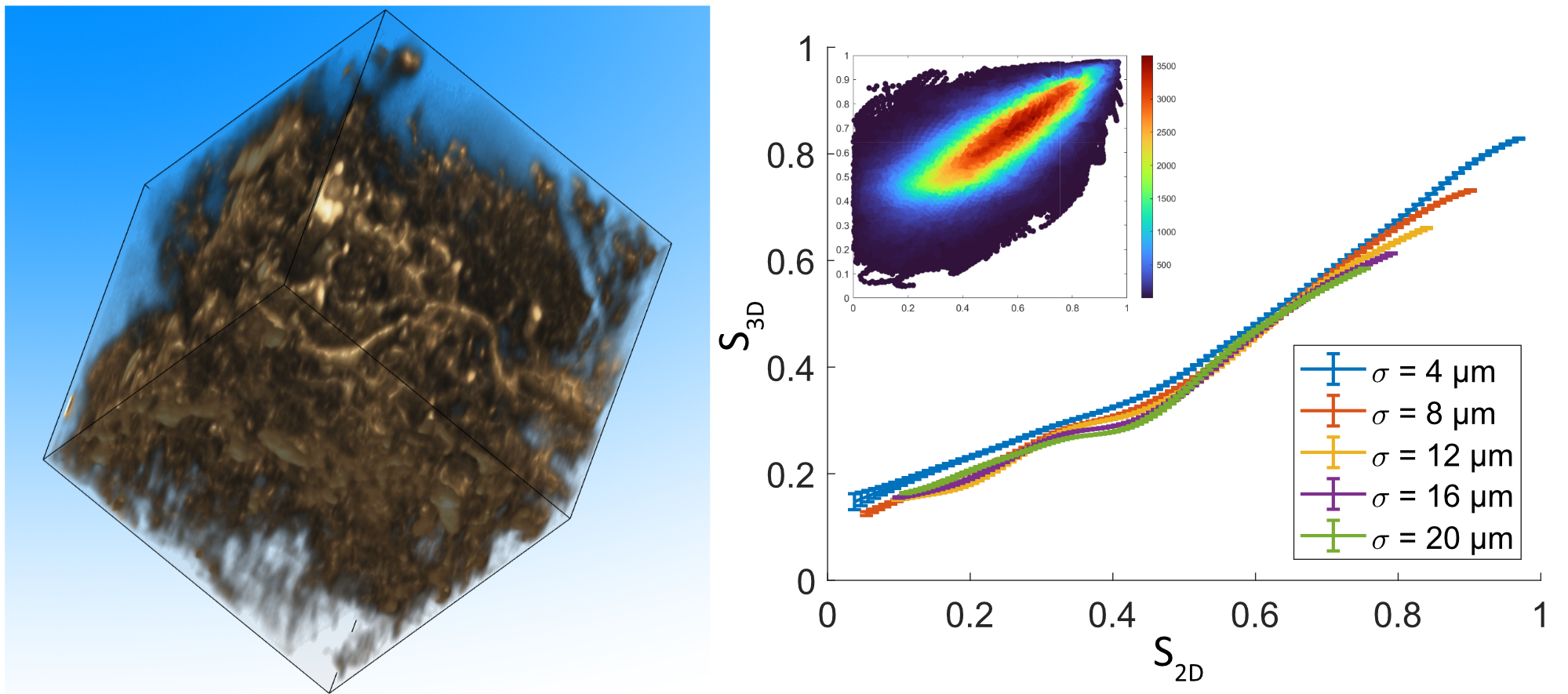}
\caption{\textbf{Relation between 2D and 3D ECM scalar nematic order parameters.} Left: Digital dECM volume of TAMRA-stained primary human liver tissue. For visibility, low intensities are transparent. The edge length of the volume corresponds to $200\mu$m. Right: scalar nematic order parameter in 2D planes $S_{2D}$ and in the volume $S_{3D}$ for different coarse graining lengths $\sigma$ of $\langle \textbf{Q} \rangle_\sigma$. Data is averaged using a Gaussian moving average with a standard deviation of 0.05. The inset shows raw data for $\sigma = 4 \mu m$. The average Pearson correlation coefficient across coarse graining length scales is $R = 0.79$.}
  \label{fig:2D3D_nematicOrderComparison}
\end{figure}

Histological H$\&$E-stained slides are thin sections from 3D tumour tissue where the ECM is stained by eosin. Histological slides with eosin-stained ECM are prepared as the standard for all operable tumours. The sections are $3-5 \mu m$ in thickness. The measurable 2D orientation field arises from projecting the eosin-absorbance onto the plane visible by light microscopy. Therefore, as long as the nematic orientation is not perfectly perpendicular to the section, 2D order parameters should also indicate the local 3D order. We tested this with decellularized ECM (dECM) of human liver tissue stained for TAMRA (Fig.~\ref{fig:2D3D_nematicOrderComparison}) and 3D spinning disk microscopy (see Methods, Sec.~\ref{sec:methods:dECM_Imaging}). The dECM was provided by the Charit\'{e} in Berlin (Ethics approval from the Ethikkommission der Charité – Universitätsmedizin Berlin exist for the use of tissue removed in the course of partial liver resection to obtain and cultivate liver cells and matrix components (EA1/289/16)). For comparison, we calculate the 2D scalar order parameter as described in Methods (Sec.~\ref{methods:imageAnalyses}) in the main text for every $xy$-plane. 3D scalar order parameters are similarly calculated by considering the respective Q-tensor $Q_{\alpha \beta} = \frac{3}{2} \,\langle n_\alpha n_\beta - \frac{1}{3} \,\delta_{\alpha\beta} \rangle_\sigma$ ($\alpha, \beta \in \{x,y,z\}$) and diagonalizing it:

\begin{align*}
	\langle Q \rangle_\sigma = \begin{bmatrix}
S & 0 & 0 \\
0 & -\frac{1}{2}S & 0 \\
0 & 0 & -\frac{1}{2}S
\end{bmatrix}.
\end{align*}
This enables us to compare $S_{2D}$ and $S_{3D}$ at every pixel of the digital volume of dECM for different coarse-graining length scales $\sigma$, see Fig. \ref{fig:2D3D_nematicOrderComparison}. As expected, we observe a good correlation between $S_{2D}$ and $S_{3D}$, with an average Pearson correlation coefficient of $\bar{R} = 0.79$. This is expected because when the section plane is not entirely orthogonal to the local nematic director, the projection onto the 2D plane yields information about the direction in the dimensions of the plane. However, if a nematic phase in 3D is completely orthogonal to the section, no information on the orientation can be obtained, which is why such cases are directly discarded by the constraints that orientation coherency must be $>0.1$ and orientation energy $>0.01$ during the structure tensor analysis (described in Methods, Sec.~\ref{methods:imageAnalyses}). However, in real images, this is very rare.\\

\subsection{Cancer cell nematic order dimensionality\label{app:cell_dimensionality}}
\begin{figure}[htp]
    \includegraphics[width=0.45 \textwidth]{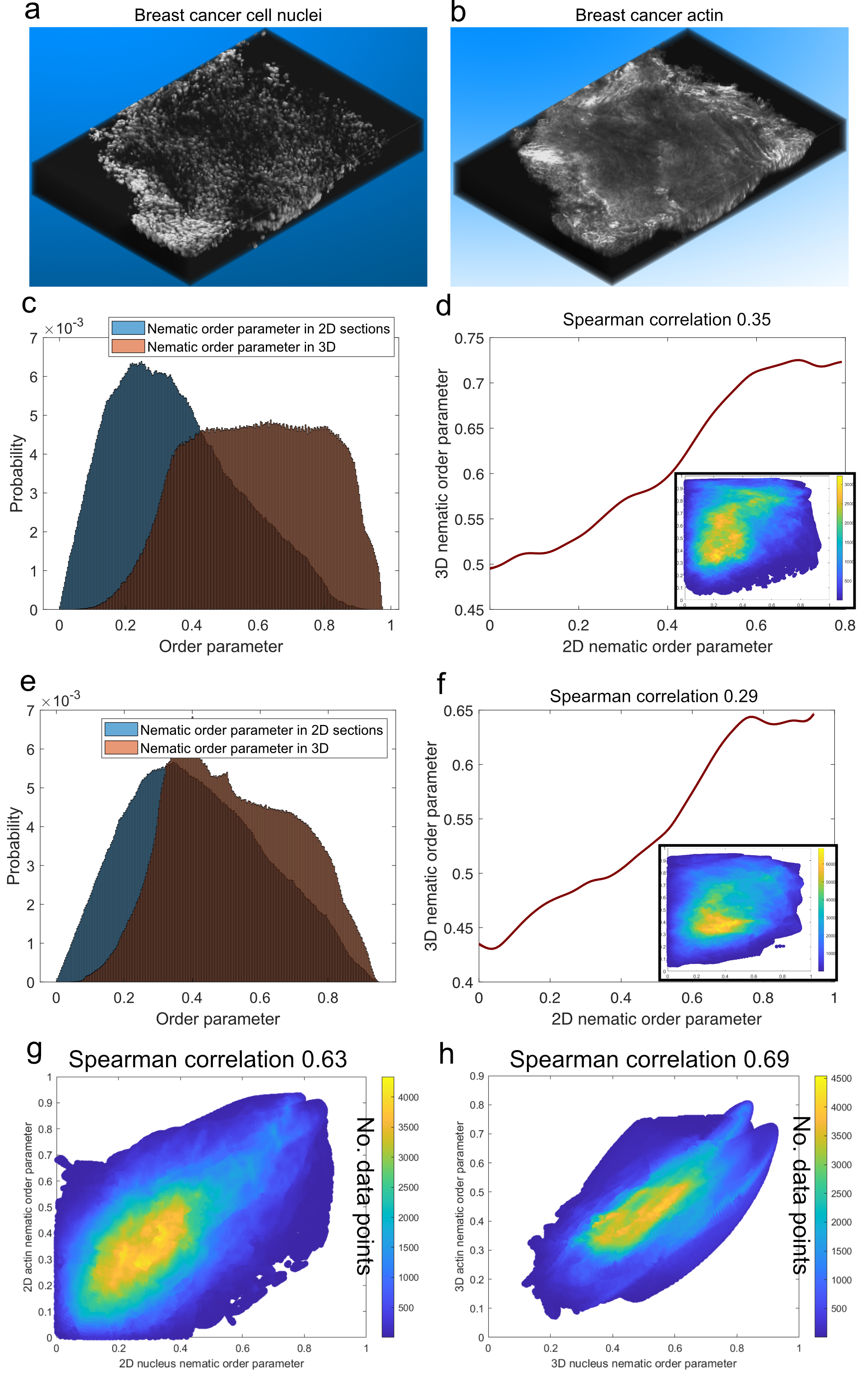}
    \caption{\textbf{Relation between 2D and 3D cell scalar nematic order parameters in primary human breast cancer.} \textbf{a} Digital tissue volume of primary human breast cancer ($489 \mu m \times 721 \mu m \times 86 \mu m$) stained for \textbf{a} DNA, \textbf{b} actin. \textbf{c} Scalar nematic order parameter distributions based on nucleus (DNA) intensities in 2D and 3D. \textbf{d} Gaussian moving average (standard deviation 0.05) of the correlation between 2D and 3D nucleus nematic order parameters. Inset shows the raw data. Spearman correlation coefficient 0.35. \textbf{e} Scalar nematic order parameter distributions based on actin intensities in 2D and 3D. \textbf{f} Gaussian moving average (standard deviation 0.05) of the correlation between 2D and 3D actin nematic order parameters. Inset shows the raw data. Spearman correlation coefficient 0.29. \textbf{g} Correlation between 2D actin and nuclei nematic order parameters. Spearman correlation coefficient is 0.63. \textbf{h} Correlation between 3D actin and nuclei nematic order parameters. Spearman correlation coefficient is 0.69.}
      \label{fig:2D3D_cellNematics}
\end{figure}

In the main text, we report 2D cell nematic ordering based on nucleus (DNA) intensities on histology slides. Here, we investigate the relation between 2D and 3D nematic cell ordering based on DNA (nucleus) and actin intensities in a 3D volume of primary human breast cancer shown in Fig.~\ref{fig:2D3D_cellNematics}a, d and in the main text in Fig.~1d. Actin can be seen as a proxy of cancer cell outlines. Specimen and imaging details can be found in the Methods, Sec.~\ref{methods}. \\
We observe cell nematic ordering in 3D and 2D based on nuclei (Fig.~\ref{fig:2D3D_cellNematics}a) and actin (Fig.~\ref{fig:2D3D_cellNematics}b).  For both cases, 2D and 3D nematic scalar order parameters are moderately correlated (Fig.~\ref{fig:2D3D_cellNematics}c-f). We see that in 2D as well as 3D, actin and nucleus nematic order parameters are strongly correlated (Fig.~\ref{fig:2D3D_cellNematics}g, h).

\subsection{Mapping sizes from three to two dimensions \label{SI:clusterMapping}}

The clinical images are two-dimensional projections of three-dimensional cancer clusters. However, the simulations consider two-dimensional clusters. The study of mapping from one to another is called stereology \cite{weibel1966practical}. An exact mapping can be derived for collections of perfect spheres \cite{wicksell1925corpuscle}, and an analogous mapping can be derived for ellipsoids oriented isotropically \cite{wicksell1926corpuscle}. For more irregular shapes, studies show a strong correlation between the 2D observed size distributions and the real 3D volume distributions \cite{sahagian19983d}.      

As a specific example, we discuss the case of spherical clusters following the analysis in Ref.~\cite{wicksell1925corpuscle} and show that the observed size distribution remains unchanged when looking at randomly chosen cross-sections. Consider a sphere of radius $R$ and central cross-section area $A_0 = \pi R^2$. We consider a cross-section of the sphere corresponding to a plane at a distance $d$ from its centre. The observed cross-sectional area is then $A = \pi(R^2 - d^2)$, and hence $
    d = \sqrt{\frac{A_0 - A}{\pi}} \: .
$

We now assume the probability distribution of the $d$ to be uniformly distributed across the radius of the sphere, 
\begin{equation}
    p_d(d) = 1/R, \qquad 0<d<R \:.
\end{equation}
Changing variables to $A$,  we find
 \begin{displaymath}
     p(A | R) =  p_d *\frac{d(d)}{dA} = \sqrt{\frac{\pi}{A_0}}\frac{1}{2\sqrt{A_0 - A}\sqrt \pi} \: ,
 \end{displaymath}
\begin{equation}
    p(A|A_0) =  \frac{1}{2} \frac{1}{\sqrt{A_0(A_0-A)}}, \qquad A<A_0.
\end{equation}
This is the probability of observing a cross-sectional area of $A$, for a sphere with a central cross-sectional area $A_0$. 

Finally, the observed distribution of $A$, for a distribution of sphere cross-sections $p(A_0)$, is given by
\begin{equation}
    p(A) = \int_A^\infty p(A|A_0) p(A_0) \, dA_0 \: .
    \label{eq:2d3daveraging_A0}
\end{equation}
If $p(A_0)$ is distributed as a power law, Eq.~(\ref{eq:2d3daveraging_A0}) implies that $p(A)$ is distributed as a power law with the same exponent. 

\subsection{Defect dimensionality\label{app:defect_dimen}}
We suspect that 3D breast ECM contains disclinations whose crossing in the histological section planes can be identified by $\pm 1/2$ defects. 

To check the consistency of this, we ran simulations in 3D (active nematics comprising disclinations \cite{ruske2021morphology}). For randomly chosen cross-sectional planes, we studied the director configurations where the 3D disclinations crossed the chosen plane. We expect from liquid crystal theory that the director configuration in the plane perpendicular to a disclination line can vary between $+1/2$-like to $-1/2$-like, each occurring with equal probability, via twist configurations where the director field predominantly points normal to the plane \cite{ruske2021morphology}.

\begin{figure}[H]
\centering
\includegraphics[width=0.5 \textwidth]{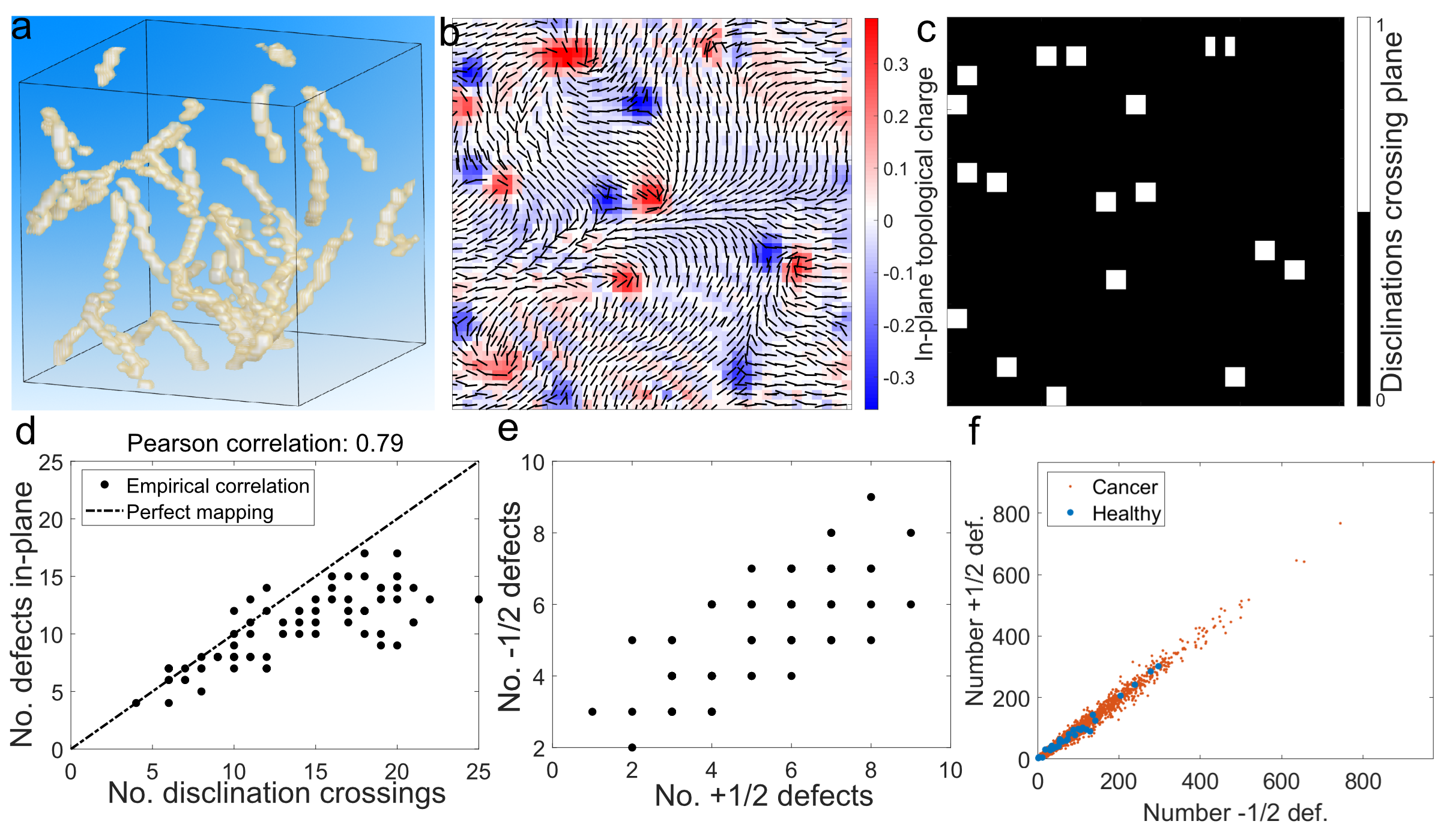}
\caption{\textbf{Analyses of simulation data of 3D director fields containing disclinations with respect to 2D sections. Half integer defects in sections indicate where disclinations are crossing the section.} \textbf{a} Disclination lines in 3D. \textbf{b} A representative section with the normalized director field $(dx, dy)$ containing no information about $dz$. The heatmap indicates the topological charge measured in-plane (see Methods Sec.~\ref{methods:imageAnalyses}). \textbf{c} Positions at which disclinations cross the plane shown in \textbf{b}, which correspond well with the defects identified from the 2D topological charge. \textbf{d} Number of $\pm 1/2$ defects identified in-plane versus the actual number of disclination lines crossing the section. Defects identified in plane and disclination crossings exhibit a robust correlation of $R = 0.79$. \textbf{e} Linear correlation between  $+1/2$ and $-1/2$ defects measured in-plane, comparable to the linear scaling in the histological sections that is shown in \textbf{f}. \textbf{f} Linear correlation between number of $+1/2$ and $-1/2$ defects ($\sigma = 5 \mu$m) in the histological samples. Each dot corresponds to the respective numbers for one patient.}
  \label{fig:2D3D_defects}
\end{figure}

We observe that the in-plane defects (ignoring out-of-plane director components; Fig.~\ref{fig:2D3D_defects}b) indeed correspond to crossings of the disclinations with the plane (Fig.~\ref{fig:2D3D_defects}c). The number of defects measured in the plane is less than the number of disclination crossings, as expected, as the twist configurations will not appear as 2D topological defects, and are not recorded in the 2D count. However, the estimated and real numbers exhibit a robust correlation with a Pearson correlation of $R = 0.79$ (Fig.~\ref{fig:2D3D_defects}d). Moreover, the number of $+1/2$ and $-1/2$ defects are, on average, equal (Fig.~\ref{fig:2D3D_defects}e) as expected, and as observed in the histological sections (Fig.~\ref{fig:2D3D_defects}f). These similarities strongly suggest that breast ECM contains 3D disclinations that can be recognized on 2D sections in the form of $\pm 1/2$ defects\\

\subsection{Dimensionality of anchoring}

In histological images of the ECM, we measure the 2D in-plane projection of the 3D orientation field. In this subsection, we estimate how projection affects the distribution of anchoring angles measured in Fig. 5 of the main text.

Consider a surface in the $y-z$ plane, with its normal vector in the $x$-direction. We describe the 3D orientation of a nematic particle at the origin using a polar angle $\theta_{3D}$ and azimuthal angle $\varphi_{3D}$, chosen such that $\theta_{3D}=0^\circ$ corresponds to the $x$- axis. Thus, in Cartesian coordinates, the components of a normalized unit vector $\mathbf{n}$ in the ($\theta_{3D}, \varphi_{3D}$) direction are given by $n_x = \cos\theta_{3D}$, $n_y = \sin\theta_{3D} \cos\varphi_{3D}$, and $n_z = \sin\theta_{3D} \sin\varphi_{3D}$. 

We further consider that we measure only the orientation field projected onto the $x-y$ plane, losing information about the $z$-component. The measured orientation field in the plane is defined to be $\theta_{2D}$, such that components of a normalized unit vector $\mathbf{m}$ are given by $m_x = \cos\theta_{2D}$, $m_y = \sin\theta_{2D}$.  

\begin{figure}[!ht]
\includegraphics[width=0.5\textwidth]{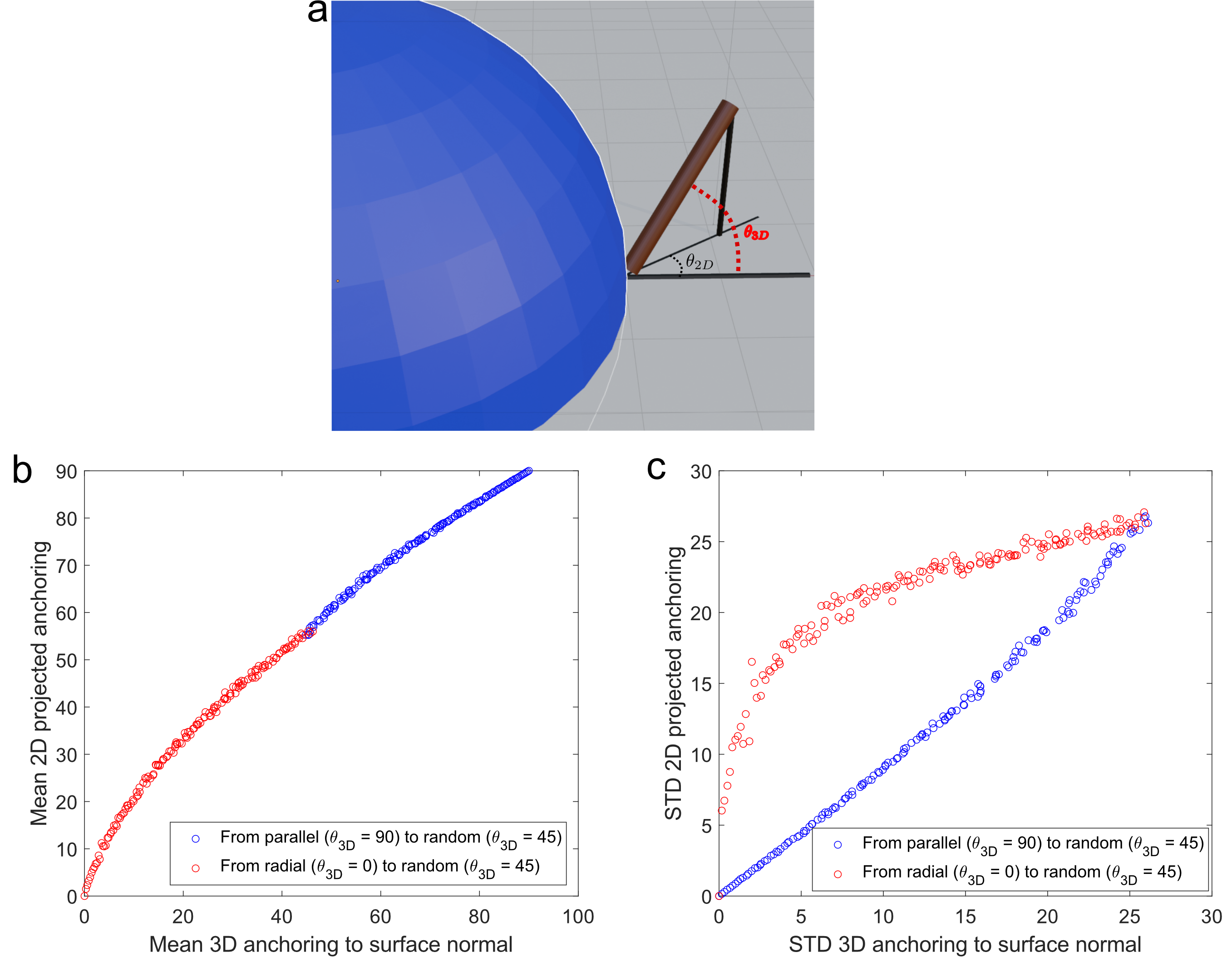}
\caption{Relation between 3D anchoring and projected 2D anchorings. \textbf{a} Schematic of the 3D and 2D anchoring angles. \textbf{b} Mapping of the average anchoring angles in 3D to projected 2D average anchoring angles. \textbf{c} Mapping of standard deviations of anchoring angles in 3D to projected 2D anchoring angles.}
  \label{fig:2D3D_anchoring}
\end{figure}

Note that $\theta_{2D}=90^\circ$ corresponds to planar anchoring, and $\theta_{2D}=0^\circ$ corresponds to homeotropic anchoring in 2D. In 3D, however, the `anchoring' angle is the polar angle $\theta_{3D}$ \cite{ruske2021morphology}. Furthermore, the angles $\theta_{3D}$ and $\theta_{2D}$ are different, unless the 3D orientation field happens to lie in the $x-y$ plane. Fig.~\ref{fig:2D3D_anchoring}a shows a schematic of the geometry under consideration.
We study the mapping $\theta_{3D}\rightarrow\theta_{2D}$ by simulating nematics anchored perpendicular ($\theta_{3D}=0^\circ$, shown in blue) or parallel ($\theta_{3D}=90^\circ$, shown in red) to the surface, with different amplitudes of noise in the orientation angle (N=1000 samples for each noise level). In Fig.~\ref{fig:2D3D_anchoring}b, we find that the average anchoring angle measured in 2D is consistent with the average anchoring angle in 3D, showing that a preference for parallel anchoring in 2D corresponds to a preference for parallel anchoring in 3D. In Fig.~\ref{fig:2D3D_anchoring}c, we compare the fluctuations of the orientation field about the measured mean value. We find that the statistics of projected 2D anchoring are similar to the 3D anchoring in the parallel (planar) case, but the fluctuations of the projected anchoring angle are magnified in the case of perpendicular (homeotropic) anchoring.  

In the main text, we report the anchoring angle $\theta_{a}$. We define this as the angle subtended by the orientation axis of the projected ECM fibre to the cancer cluster tangent vector. Thus, $\theta_a=|90^{\circ}-\theta_{2D}|$.

\section{Atomic force microscopy on primary breast cancer tissue \label{app:AFM}}
We performed AFM experiments on a primary breast cancer specimen after surgery to obtain order of magnitude elastic and viscous stress scales of cancer clusters and ECM (see Methods). With aid from pathologists, we roughly annotated the DNA image that is aligned to the mechanical measurements into dense cancer clusters and ECM. We group the mechanical data obtained from the AFM measurements into these regions.

\begin{figure}[htp]
\centering
\includegraphics[width=0.5 \textwidth]{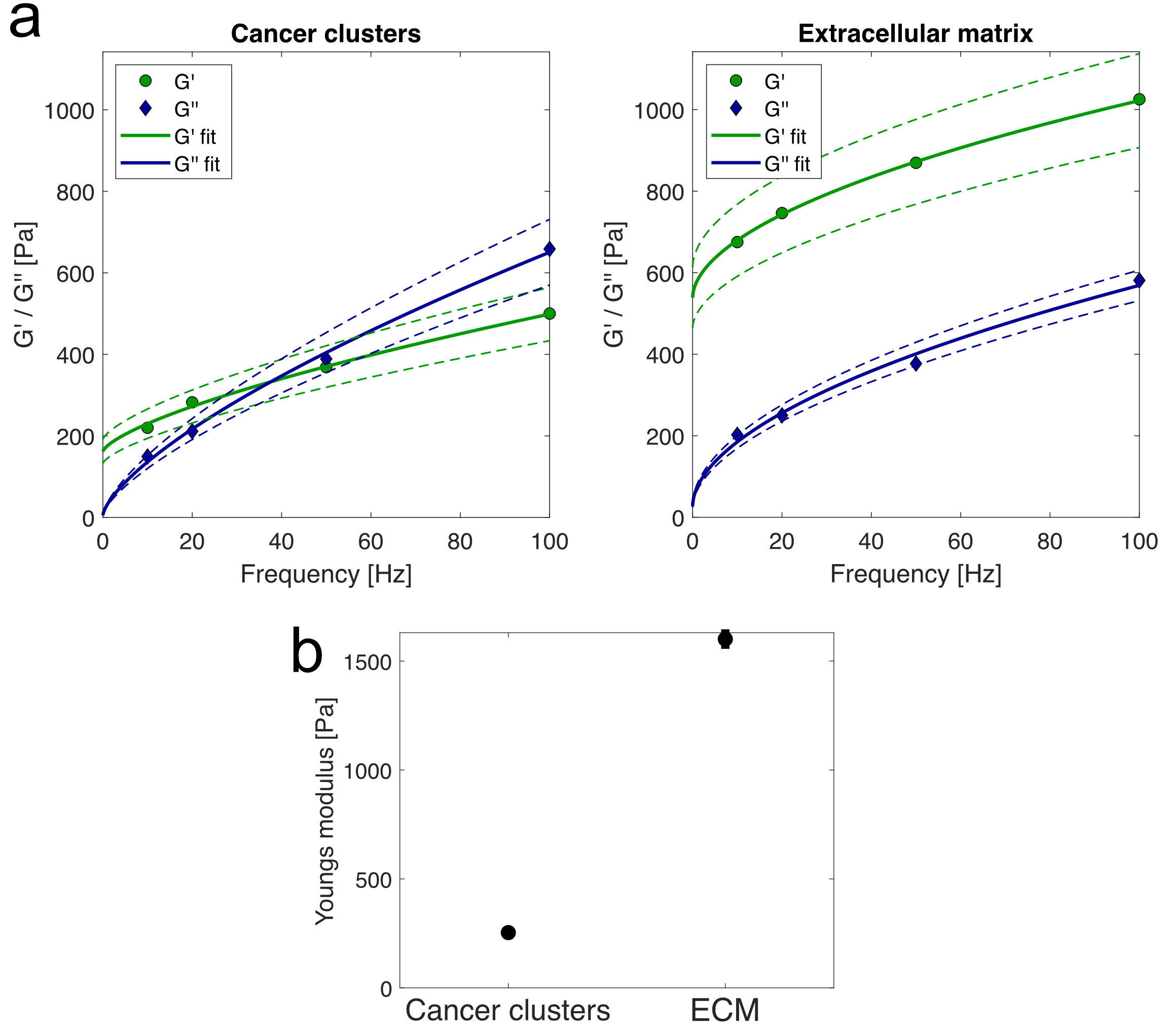}
\caption{\textbf{a} Frequency-dependent AFM measurements of a primary human breast cancer sample (internal ID MCA 239) for different tissue regions: Cancer clusters, where DNA signal is very dense. Extracellular matrix: where no DNA signal is found. Frequency scans of the shear storage and shear loss ($G'_f; \; G''_f$) were fitted by the fractional derivative Kelvin-Voigt model \cite{jozwiak2015fractional} to extrapolate the low frequency regime of relevance $f = 0.01$Hz. For the ECM we estimate $G''_\text{f=0.01Hz} \sim 30$Pa and $G'_\text{f=0.01Hz} \sim 500$Pa and for dense cancer cluster: $G''_\text{f=0.01Hz} \sim 10$Pa and $G'_\text{f=0.01Hz} \sim 200$Pa. Dashed lines represent the $95 \%$ confidence intervals of the fits. \\ \textbf{b} Measured Young's moduli in the different regions ($95\%$ CI for cancer is $\approx 10$Pa and for ECM $\approx 60$Pa). }
  \label{fig:SI_AFM}
\end{figure}

We then look at the frequency scans of the shear storage and shear loss moduli ($G'_f; \; G''_f$) which we fitted by the fractional derivative Kelvin-Voigt model \cite{jozwiak2015fractional} to extrapolate the low frequency regime of $f \sim 0.01$ Hz relevant to the timescale of cancer activity (motility). For the ECM we estimate $G''_{f=0.01\text{Hz}} \sim 30$ Pa and $G'_{f=0.01\text{Hz}} \sim 500$ Pa and for dense cancer cluster: $G''_{f=0.01\text{Hz}} \sim 10$ Pa and $G'_{f=0.01\text{Hz}} \sim 200$ Pa (Fig. \ref{fig:SI_AFM}a). 

We also measured the Young's modulus $Y$ in the respective regions (Fig. \ref{fig:SI_AFM}b) to approximate the frequency-independent shear modulus of the ECM for the simulations. We observe that $\langle Y \rangle \sim 1600$ Pa. Since the measured Young's modulus estimates the time-independent material elasticity when all relaxations have already taken place, we can estimate the respective shear modulus $G$ via $G = Y/(2(1+\nu))$, where $\nu$ is the Poisson ratio, which is around $\nu \approx 0.5$ for biological materials 
\cite{Kubitschke2018, mahaffy2000scanning, raveh2004elastic}. Thus, we arrive at a rough estimate of the shear modulus of the ECM: $G \sim 500$ Pa which we use for our model.

\section{Robustness of size and shape distributions to changing simulation parameters \label{app:robustness}}

\begin{figure}[H]
\includegraphics[width=0.45 \textwidth]{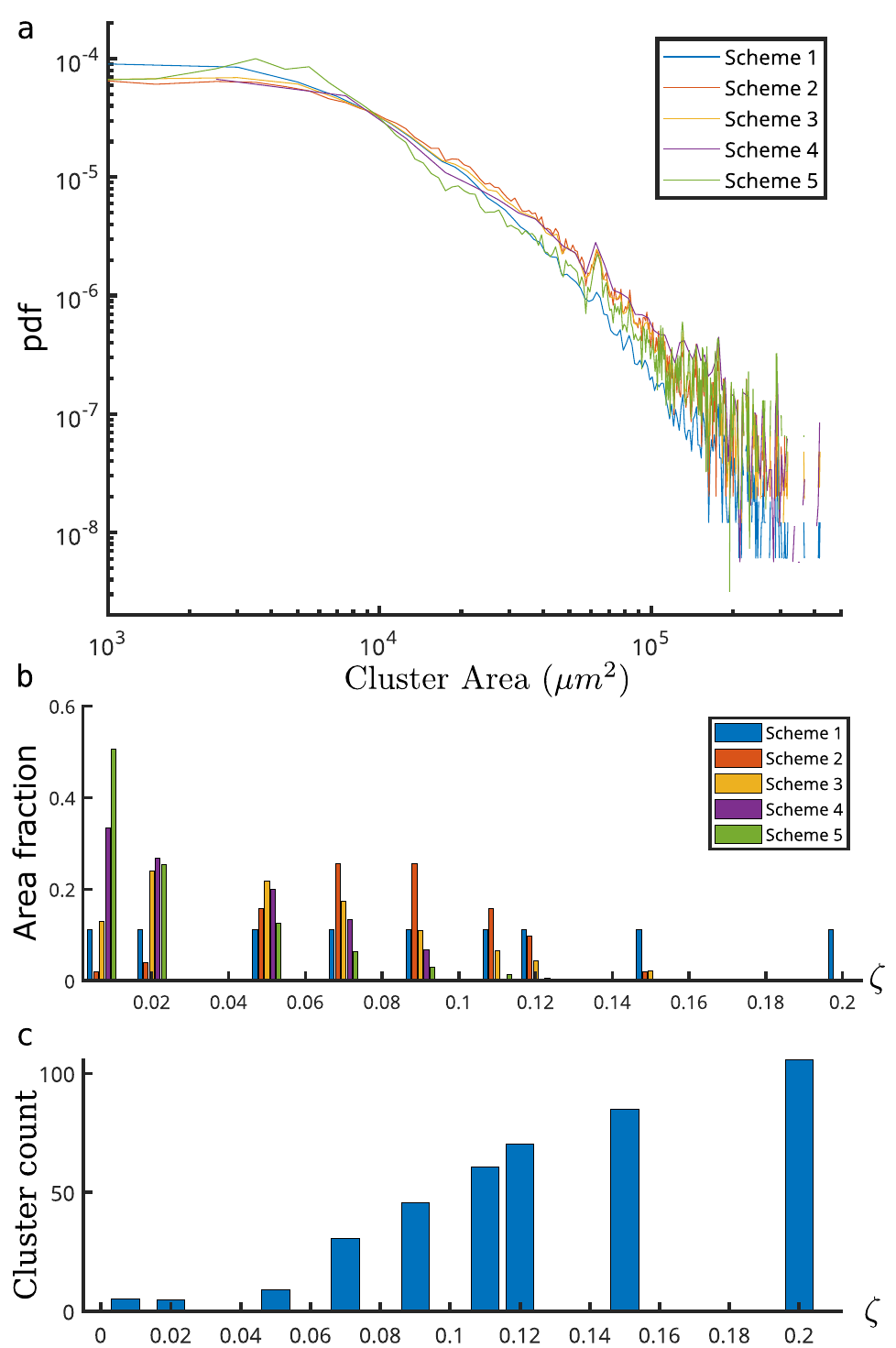}
\caption{\textbf{a} Observed size distributions of cancer clusters for five different schemes of activity averaging, represented by different colours, all showing a similar power-law distribution. \textbf{b} Different activity averaging schemes, showing relative fractions of different activities in the system. \textbf{c} Bar plot showing the number of cancer clusters increases with activity, for a fixed area fraction of cancer cells.}
  \label{fig:activityAveraging}
\end{figure}

When studying the distribution of cancer cluster sizes, we need to account for the fact that the motility of cancer clusters may be different across different clusters and different patients. Therefore, we average simulation results over a distribution of activity to compare to the histological data. In Fig. \ref{fig:activityAveraging}a, we find that changing the exact form of the activity distribution does not make a qualitative difference in the distribution of size data, as long as the sampling includes clusters with intermediate activity. Different sampling schemes have different area fractions of cancer clusters with a given activity. These are labelled by different colours and the relative weights for different activity values are shown in Fig. \ref{fig:activityAveraging}b. Systems with higher activity have a significantly higher number of cancer clusters, which are smaller in size (see Fig. 2c in the main text, and also Fig. \ref{fig:activityAveraging}c). These form the large majority of the observed clusters, dominating the statistics of the overall distribution. In the results presented in the text, we have used `Scheme 1',  which samples a broad spread of activity in cancer clusters. 

It is important to emphasize that throughout the text, we use sampling schemes with different area fractions of each activity. We discuss number fractions only in Fig.~\ref{fig:activityAveraging} to justify why our results do not depend strongly on the exact distribution of activity chosen.

\begin{figure}[htp]
\includegraphics[width=0.5 \textwidth]{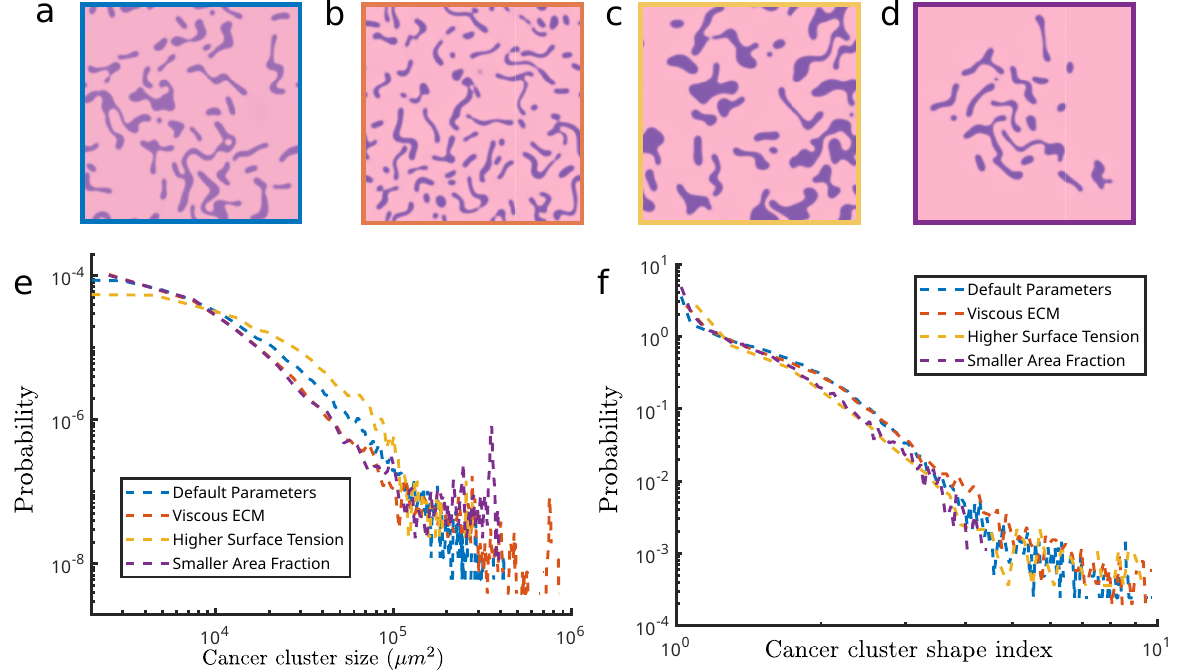}
\caption{Robustness of cancer cluster sizes and shapes to simulation parameters. \textbf{a} Simulation snapshot with default simulation parameters. \textbf{b} Simulation snapshot with a viscous ECM. \textbf{c} Simulation snapshot for a higher surface tension. \textbf{d} Simulation snapshot for a smaller area fraction. \textbf{e} Distribution of size and \textbf{f} shape after activity averaging. The blue line shows results for parameters reported in the text, the red line shows results for purely viscous ECM, the yellow line shows results upon doubling the surface tension parameters, and the purple line shows a different area fraction.}
  \label{fig:diffParameters}
\end{figure}

We also looked at the sensitivity of our results to the chosen parameters of the model. The effect of changing activity is already shown in Fig 2b of the main text. Fig. \ref{fig:diffParameters}a shows a snapshot of a dynamical steady state with the default parameters used in the text ($\zeta=0.12$). Fig. \ref{fig:diffParameters}b shows the dynamical steady state when the ECM is modelled as a viscous fluid instead of a Kelvin-Voigt fluid ($E^0 = 0$), in which case the cancer clusters are more elongated. Fig. \ref{fig:diffParameters}c shows the dynamical steady state with the surface tension parameters doubled ($A_\phi = K_\phi = 2$). In this case, we find that the average cancer clusters are less elongated. In Fig. \ref{fig:diffParameters}d, we look at the default parameters but with a much smaller area fraction ($10\%$ cancer cluster coverage instead of $25\%$). In this case, the dynamical steady state looks similar to the default parameters, but the area covered by the cancer clusters is much smaller.

Although the sizes and shapes of cancer clusters change upon changing parameters in individual simulations, the results for the resulting size and shape distributions show very little difference after averaging over a range of activities (see Fig. \ref{fig:diffParameters}e and f). This is because activity is a critical parameter in all these simulations, and averaging over activity scans over the range of shapes and sizes observed with different activities. 

The cluster size distribution presented for a large cluster breaking up (deep blue curve in Fig.~\ref{fig:sizeshape}a) shows a peak at large cluster sizes due to the inclusion of snapshots of intermediate stages where the initial cluster is not fully broken apart. Considering simulation snapshots only at very late times, when the initial droplet has fully broken apart, we recover a size distribution identical to the coarsening case. The first three bins from the simulations do not exhibit the power law scaling because of the difficulty of resolving the smallest clusters in the simulation phase field. 

\section{Comparing healthy and cancer ECM \label{app:compareHealthyCancer}}
\begin{figure}[htp]
\centering
\includegraphics[width=0.5 \textwidth]{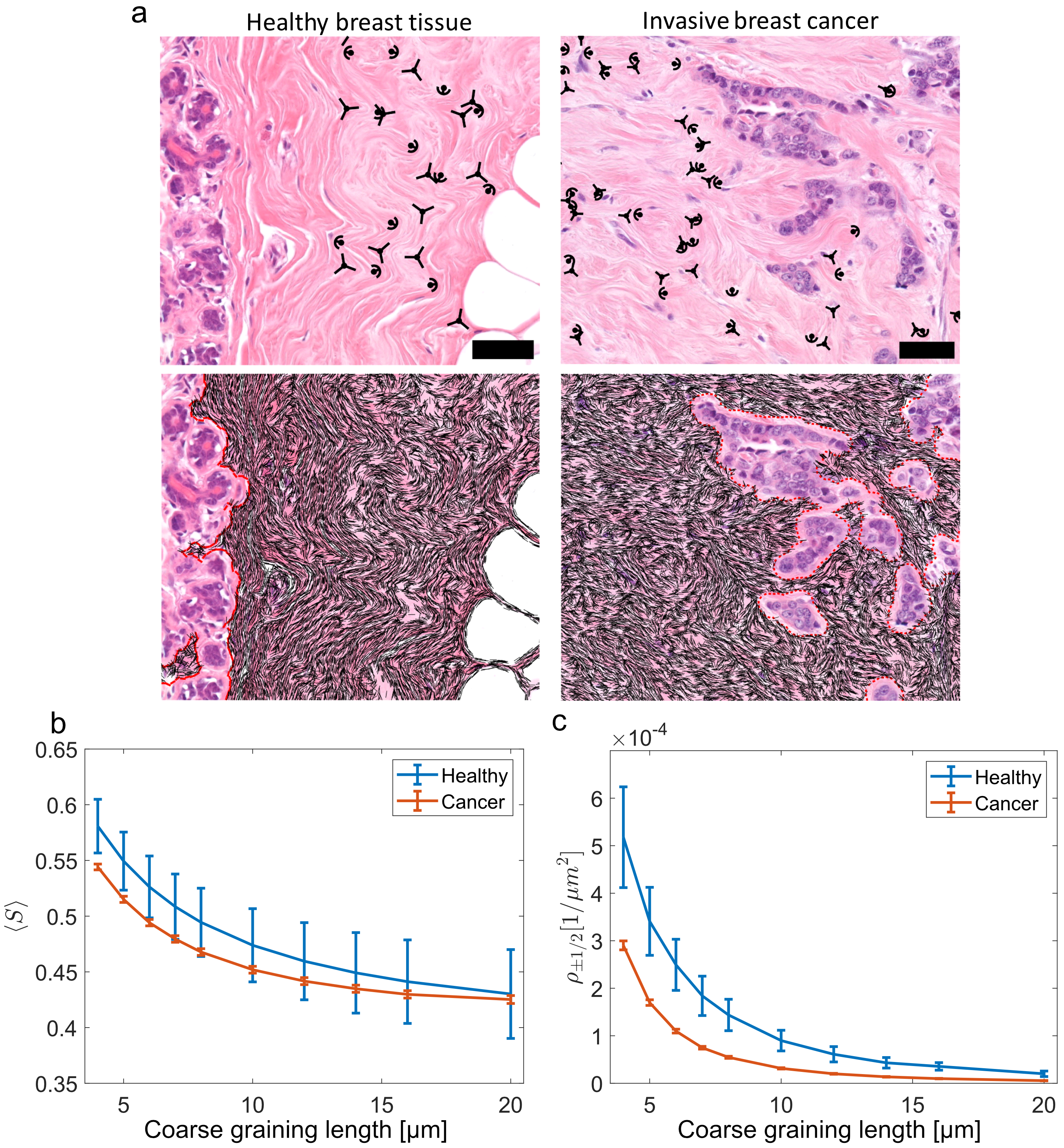}
\caption{\textbf{a} Examples of histological sections of healthy breast tissue and invasive breast cancer tissue with topological defects and director fields overlaid. Scale bars are $100 \mu$m. \textbf{b} Average nematic order $\langle S \rangle$ for different coarse-graining lengths $\sigma$. \textbf{c} Defect densities $\rho_{\pm 1/2}$ for different coarse-graining lengths $\sigma$. Defect densities are shown for healthy ECM ($N = 32$ patients) and invasive breast cancer ECM ($N = 2012$ patients). Number densities of $\pm 1/2$ defects scale with $\rho_{\pm 1/2} = a \sigma^{-2}$. Healthy: $a = 0.0085 \; (R^2 = 0.997)$; cancer: $a = 0.0043 \; (R^2 = 0.983)$.}
  \label{fig:healthyCancer_avergeNematicOrder}
\end{figure}

In Fig.~\ref{fig:healthyCancer_avergeNematicOrder}b, we compare the scalar nematic order parameter between the ECM of healthy breast tissue and the ECM of invasive breast cancer for different coarse-graining lengths $\sigma$. We observed no statistical differences for $\sigma > 7 \mu$m. Below this threshold, healthy ECM exhibits slightly higher nematic order on small length scales and a higher topological defect density $\rho_{\pm 1/2}$ (Fig.~\ref{fig:healthyCancer_avergeNematicOrder}c).

\section{Patient data \label{app:patientData}}
For our retrospective histopathological study, we investigated 2170 breast cancer patients, of which 2012 had  usable clinical follow-up. The analyses were based on H$\&$E-stained digitized tissue sections of primary tumours. The cases considered were classified according to the WHO histological classification of breast carcinomas (3rd edition of 2003) which follows the TNM classification of the UICC (Union Internationale Contre le Cancer) in terms of grading and staging. All data used for this study comes from female patients. The clinic evaluated the sex of the patients by specification.  Patients were informed and
gave ethical, informed consent, approved by the ethics
committee of the Ärztekammer Hamburg (Medical Council
Hamburg) with Processing No. PV2946. The descriptive statistics of the patient collective are shown below.\\
In the following, the descriptive statistics are shown for the complete patient collective (Table~\ref{tab:All}), the training set (Table~\ref{tab:trainingSet}), and the test set (Table~\ref{tab:testSet}). For the defect-cancer distance observable patients without any defects in the ECM were neglected resulting in an overall 1982 patients for analysis.\\
Explanation of all of the characteristics:\\
\begin{itemize}
    \item \textbf{Age}: Source: patient file. Explanation: Age at the time of surgery.
    \item \textbf{Grade}: Source: histological sample, defined by the examining pathologist following the Nottingham grading system in every breast cancer case. Explanation: Cancerous tissue can be morphologically assessed for the aggressiveness of the tumour based on the degree of differentiation (extent of degeneration and mitosis).\\
    Grading following the internationally accepted standard by UICC (Union Internationale Contre le Cancer):\\
G1: well-differentiated\\
G2: moderately differentiated\\
G3: poorly differentiated 
\item \textbf{Size}: Source: histological slide, measured by the examining pathologist. Explanation: The tumour size affects prognosis as it is an indicator of tumour growth. 
\item \textbf{Lymph Node Status}: Source: histological preparation, measured by the examining pathologist. Explanation: Contains the information if any lymph nodes are invaded (Lymph Node Status = 1) or not (Lymph Node Status = 0). This is a strong and independent negative prognostic factor.\\ 
When breast cancer begins to spread, the first metastatic site is often one of the neighbouring lymph nodes. From there, the tumour is able to spread further via the lymphatic system.
\item \textbf{Hormone Receptor Status}: Source: Immunohistochemical staining, performed and analysed by the pathologist. Explanation: The hormones oestrogen and progesterone can affect the growth of breast cancer. They bind to hormone receptors of the cell, which then transmit the growth signal into the cell. To estimate whether a tumour grows hormone-dependent, the proportion of cells and the amount of corresponding hormone receptor-positive cells are analysed. If more than one percent of all tumour cells are receptor-positive, it is assumed that the tumour is hormone-sensitive. This is indicated by ER+ (oestrogen receptor-positive) and/or PR+ (progesterone receptor-positive).
\item \textbf{Chemotherapy Status}: Source: Follow-up patient questionnaire: Information provided by the patient and/or treating physician. Explanation: Through chemotherapy, the cancer cells are attacked. The therapy is mainly used for tumours with aggressive growth. Before surgery, neoadjuvant chemotherapy can be used to shrink the size of the tumour. 
\end{itemize}

\begin{table}[ht]
    \centering
    \caption{Clinical characteristics of complete patient collective (N = 2012) \label{tab:All}.}
    \begin{tabular}{ll}
    \toprule
    \textbf{Characteristic} & \textbf{Value (Proportion)} \\
    \midrule
    \textbf{Age} & \\
    \hspace{1em} mean $\pm$ std [range] & 58.77 $\pm$ 11.74 [25, 98] \\
    \addlinespace
    \textbf{Grade} & \\
    \hspace{1em} 1 & 463 (0.23) \\
    \hspace{1em} 2 & 1072 (0.53) \\
    \hspace{1em} 3 & 470 (0.23) \\
    \hspace{1em} missing & 7 (0) \\
    \addlinespace
    \textbf{Size [cm]} & \\
    \hspace{1em} mean $\pm$ std [range] & 1.95 $\pm$ 1.41 [0.2, 16] \\
    \hspace{1em} missing & 4 (0) \\
    \addlinespace
    \textbf{Nodal Status} & \\
    \hspace{1em} pN0 & 1470 (0.73) \\
    \hspace{1em} pN+ & 520 (0.26) \\
    \hspace{1em} missing & 22 (0.01) \\
    \addlinespace
    \textbf{ER Status} & \\
    \hspace{1em} ER 0 & 215 (0.11) \\
    \hspace{1em} ER 1 & 1791 (0.89) \\
    \hspace{1em} missing & 6 (0) \\
    \addlinespace
    \textbf{PR Status} & \\
    \hspace{1em} PR 0 & 402 (0.2) \\
    \hspace{1em} PR 1 & 1599 (0.79) \\
    \hspace{1em} missing & 11 (0.01) \\
    \addlinespace
    \textbf{Chemo Status} & \\
    \hspace{1em} Chemo 0 & 949 (0.47) \\
    \hspace{1em} Chemo 1 & 493 (0.25) \\
    \hspace{1em} missing & 570 (0.28) \\
    \bottomrule
    \end{tabular}
\end{table}

\begin{table}[ht]
    \centering
    \caption{Clinical characteristics of training set population (N = 600). \label{tab:trainingSet}}
    \begin{tabular}{ll}
    \toprule
    \textbf{Characteristic} & \textbf{Value (Proportion)} \\
    \midrule
    \textbf{Age} & \\
    \hspace{1em} mean $\pm$ std [range] & 57.78 $\pm$ 10.83 [25, 98] \\
    \addlinespace
    \textbf{Grade} & \\
    \hspace{1em} 1 & 156 (0.26) \\
    \hspace{1em} 2 & 323 (0.54) \\
    \hspace{1em} 3 & 120 (0.2) \\
    \hspace{1em} missing & 1 (0) \\
    \addlinespace
    \textbf{Size [cm]} & \\
    \hspace{1em} mean $\pm$ std [range] & 1.87 $\pm$ 1.41 [0.2, 15] \\
    \hspace{1em} missing & 1 (0) \\
    \addlinespace
    \textbf{Nodal Status} & \\
    \hspace{1em} pN0 & 464 (0.77) \\
    \hspace{1em} pN+ & 129 (0.22) \\
    \hspace{1em} missing & 7 (0.01) \\
    \addlinespace
    \textbf{ER Status} & \\
    \hspace{1em} ER 0 & 56 (0.09) \\
    \hspace{1em} ER 1 & 543 (0.91) \\
    \hspace{1em} missing & 1 (0) \\
    \addlinespace
    \textbf{PR Status} & \\
    \hspace{1em} PR 0 & 112 (0.19) \\
    \hspace{1em} PR 1 & 485 (0.81) \\
    \hspace{1em} missing & 3 (0.01) \\
    \addlinespace
    \textbf{Chemo Status} & \\
    \hspace{1em} Chemo 0 & 287 (0.48) \\
    \hspace{1em} Chemo 1 & 135 (0.23) \\
    \hspace{1em} missing & 178 (0.3) \\
    \bottomrule
    \end{tabular}
\end{table}

\begin{table}[ht]
    \centering
    \caption{Clinical characteristics of test set population (N = 1412).  \label{tab:testSet}}
    \begin{tabular}{ll}
    \toprule
    \textbf{Characteristic} & \textbf{Value (Proportion)} \\
    \midrule
    \textbf{Age} & \\
    \hspace{1em} mean $\pm$ std [range] & 59.19 $\pm$ 12.08 [25, 98] \\
    \addlinespace
    \textbf{Grade} & \\
    \hspace{1em} 1 & 307 (0.22) \\
    \hspace{1em} 2 & 749 (0.53) \\
    \hspace{1em} 3 & 350 (0.25) \\
    \hspace{1em} missing & 6 (0) \\
    \addlinespace
    \textbf{Size [cm]} & \\
    \hspace{1em} mean $\pm$ std [range] & 1.99 $\pm$ 1.42 [0.4, 16] \\
    \hspace{1em} missing & 3 (0) \\
    \addlinespace
    \textbf{Nodal Status} & \\
    \hspace{1em} pN0 & 1006 (0.71) \\
    \hspace{1em} pN+ & 391 (0.28) \\
    \hspace{1em} missing & 15 (0.01) \\
    \addlinespace
    \textbf{ER Status} & \\
    \hspace{1em} ER 0 & 159 (0.11) \\
    \hspace{1em} ER 1 & 1248 (0.88) \\
    \hspace{1em} missing & 5 (0) \\
    \addlinespace
    \textbf{PR Status} & \\
    \hspace{1em} PR 0 & 290 (0.21) \\
    \hspace{1em} PR 1 & 1114 (0.79) \\
    \hspace{1em} missing & 8 (0.01) \\
    \addlinespace
    \textbf{Chemo Status} & \\
    \hspace{1em} Chemo 0 & 662 (0.47) \\
    \hspace{1em} Chemo 1 & 358 (0.25) \\
    \hspace{1em} missing & 392 (0.28) \\
    \bottomrule
    \end{tabular}
\end{table}
\clearpage
\section{4-fold cross validation of patient prognosis\label{app:crossval}}
We validate the prognostic results shown in the main text via 4-fold cross validation below. As performance metrics we use the standard parameters accuracy ($\frac{\text{TP} + \text{TN}}{\text{TP} + \text{TN} + \text{FP} + \text{FN}}$), Precision ($\frac{\text{TP}}{\text{TP} + \text{FP}}$), Recall ($\frac{\text{TP}}{\text{TP} + \text{FN}}$) and F1-score ($2 \cdot \frac{\text{Precision} \cdot \text{Recall}}{\text{Precision} + \text{Recall}} = \frac{2 \cdot \text{TP}}{2 \cdot \text{TP} + \text{FP} + \text{FN}}$), where $TP$ is the number to true positives, $TN$ true negatives and "false" numbers for $F$ respectively. In addition, we record the concordance index (c-Index, Harrel’s c-Index) \cite{harrell1996multivariable} that measures how well a score can sort patient pairs with respect to survival time. A value of 0.5 indicates a random score and a value of 1 a perfect score. Finally, the p-value of the log-rank test between high- and low-risk groups is reported alongside the found threshold.\\
We see that for both, the cluster irregularities $\langle SI \rangle$ (Table \ref{tab:CV_Ir}) and defect-to-cancer distances $\max(\Delta_{DC})/\xi_{ECM}$ (Table \ref{tab:DCD}), the prognostic results of the main text are reproduced and therefore robust.

\begin{table}[ht]
\centering
\caption{4-fold cross validation for the average cluster irregularities $\langle SI \rangle$.}
\begin{tabular}{lccccc}
\toprule
\textbf{Metric} & \textbf{k=1} & \textbf{k=2} & \textbf{k=3} & \textbf{k=4} & \textbf{Average} \\
\midrule
Accuracy     & 0.57 & 0.54 & 0.75 & 0.72 & 0.65 \\
Precision    & 0.65 & 0.64 & 0.25 & 0.34 & 0.47 \\
Recall       & 0.25 & 0.22 & 0.27 & 0.28 & 0.26 \\
F1-score     & 0.36 & 0.34 & 0.26 & 0.30 & 0.32 \\
C-index      & 0.62 & 0.60 & 0.60 & 0.62 & 0.61 \\
Log-rank p   & 3.00E-09 & 4.00E-06 & 4.00E-06 & 5.00E-08 & 2.53E-06 \\
Threshold    & 1.60 & 1.60 & 1.80 & 1.76 & 1.69 \\
\bottomrule
\end{tabular}
\label{tab:CV_Ir}
\end{table}

\begin{table}[ht]
\centering
\caption{4-fold cross validation for the normalized maximum defect-to-cancer distances $\max(\Delta_{DC})/\xi_{ECM}$.}
\begin{tabular}{lccccc}
\toprule
\textbf{Metric} & \textbf{k=1} & \textbf{k=2} & \textbf{k=3} & \textbf{k=4} & \textbf{Average} \\
\midrule
Accuracy     & 0.74 & 0.75 & 0.58 & 0.72 & 0.70 \\
Precision    & 0.25 & 0.20 & 0.46 & 0.27 & 0.30 \\
Recall       & 0.27 & 0.27 & 0.20 & 0.24 & 0.25 \\
F1-score     & 0.26 & 0.23 & 0.28 & 0.26 & 0.26 \\
C-index      & 0.56 & 0.56 & 0.57 & 0.57 & 0.57 \\
Log-rank p   & 1.00E-05 & 7.00E-04 & 2.00E-02 & 4.00E-04 & 6.83E-03 \\
Threshold    & 0.80 & 0.70 & 1.37 & 0.88 & 0.94 \\
\bottomrule
\end{tabular}
\label{tab:DCD}
\end{table}

\section{Dominant mechanical activity and vital primary breast cancer cell tracking \label{app:dominant_activity}}
\begin{figure}[htp]
\centering
\includegraphics[width=0.5 \textwidth]{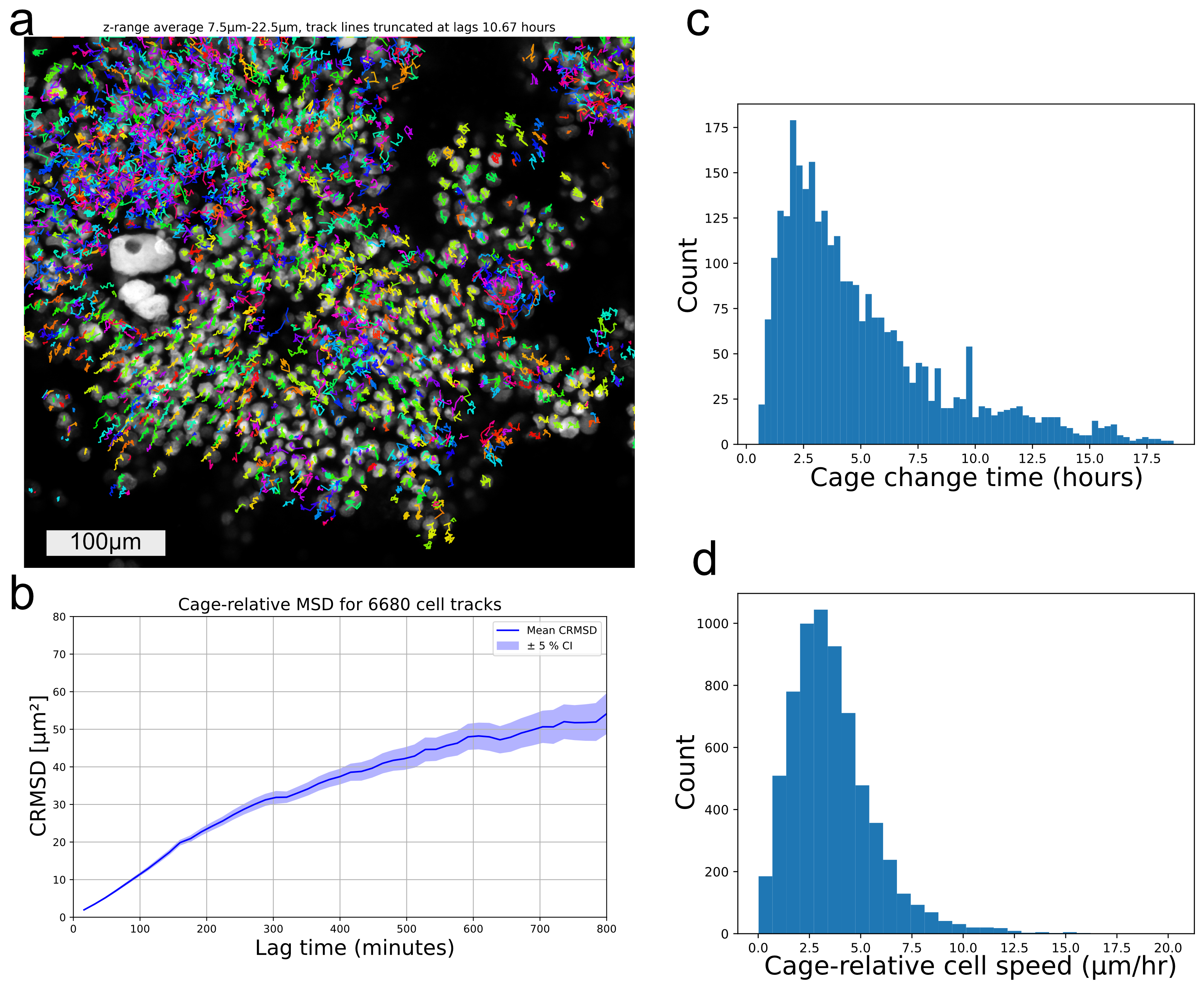}
\caption{\textbf{a} Cell tracking in a vital primary breast cancer tissue explant after surgery. Coloured lines: 2D projection of cell trajectories in 3D within $\approx 11$ hours. Large domains consist of motile cells. \textbf{b} Cage-relative mean squared displacement (CRMSD) versus different lag times. Here, "cage-relative" refers to the motion of cells relative to their nearest neighbours in 3D. \textbf{c} Cage change time of cells in vital primary breast cancer explants calculated as the time $t$, where $\sqrt{\text{CRMSD(t)}} = \langle R_{\text{eff}} \rangle$, where $\langle R_{\text{eff}} \rangle$ is the average effective cell radius. \textbf{d} Distribution of the cage-relative cell speed for lag times of 2 hours.}
  \label{fig:SI_vital}
\end{figure}

In the main text we conclude that the active energy input from mechanical stresses resulting from motility dominate over that induced by cell proliferation. Here, we want to support this conjecture by comparing the active energy intakes from proliferation and motility respectively:

Proliferation is a mechanical activity that generates an active nematic stress \cite{saw2018biological, dell2018growing}. An upper bound for the localized speeds can be estimated by calculating the distance a cell is pushed during a cell division, approximately a cell diameter ($\sim$10$\mu$m in the histology slides), divided by the time that a cell division needs (literature $\sim2$ hours \cite{ibrahim2022assessment}), which gives as a maximum proliferation induced speed $\sim 5 \mu$m/hour.  This is comparable to the average cell motility speed of $\sim 3\mu$m/hour that we have found by vital cell tracking (compare Fig. \ref{fig:SI_vital}d and references \cite{gottheil2023state, grosser2021cell}). Other evidence that leads to the same conclusion is that traction force microscopy has shown that cell proliferation generates active stresses of approximately 50–150 Pa \cite{jahan2017traction, liu2021quantifying, nam2021cellular}, while cancer cell migration produces active stresses of around 50–400 Pa \cite{rossetti2024optogenetic, peschetola2013time, trepat2009physical, kraning2012cellular}. These values are therefore of the same order of magnitude.
However, to compare the active energy input to the clusters from cell motion and proliferation we also need to know the fractions of dividing and motile cells. Therefore we measured the mitotic marker Phospho-histone H3 in our patient collective (compare main text and Methods Sec.~\ref{methods:Proliferation}). With this marker we calculated the fraction of cells that are actively dividing (mitosis; patient median is $0.46\%$ and 99 percent of patients exhibit fractions $<4\%$). By contrast, vital tracking experiments within primary human breast tumour explants after surgery (Patients were asked to participate through approved ethics votes (breast ethics vote No. 073-13-11032013). The samples were classified as tumour tissue by pathologists at Leipzig University Hospital) show that a much larger fraction of $\approx 30\%$ of cells changes their neighbours by motility within similar time span compared to mitosis (compare Fig. \ref{fig:SI_vital}c). \\
This shows that motility contributes substantially more active energy to the system than proliferation.

\section{Relation between cluster irregularity and aspect ratio\label{app:AR_ir}}
In the main text we report shape distributions based on the \textit{cluster shape index}, or \textit{cluster irregularity} defined as $SI = P/\sqrt{4 \pi A}$, where $P$ is the perimeter and $A$ the area. \\
Here, we compare this metric to the cluster aspect ratio, defined as the ratio of the major to minor axis lengths of the best-fit ellipse derived from the second spatial moments of the cluster's pixel distribution. An aspect ratio of 1 corresponds to a perfectly circular (isotropic) shape, while larger values indicate increasingly elongated, anisotropic shapes.

\begin{figure}[H]
\centering
\includegraphics[width=0.5 \textwidth]{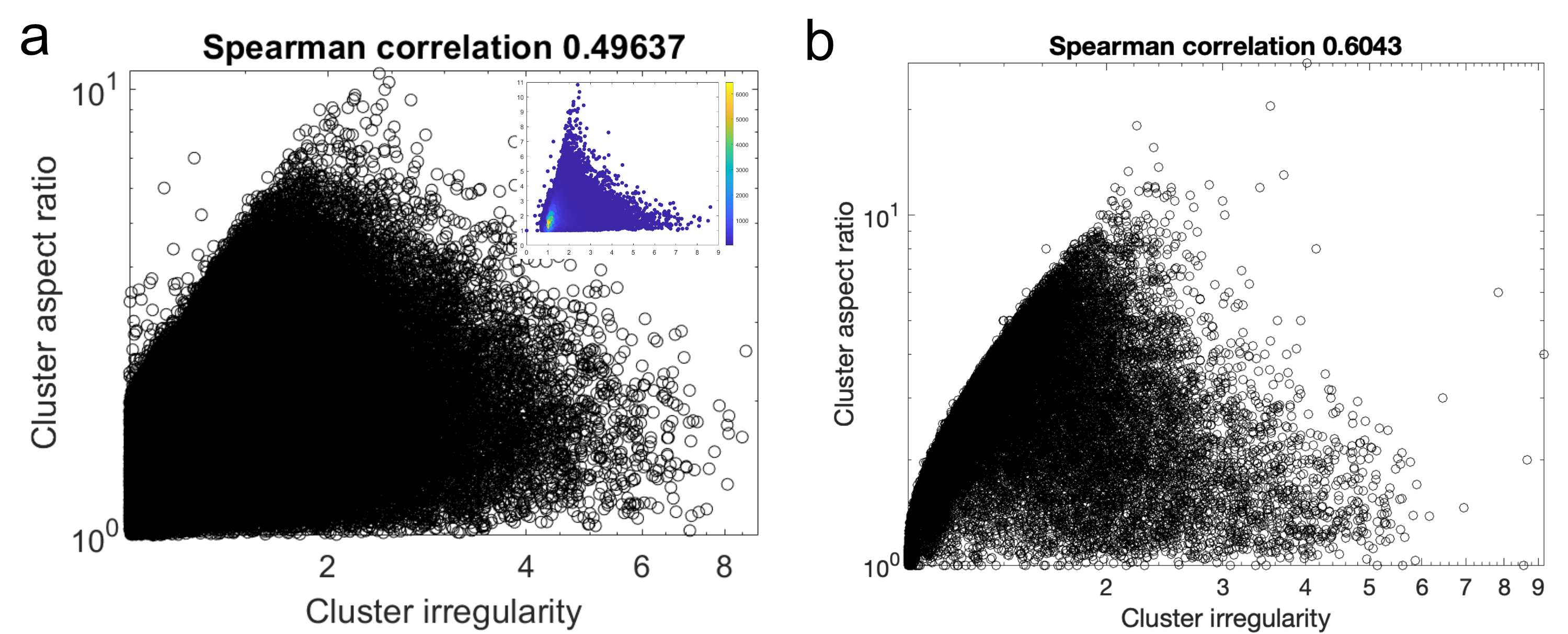}
\caption{Comparison of the relationship of the cluster shape index presented in the main text and the cluster aspect ratio. \textbf{a} Clusters observed in histology images. Inset shows the point density for the log-transformed axes. \textbf{b} Clusters observed in simulations.}
  \label{fig:irr_AR}
\end{figure}

Unlike the shape index, the aspect ratio is insensitive to high-frequency fluctuations along the boundary and primarily captures large-scale elongation—i.e., the lowest mode of shape deformation. As such, while the shape index provides a more detailed measure of boundary roughness or irregularity, the aspect ratio offers a coarse-grained view focused on the global geometry of the cluster. \\
In Fig. \ref{fig:irr_AR}, we compare their correlations in the histological images and simulation images and find that their relation is highly comparable with similar Spearman correlation coefficients and data distributions, indicating a similar relation of high-to-low frequency shape instabilities in simulations and real breast cancer.
\section{3D topological disclinations in primary breast ECM \label{app:disclinations}}
\begin{figure}[H]
\centering
\includegraphics[width=0.5 \textwidth]{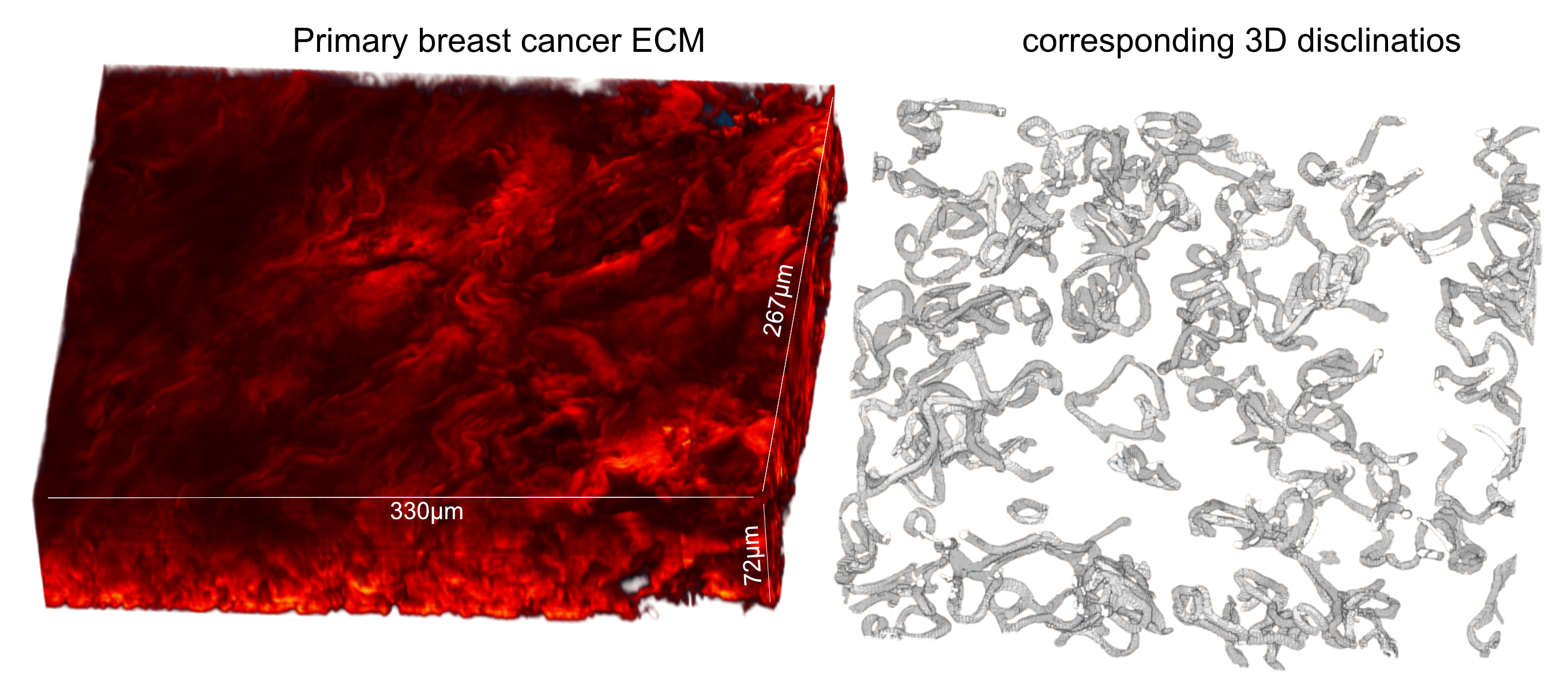}
\caption{3D disclination lines and loops in the ECM of a human breast cancer patient.}
  \label{fig:ECMDisclinations}
\end{figure}

In the main text we report statistics on 2D $\pm 1/2$ topological defects and argue that they correspond to 3D disclination lines and loops in 3D, as shown in Appendix \ref{app:defect_dimen} above. \\
Here, we show direct evidence of 3D disclination loops and lines in 3D in decellularized ECM from a human breast cancer patient in Fig.\ref{fig:ECMDisclinations}. We find the disclinations according to the procedure described in \cite{ruske2021morphology}. 

\section{Nematic screening length \label{app:nematic_screening}}
\begin{figure}[h]
\centering
\includegraphics[width=0.5 \textwidth]{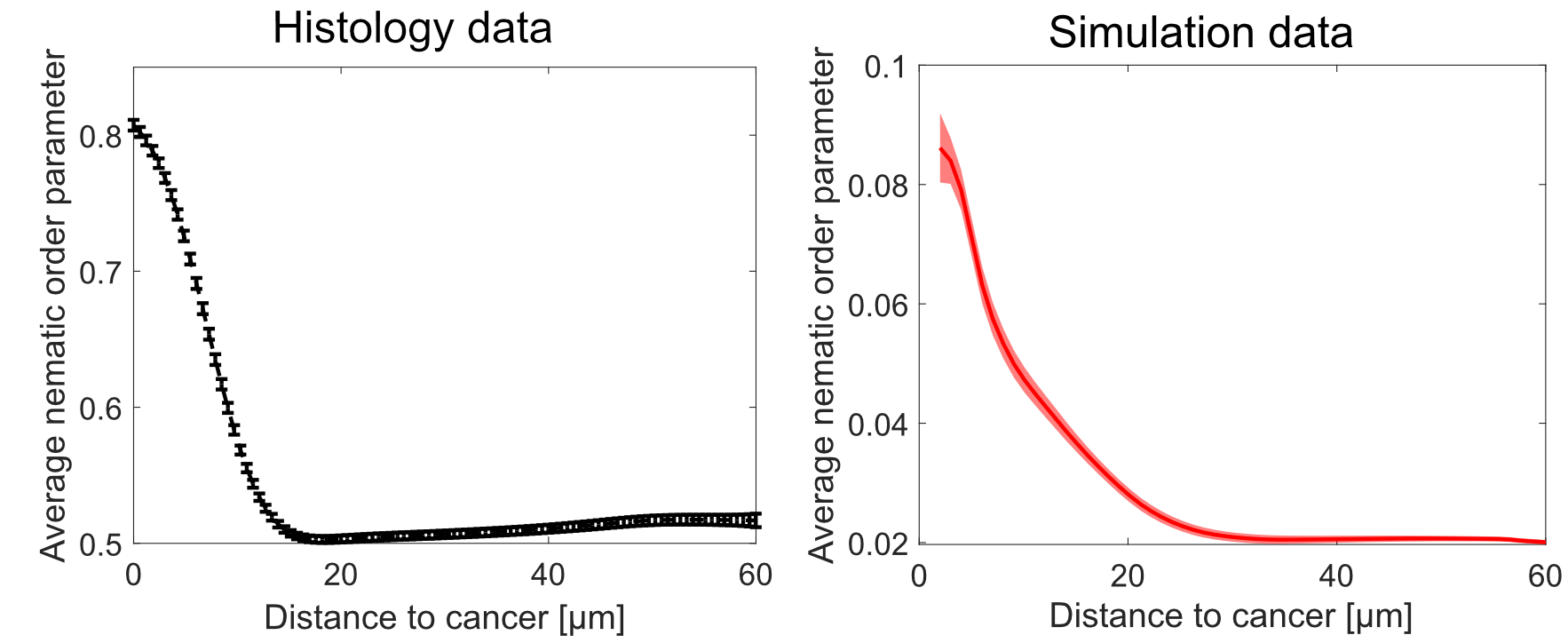}
\caption{Average ($\pm 95 \%$ CI) nematic order parameter in the ECM plotted against the distance to the cancer clusters. For both simulation and clinical images we observe a screening length of $\sim 10 \mu$m.}
  \label{fig:nematicScreening}
\end{figure}

In the main text we argue that in invasive lesions the active forces induced by collective, nematic cell migration align the ECM in the vicinity of the cancer cluster boundaries, leading to a more parallel ECM-cancer alignment. Another fingerprint of these active forces in the neighbouring ECM is an increased nematic order parameter, as shown in Fig. \ref{fig:nematicScreening}. We see that the nematic order in the ECM is high close to the cancer clusters and decreases to the bulk value farther away from the clusters. For both simulation and clinical images we observe a screening length of $\sim 10 -15\mu$m.

\section{3D simulations \label{app:3D_Sims}}

\begin{figure}[ht]
\centering
\includegraphics[width=0.5 \textwidth]{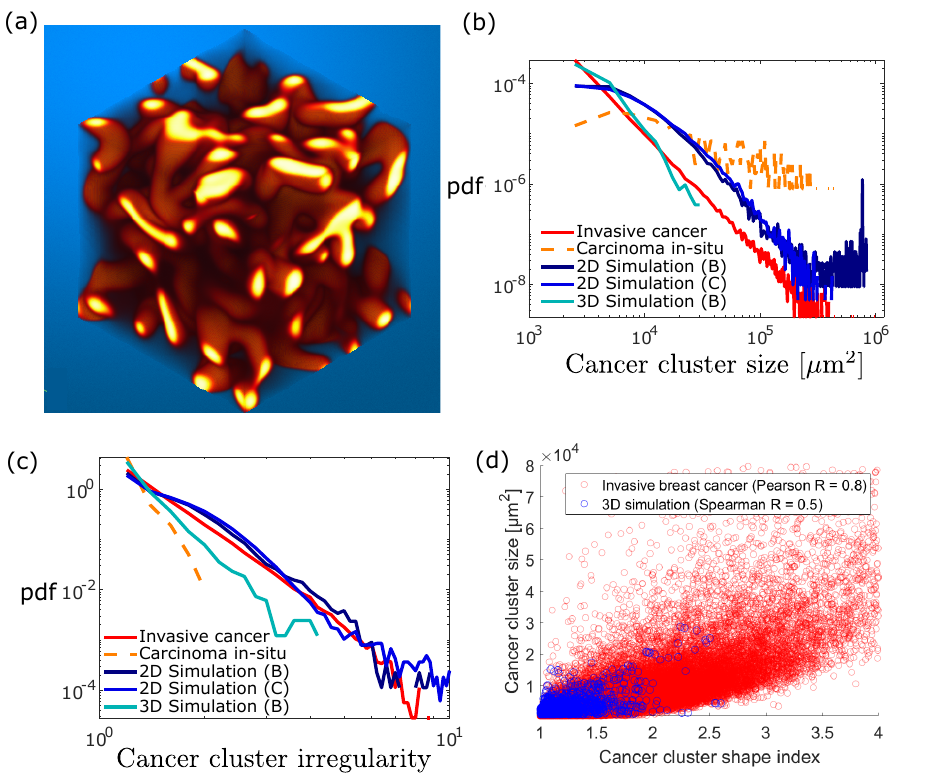}
\caption{(a) Typical snapshot of a simulated 3D active cluster in dynamical steady state. The phase field denoting the cancer clusters is shown in red in the bulk, and the yellow regions show the clusters at the boundary. (b) Size distribution of simulated 3D clusters shows good agreement with histological data for invasive breast cancer. In the legend, (B) and (C) denote different initial conditions for the simulations -- starting from a single droplet large droplet breaking apart, and coarsening from multiple small droplets evenly distributed, respectively. (c) Shape distribution of simulated 3D clusters shows a power law over a large range. The 3D simulation results underestimate the prevalence of clusters with large shape indices. (d) 3D simulations also show a strong correlation between cluster shapes and cluster sizes. However, our size limitations in 3D result in a smaller range of sizes and irregularity compared to the histological samples.}
  \label{fig:3D_sim_results}
\end{figure}

We performed 3D simulations of a single large active nematic droplet breaking up into a dynamical steady state. For simplicity, the confinement was taken to be a viscous fluid.

We used analogous parameter values to the 2D simulations. Simulations were run in a cubic domain of size $128 \times 128 \times 128$ LB units with periodic boundary conditions, starting from a single large active drop of radius 45 LB units (corresponding to an area coverage of $18\%$). After a burn-in time of 500 LB timesteps, the simulations were run for 80000 LB timesteps, with snapshots taken every 4000 steps. The activity values averaged over correspond to $\zeta = \{0.08, 0.1, 0.12, 0.14, 0.16\}$, which is comparable to the activity range used for the 2D simulations. Due to the long running time of 3D simulations, the box edge length and the number of simulated timesteps are smaller than those used for the 2D simulations.

A 3D simulation snapshot from the simulation is shown in Fig.~\ref{fig:3D_sim_results}a (see also Supplementary Movie S6 \cite{SM} for the dynamics). We then took 2D cross-sections of the full 3D simulations at regular intervals, along all three axes, and then calculated the area and shape indices for each cluster projection. We find that the cross-section of 3D clusters show similar distributions of shapes and sizes as the purely 2D simulations and the clinical data (Fig.~\ref{fig:3D_sim_results} b--c). There is a strong positive correlation between cluster shapes and sizes (Fig.~\ref{fig:3D_sim_results}d). 

Due to the limitation of 3D simulation runtime, we were only able to simulate box sizes in 3D smaller than the largest clusters observed histologically (Fig.~\ref{fig:3D_sim_results} d). Therefore, we do not observe large clusters in our current 3D simulations. This also leads to underestimating the distribution of shape indices (irregularity), since larger clusters are associated with more irregular shapes in both histology and simulations.

\section{Nematic Order Parameter of the ECM \label{SI:NonZero_SECM}}

\begin{figure}[h]
\centering
\includegraphics[width=0.5 \textwidth]{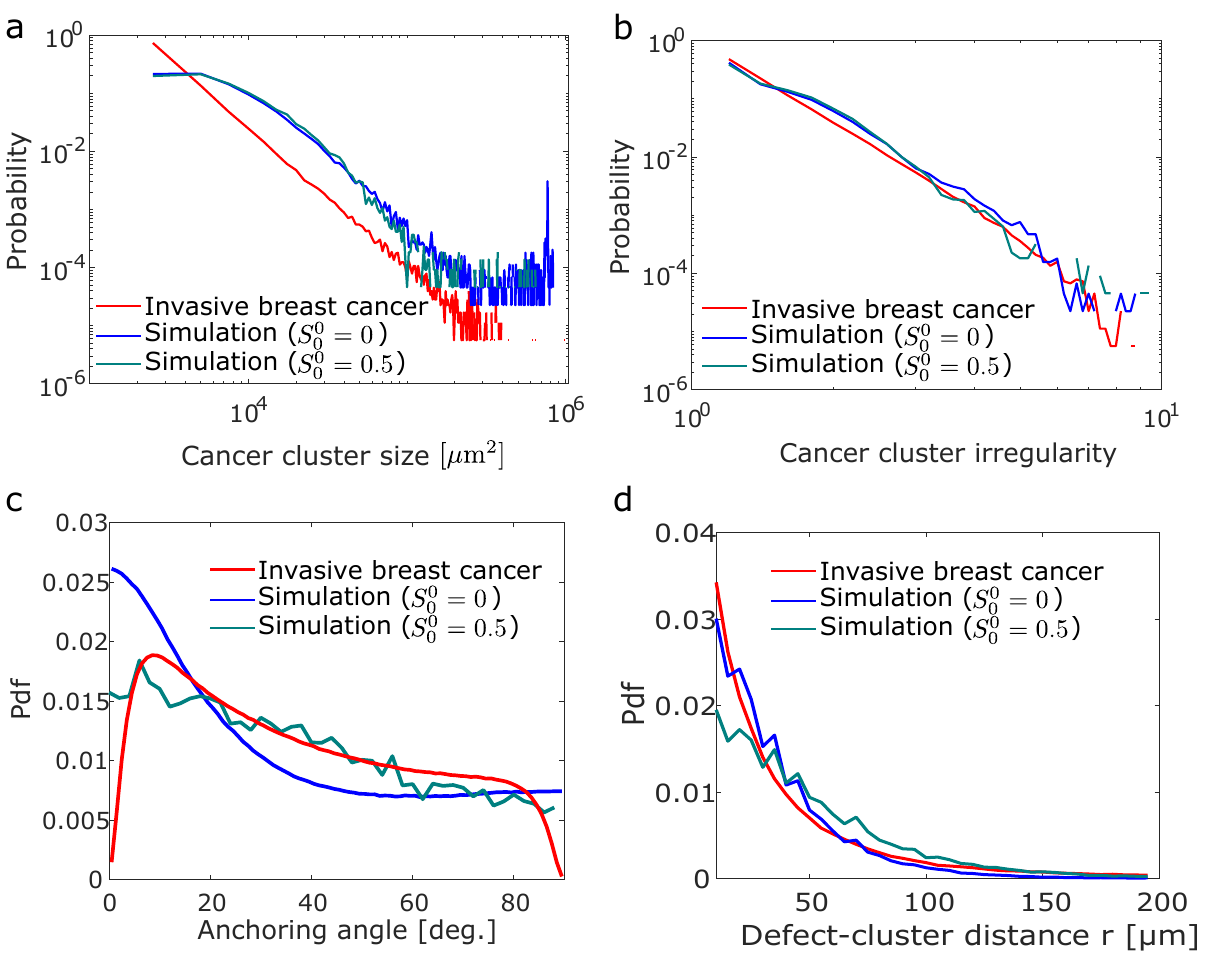}
\caption{(a) Cancer cluster size distribution for simulations with $S_0^0 = 0.5$ and (b) distribution of cancer cluster shape (irregularity)  for simulations with $S_0^0 = 0.5$ are very similar to results presented in the main text ($S_0^0 = 0$) (c) The angle made by the ECM at the cancer cluster boundary for $S_0^0 = 0.5$ follows the same qualitative trends as the main text ($S_0^0=0$), but the magnitude of the variation is reduced. (d) Topological defects in the ECM are also distributed exponentially away from the cancer cluster, but the distribution is flatter with a larger cutoff length.}
  \label{fig:SECM_0p5}
\end{figure}
In the main text, we use no inherent nematic order in the ECM ($S_0^0=0$) to isolate the effect of active flows in the ECM. In this section, we show that adding an inherent nematic order ($S_0^0\neq 0$) due to a free energy does not change the results significantly for the given parameters. 

We find that the shape and size distributions of the cancer clusters remain unchanged when using $S_0^0 = 0.5$ (Fig.~\ref{fig:SECM_0p5}a--b). However, while the ECM anchoring at cancer clusters still has a preference for parallel anchoring (Fig.~\ref{fig:SECM_0p5}c), the magnitude of the variation is reduced. The defect density decays exponentially away from the cancer clusters as before, but a higher $S_0^0$ leads to a smaller overall number of topological defects. This is consistent with larger active flows required to bend the nematic field, since a non-zero $S_0^0$ introduces a higher  energy penalty for distortions in the ECM ($\sim K {S_0^0}^2$).  

One important caveat for these results is that the chosen free-energy parameters (as mentioned in the main text, except $S_0^0 = 0.5$) are weak i.e. the thermodynamic response of the ECM is low, and the active flows play a dominant role in the dynamics of the ECM in the vicinity of the cancer cluster.

\section{Cancer cluster size and shape correlation \label{app:sizeshapecorrelation}}

To characterise the relationship between cluster size and shape index beyond the Spearman/distance coefficient reported in the main text, we evaluate five complementary association measures (Table \ref{tab:sizeShapeAssoc}). The Spearman and Kendall coefficients are rank-based measures of monotonic association, the Pearson coefficient measures linear association in log-size, and the linear-fit $R^2$ quantifies how well a straight line in log-size describes the shape index. These four measures are all sensitive only to monotonic (or linear) relationships. The distance correlation \cite{szekely2007measuring} complements them since it captures non-monotonic dependencies. Were the size–shape relationship substantially non-monotonic, the distance correlation would exceed the rank-based measures; instead it closely matches them for both invasive breast cancer (0.82 vs. Spearman 0.80) and the simulation (0.69 vs. 0.71), confirming that the coupling is strong and monotonic both in invasive breast cancer patients and simulation. \\
The carcinoma \textit{in situ} control shows no association by any measure (Table \ref{tab:sizeShapeAssoc}), as expected for clusters whose sizes are set by proliferation and confinement rather than motility-driven fission.

\begin{table}[ht]
    \centering
    \caption{Association between cluster size and shape index for invasive breast cancer, simulation, and carcinoma \textit{in situ}. All rank/linear coefficients are highly significant ($p<10^{-3}$) except for carcinoma \textit{in situ} ($p>0.05$). Kendall's $\tau$ and the distance correlation are evaluated on a random subsample ($n_{\text{sub}}=4000$ for invasive and simulation, $500$ for \textit{in situ}). \label{tab:sizeShapeAssoc}}
    \begin{tabular}{lccc}
    \toprule
    & \textbf{Invasive} & \textbf{Simulation} & \textbf{\textit{in situ}} \\
    \midrule
    Spearman $\rho$        & $0.80$ & $0.71$ & $0.06$ \\
    Kendall $\tau$         & $0.61$ & $0.52$ & $0.04$ \\
    Pearson (log-size)     & $0.84$ & $0.68$ & $0.09$ \\
    Linear fit $R^2$       & $0.70$ & $0.47$ & $0.01$ \\
    Distance correlation   & $0.82$ & $0.69$ & $0.13$ \\
    \bottomrule
    \end{tabular}
\end{table}

\section{Dependence of the average cluster shape index on the activity-to-elasticity ratio \label{app:shape_ratio}}

\begin{figure}[h]
\centering
\includegraphics[width=0.2\textwidth]{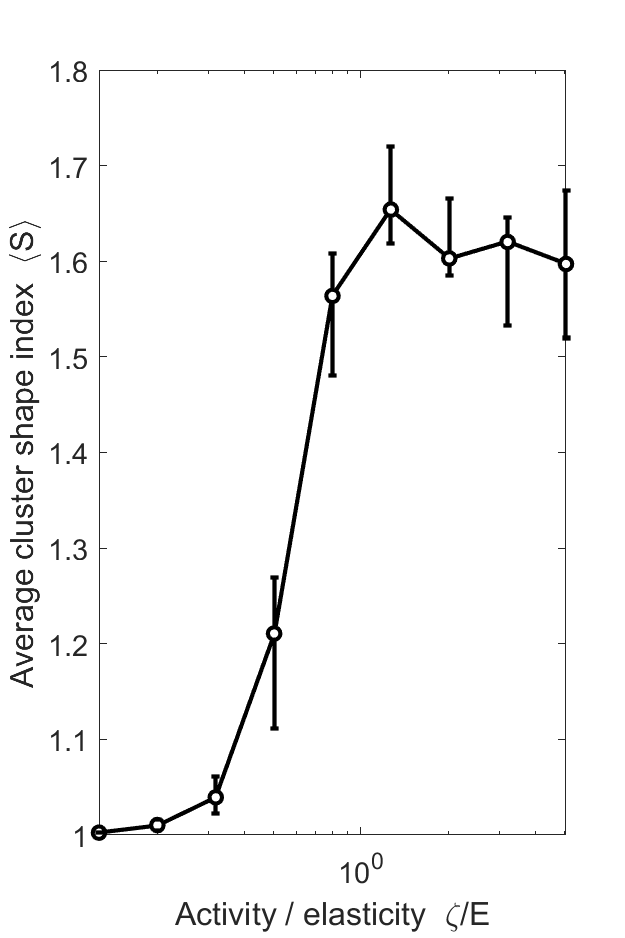}
\caption{\textbf{Average cluster shape index as a function of the activity-to-elasticity ratio $\zeta/E$.} Binned median shape index ($\pm 95\%$ bootstrap confidence interval) of single breaking clusters from the phase-diagram simulations (Fig.~\ref{fig:simulations}). The shape index is maximal near $\zeta/E \approx 1$, where active stresses balance the restoring forces from ECM elasticity and surface tension, and decreases at higher $\zeta/E$ as clusters fragment into smaller, rounder daughter clusters.}
\label{fig:shapeRatio}
\end{figure}

To quantify how the average cluster shape index depends on the two competing mechanical stresses, we collapse the phase-diagram data (Fig.~\ref{fig:simulations}) onto the single control parameter $\zeta/E$, the ratio of the cluster activity $\zeta$ to the ECM elasticity $E$. Fig.~\ref{fig:shapeRatio} shows the binned median shape index ($\pm 95\%$ bootstrap confidence interval) as a function of $\zeta/E$. The shape index rises with increasing $\zeta/E$, reaches a maximum near $\zeta/E \approx 1$, and decreases slightly at higher ratios. \\
This non-monotonic behaviour follows from the balance of stresses acting on a single cluster. Active stresses generated by collective cell motility ($\propto \zeta$) deform the cluster and increase its shape index, while the surface tension imposed by the ECM elasticity ($\propto E$) restores a compact shape. The maximum shape index is realized where, locally in time and space, the active stresses are balanced against these restoring forces: at $\zeta/E \approx 1$ the clusters attain their most deformed, high-shape-index configurations without yet breaking. Beyond this balance point, the active stresses overcome surface tension and fragment the cluster into smaller, individually rounder daughter clusters, which lowers the per-cluster average shape index; at sufficiently high activity this reduction is modest, since the daughter clusters are themselves driven towards deformation. An analogous balance, expressed through an active capillary number, governs the shape of an unconfined active nematic droplet \cite{giomi2014spontaneous}; here the elastic restoring force of the ECM introduces an additional memory effect that shifts the balance point and precludes a closed-form treatment. \\
We emphasise that Fig.~\ref{fig:shapeRatio} and Fig.~\ref{fig:simulations} report the \textit{average} shape index of a single breaking cluster across parameter space, a quantity distinct from the single-cluster steady-state size--shape relationship of Fig.~\ref{fig:sizeshape}c.

\section{Analyses of 3D histology of invasive breast cancer\label{app:3DClusterShapeSize}}
To assess whether the two-dimensional cluster size and shape distributions
reported in the main text are representative of the underlying three-dimensional
tumour architecture, we reconstructed a 3D volume from 100 serial histological
sections of an invasive breast cancer sample (final isotropic voxel size
$4\,\mu$m; see Methods, Sec.~\ref{methods:3Dreconstruction}).
Cancer-cell clusters were segmented from the haematoxylin channel and identified
as 26-connected components in 3D (compare Methods, Sec.~\ref{methods:3DClusterAnalyses}). Clusters touching any face of a subvolume were discarded to avoid truncation artefacts, and clusters with volumes below a sphere of radius $10\,\mu$m (single-cell scale) were excluded, leaving $N = 34385$ clusters pooled across subvolumes. For each cluster we measured its volume as the voxel count and its 3D shape index as $S_{3D}=A^{3/2}/(6\sqrt{\pi}\,V)$, where the surface area $A$ was computed from a smoothed isosurface mesh to suppress voxel-staircase artefacts; $S_{3D}=1$ for a sphere. In parallel we measured the corresponding 2D cross-sections on every section plane, using the same shape definition $S_{2D}=P/\sqrt{4\pi A}$ as in the main text.\\

\begin{figure}[htp]
\centering
\includegraphics[width=0.5\textwidth]{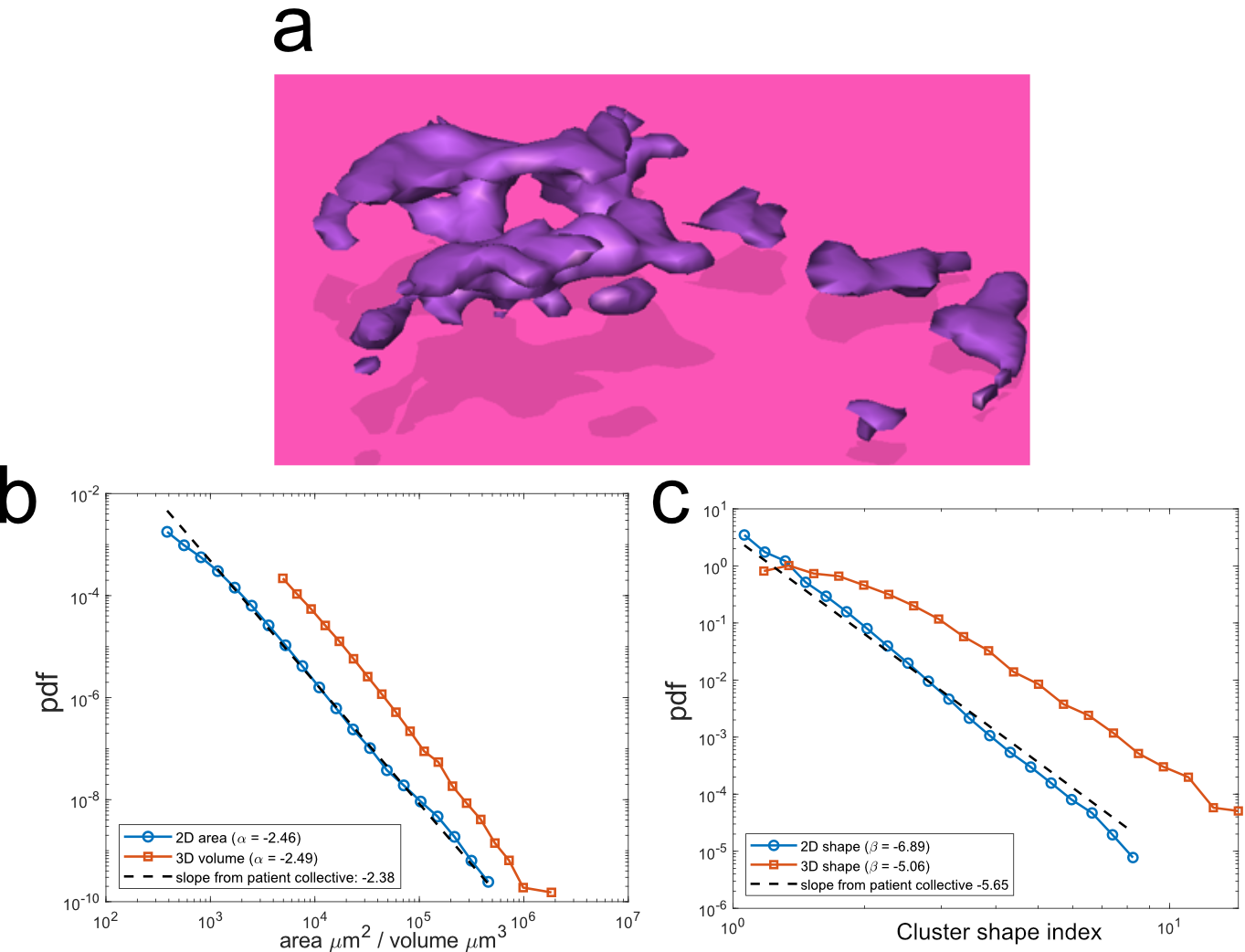}
\caption{2D and 3D cluster size and shape distributions of invasive breast cancer. \textbf{a} Rendering of disconnected 3D clusters from 3D reconstruction of an invasive breast cancer histological data. Image width is $\approx 400 \mu$m. \textbf{b} Size distributions of the 2D and 3D clusters. \textbf{c} Distribution of shape index for the 2D and 3D clusters.}
\label{fig:3DHistologyAppendix}
\end{figure}

The 3D cluster-volume distribution follows an approximate power law
$P(V)\sim V^{\alpha_{3D}}$ with $\alpha_{3D}=-2.49$ (Fig. \ref{fig:3DHistologyAppendix}b, fit over
$5\times10^{3}$--$1\times10^{5}\,\mu$m$^3$, $R^2=0.99$). Measured on the same
reconstructed volume, the 2D cross-sectional area distribution gives
$\alpha_{2D}=-2.46$ ($R^2=0.99$), close to the 3D exponent and
in line with the patient-cohort value reported in the main text ($\alpha=-2.38$).\\
The 2D shape index distribution decays with exponent $\beta_{2D}=-6.28$ ($R^2=0.99$), consistent with the value ($\beta=-5.65$) from the patient collective, while the 3D shape index decays more slowly, with $\beta_{3D}=-2.66$ ($R^2=0.97$). \\
Overall, these analyses show that the invasive lesions indeed consist of individual cancer clusters within the ECM, disconnected in 3D. The statistics are governed by small clusters.

\section{Supplementary Movies}

\subsection{Movie S1}
Caption: Simulation data showing the dynamics of a single large active cancer droplet (dark pink) in a viscoelastic ECM (light pink). Active forces cause the cancer droplet to split apart into many tiny clusters, which form a dynamical steady state. In this state, individual cancer clusters move around, split and merge, however, the statistical properties of the clusters overall do not change with time. 

\subsection{Movie S2}
Caption: Simulation data showing the dynamics of many small active cancer droplets (dark pink) in a viscoelastic ECM (light pink). The droplets move and grow until they reach a dynamical steady state, identical to the one reached from a single large droplet initial condition. In this state, individual cancer clusters move around, split and merge, however, the statistical properties of the clusters overall do not change with time. 

\subsection{Movie S3a--e}
Caption: Simulation data showing the dynamics of a single active cancer droplet (dark pink) in a viscoelastic ECM (light pink), with parameters corresponding to the representative points shown in Fig. 4 in the main text. All simulations are run from the same initial configuration. 

\subsection{Movie S4}
Caption: Simulation data showing the merging of two touching circular droplets of identical radius. The left subpanel shows the two clusters (deep pink) and the ECM (light pink) at any given time. The right subpanel plots the evolution of the aspect ratio of the bounding box as a function of time.

\subsection{Movie S5}
Caption: Movie of a vital primary breast cancer explant after surgery. The nuclei are stained and tracked as described in Methods, Sec.~\ref{sec:Methods:ImagingStainingWorkflow}. The movie represents a 2D plane from 3D data for better visualization of the data. It is clearly visible that a large fraction of cells is motile due to cell migration and not proliferation (we do not find a proliferation event in the 3D data by visual inspection, which is in accordance with the proliferation fractions we measured in our breast cancer patient collective).

\subsection{Movie S6}
Caption: Movie showing a simulated 3D cancer cluster breaking apart into many small clusters, exhibiting a dynamical steady state. Red regions show the cancer clusters, and yellow regions highlight the 2D cross-sections of the cancer clusters at the faces of the (periodic) box.

\subsection{Movie S7}
Caption: Movie showing a simulated cancer cluster breaking apart, along with ECM nematic order, velocity field and vorticity field. In the top left panel, $\phi=1$ represents the cancer cluster and $\phi=0$ represents the ECM. In the top right panel, red and green dots label $+1/2$ and $-1/2$ nematic topological defects in the ECM, respectively.

\bibliography{main_methods_bib}

\end{document}